\documentclass[11pt]{article}

\usepackage{fourier}
\usepackage[margin=1in,footskip=0.25in]{geometry} \usepackage[dvipsnames,table]{xcolor}             \DeclareMathAlphabet{\mathcal}{OMS}{cmsy}{m}{n}

\usepackage[sort&compress,square,comma,numbers]{natbib} \usepackage[pagebackref]{hyperref}                \usepackage{url}                                  \usepackage{breakcites}                           

\usepackage[nosumlimits]{amsmath} \usepackage{amsthm}               \usepackage{aliascnt}             \usepackage{amssymb}              \usepackage{amsfonts}             \usepackage{bm}                   \usepackage{bbm}                  \usepackage{mathtools}            \usepackage{nicefrac}             \usepackage{stmaryrd}             \usepackage{pifont}              

\usepackage{graphicx}       \usepackage{float}          \usepackage{caption}        \usepackage{subcaption}     \usepackage{booktabs}       \usepackage[most]{tcolorbox}  \usepackage{pgfplots}
\pgfplotsset{compat=1.18}
\usetikzlibrary{arrows.meta,patterns,patterns.meta,decorations.markings,decorations.pathreplacing,positioning,calc}

\usepackage{algorithm}      \usepackage{algpseudocode}  \usepackage{algorithmicx}   

\usepackage{enumitem}       \usepackage{xspace}         \usepackage{relsize}        \usepackage{titlesec}       \usepackage[normalem]{ulem} \usepackage{soul}           \usepackage{contour}        \usepackage{cancel}         \usepackage{todonotes}      \usepackage[capitalize,nameinlink,noabbrev]{cleveref} \usepackage{crossreftools}

\hypersetup{
    colorlinks=true,        linkcolor=PineGreen,    citecolor=NavyBlue,     filecolor=Purple,       urlcolor=Purple         }

\renewcommand*{\backref}[1]{}
\renewcommand*{\backrefalt}[4]{\ifcase #1 (No citations.)\or
        (Cited p.~#2.)\else
        (Cited pp.~#2.)\fi
}

\titlespacing*{\section}{0pt}{12pt}{5pt}
\titlespacing*{\subsection}{0pt}{11pt}{5pt}
\titlespacing*{\subsubsection}{0pt}{11pt}{5pt}
\titlespacing*{\paragraph}{0pt}{6pt}{1em}
\floatstyle{plaintop}
\restylefloat{table}

\algrenewcommand\alglinenumber[1]{\sf\scriptsize\color{NavyBlue}{#1}} 

\let\oldthebibliography\thebibliography
\renewcommand{\thebibliography}[1]{\oldthebibliography{#1}\setlength{\itemsep}{0pt}\setlength{\parskip}{0pt}}

\makeatletter
\newcommand{\sharedtheorem}[2]{\newaliascnt{#1}{thm}\newtheorem{#1@orig}[#1]{#2}\NewDocumentEnvironment{#1}{o +b}{\IfNoValueTF{##1}{\rep@typeset{#1@orig}{}{##2}}{\rep@typeset{#1@orig}{##1}{##2}}}{}\aliascntresetthe{#1}\crefname{#1}{#2}{#2s}\Crefname{#1}{#2}{#2s}\crefname{#1@orig}{#2}{#2s}\Crefname{#1@orig}{#2}{#2s}}

\newcommand{\rep@typeset}[3]{\def\rep@note{#2}\def\rep@body{#3}\edef\rep@eqstart{\the\c@equation}\protected@edef\rep@sec{\thesection}\let\rep@savedlabel\label
  \def\label##1{\rep@record{##1}\rep@savedlabel{##1}}\ifx\rep@note\@empty\begin{#1}\else\begin{#1}[#2]\fi
  #3\end{#1}}
\newcommand{\rep@record}[1]{\immediate\write\@auxout{\string\rep@store{#1}{\unexpanded\expandafter{\rep@note}}{\unexpanded\expandafter{\rep@body}}{\rep@eqstart}{\unexpanded\expandafter{\rep@sec}}}}
\long\def\rep@store#1#2#3#4#5{\global\@namedef{rep@note@#1}{#2}\global\@namedef{rep@body@#1}{#3}\global\@namedef{rep@eq@#1}{#4}\global\@namedef{rep@sec@#1}{#5}}
\makeatother
\sharedtheorem{theorem}{Theorem}
\sharedtheorem{proposition}{Proposition}
\sharedtheorem{conjecture}{Conjecture}
\sharedtheorem{corollary}{Corollary}
\sharedtheorem{lemma}{Lemma}
\sharedtheorem{fact}{Fact}
\sharedtheorem{claim}{Claim}
\sharedtheorem{assumption}{Assumption}
\sharedtheorem{setting}{Setting}
\sharedtheorem{observation}{Observation}
\sharedtheorem{problem}{Problem}
\sharedtheorem{example}{Example}
\sharedtheorem{importedtheorem}{Imported Theorem}
\sharedtheorem{importedlemma}{Imported Lemma}
\sharedtheorem{incompletetheorem}{Incomplete Theorem}

\theoremstyle{definition}
\sharedtheorem{definition}{Definition}

\theoremstyle{remark}
\sharedtheorem{remark}{Remark}

\crefname{theorem}{Theorem}{Theorems}
\Crefname{theorem}{Theorem}{Theorems}\crefname{corollary}{Corollary}{Corollaries}\Crefname{corollary}{Corollary}{Corollaries}\crefname{corollary@orig}{Corollary}{Corollaries}\Crefname{corollary@orig}{Corollary}{Corollaries}
\Crefname{equation}{Equation}{Equations}

\theoremstyle{theorem}
\makeatletter
\newtheorem*{rep@theorem}{\rep@title}
\newcommand{\newreptheorem}[2]{\newenvironment{rep#1}[1]{\def\rep@title{\Cref{##1}, restated}\begin{rep@theorem}}{\end{rep@theorem}}}
\makeatother

\newreptheorem{theorem}{Theorem}
\newreptheorem{lemma}{Lemma}
\newreptheorem{definition}{Definition}
\newreptheorem{corollary}{Corollary}

\makeatletter
\newcount\rep@uses \newcommand{\restatetheorem}[1]{\global\advance\rep@uses\@ne
  \@ifundefined{rep@body@#1}{\@latex@warning{Restated result `#1' not found; rerun LaTeX}\begin{reptheorem}{#1}\textbf{??}\end{reptheorem}}{\edef\rep@eqsaved{\the\c@equation}\begingroup
       \def\label##1{}\def\ltx@label##1{}\edef\thesection{\@nameuse{rep@sec@#1}}\setcounter{equation}{\@nameuse{rep@eq@#1}}\edef\theHequation{restated.\the\rep@uses.\noexpand\arabic{equation}}\expandafter\ifx\csname rep@note@#1\endcsname\@empty
         \begin{reptheorem}{#1}\else
         \begin{reptheorem}{#1}[\@nameuse{rep@note@#1}]\fi
       \@nameuse{rep@body@#1}\end{reptheorem}\endgroup
     \setcounter{equation}{\rep@eqsaved}}}
\makeatother

\newcommand{\actionbox}[1]{\begin{tcolorbox}[colback=white,colframe=black,width=\columnwidth,boxsep=5pt,arc=4pt] \emph{#1} \end{tcolorbox}}

\contourlength{0.8pt}
\newcommand{\unline}[1]{\uline{\phantom{#1}}\llap{\contour{white}{#1}}}

\makeatletter
\def\hlinewd#1{\noalign{\ifnum0=`}\fi\hrule \@height #1 \futurelet
	\reserved@a\@xhline}
\makeatother

\DeclareMathOperator{\diag}{diag}

\DeclareMathOperator{\Span}{span}

\newcommand{\defeq}{\ensuremath{\;{\vcentcolon=}\;}\xspace}
\renewcommand{\Re}{\mathrm{Re}} \renewcommand{\Im}{\mathrm{Im}} \newcommand{\eps}[0]{\varepsilon} \let\epsilon\eps

\DeclarePairedDelimiter{\norm}{\lVert}{\rVert}
 \newcommand{\lrnorm}[1]{\left\|#1\right\|}

\newcommand{\mat}[1]{\bm{#1}}    \renewcommand{\vec}[1]{\bm{#1}}

\newcommand{\mA}{\ensuremath{\mat{A}}\xspace}

\newcommand{\mI}{\ensuremath{\mathbf{I}}\xspace}

\newcommand{\mU}{\ensuremath{\mat{U}}\xspace}

\newcommand{\mX}{\ensuremath{\mat{X}}\xspace}

\newcommand{\mLambda}{\ensuremath{\mat{\varLambda}}\xspace}
\newcommand{\mDelta}{\ensuremath{\mat{\varDelta}}\xspace}

\newcommand{\mPi}{\ensuremath{\mat{\Pi}}\xspace}

\newcommand{\vb}{\ensuremath{\vec{b}}\xspace}

\newcommand{\ve}{\ensuremath{\mathbf{e}}\xspace}

\newcommand{\vg}{\ensuremath{\vec{g}}\xspace}
\newcommand{\vh}{\ensuremath{\vec{h}}\xspace}

\newcommand{\vx}{\ensuremath{\vec{x}}\xspace}
\newcommand{\vy}{\ensuremath{\vec{y}}\xspace}
\newcommand{\vz}{\ensuremath{\vec{z}}\xspace}

\newcommand{\cA}{\ensuremath{{\mathcal A}}\xspace}
\newcommand{\cB}{\ensuremath{{\mathcal B}}\xspace}
\newcommand{\cC}{\ensuremath{{\mathcal C}}\xspace}
\newcommand{\cD}{\ensuremath{{\mathcal D}}\xspace}

\newcommand{\cI}{\ensuremath{{\mathcal I}}\xspace}

\newcommand{\cK}{\ensuremath{{\mathcal K}}\xspace}

\newcommand{\cO}{\ensuremath{{\mathcal O}}\xspace}
\newcommand{\cP}{\ensuremath{{\mathcal P}}\xspace}
\newcommand{\cQ}{\ensuremath{{\mathcal Q}}\xspace}
\newcommand{\cR}{\ensuremath{{\mathcal R}}\xspace}
\newcommand{\cS}{\ensuremath{{\mathcal S}}\xspace}
\newcommand{\cT}{\ensuremath{{\mathcal T}}\xspace}

\newcommand{\cW}{\ensuremath{{\mathcal W}}\xspace}

\newcommand{\bbC}{\ensuremath{{\mathbb C}}\xspace}

\newcommand{\bbR}{\ensuremath{{\mathbb R}}\xspace}

\newcommand{\rC}{\ensuremath{\mathrm{C}}\xspace}

\newcommand{\rM}{\ensuremath{\mathrm{M}}\xspace}

\newcommand{\ri}{\ensuremath{\mathrm{i}}\xspace}

\newcommand{\e}{\mathrm{e}}
\renewcommand{\top}{\intercal}
\numberwithin{equation}{section}  \newcommand{\order}{\cO}

\renewcommand{\tilde}[1]{\widetilde{#1}}
\renewcommand{\hat}[1]{\widehat{#1}}

\NewDocumentCommand{\uinorm}{sO{}m}{\IfBooleanTF{#1}{\left|\opnormkern\left|\opnormkern\left|
    #3
    \right|\opnormkern\right|\opnormkern\right|
  }{
    \mathopen{#2|\opnormkern #2|\opnormkern #2|}
    #3
    \mathclose{#2|\opnormkern #2|\opnormkern #2|}
  }}
\newcommand{\opnormkern}{\mkern-1.5mu\relax}

\renewcommand{\d}[1]{\,\mathrm{d}#1}

\DeclareMathOperator{\cond}{cond}
\newcommand{\scond}{\operatorname{cond}_{\rm eig}}

\makeatletter
\newcommand{\@measuredangle@draw}[1]{\begin{tikzpicture}[baseline=0pt, line width=0.06em, line cap=round]
    \draw (#1*0.72,0) -- (0,0) -- (#1*0.66,#1*0.68);
    \draw (-12:#1*0.42) arc[start angle=-12, end angle=58, radius=#1*0.42];
  \end{tikzpicture}}
\DeclareRobustCommand{\measuredangle}{\mathord{\mathchoice
  {\@measuredangle@draw{1em}\kern0.06em}
  {\@measuredangle@draw{1em}\kern0.06em}
  {\@measuredangle@draw{0.7em}\kern0.04em}
  {\@measuredangle@draw{0.5em}\kern0.03em}}}
\makeatother

\renewcommand{\defeq}{\coloneqq}
\newcommand{\sector}{circular sector\xspace}

\title{A new angle on matrix--vector complexity}

\newcommand{\affilsep}{\textsuperscript{,}}
\AtBeginDocument{\let\origmaketitle\maketitle
  \renewcommand{\maketitle}{\begingroup\let\rlap\mbox\origmaketitle\endgroup}}
\author{Cecilia Chen\thanks{University of California, Berkeley, Berkeley, CA 94720, USA}
  \and
  Ethan N. W. Epperly\footnotemark[1]\affilsep\thanks{Lawrence Berkeley National Laboratory, 1 Cyclotron Road, Berkeley, CA 94720, USA}
  \and
  Michael W. Mahoney\footnotemark[1]\affilsep\footnotemark[2]\affilsep\thanks{International Computer Science Institute, Berkeley, CA 94704, USA}
  \and
  Raphael A. Meyer\footnotemark[1]\affilsep\footnotemark[2]\affilsep\footnotemark[3]
  \and
  John Urschel\thanks{Massachusetts Institute of Technology, 77 Massachusetts Avenue, Cambridge, MA 02139, USA}
  \end{tabular}\\[1ex]
  \begin{tabular}[t]{c}
    \normalsize\texttt{\{cdchen,\,eepperly,\,ram900\}@berkeley.edu}, \texttt{mmahoney@stat.berkeley.edu}, \texttt{urschel@mit.edu}}
\date{\today}

\begin{document}

\maketitle

\begin{abstract}
    The theory of worst-case matrix--vector complexity has been useful for guiding the design of fast linear algebra algorithms.
    However, there is a large theory--practice gap in the study of \emph{one-sided} methods, which use matrix--vector products with $\mA$ but not $\mA^\top$.
    In practice, for fundamental problems like solving linear systems, one-sided methods like GMRES are often the fastest.
    By contrast, worst-case analysis often show pessimistic results, for instance showing that one-sided methods cannot solve a linear system with ideal condition number 1 unless they read the whole matrix $\mA$.
    Such worst-case complexity results do not reflect real-world experience.
    
    To address this issue, this paper proposes a different theory of worst-case matrix--vector complexity that includes information on the positioning of the eigenvalues of $\mA$ \emph{in the complex plane}.
    Specifically, we characterize the ``well-behavedness'' of a matrix \mA through its \emph{spectral condition number} $\kappa = |\lambda_{\rm max}(\mA)| / |\lambda_{\rm min}(\mA)|$ and its \emph{angular eigenvalue gap $\theta$}, defined as the angle of the largest sector not containing an eigenvalue.
    Under these parameters, we establish matching complexity upper and lower bounds, showing the one-sided complexity of the linear system problem is $\Theta(\theta^{-2}\kappa^{\pi/\theta})$.
    These results interpolate between known complexity results for positive definite matrices ($\theta = 2\pi$), symmetric indefinite matrices ($\theta = \pi$), and general matrices ($\theta = 0$). 
    This complexity result provides a rigorous explanation for the success of one-sided methods in practice, as it shows that one-sided methods beat the ``optimal'' two-sided algorithm (CGN, complexity $\Theta(\kappa)$) when the eigenvalues of $\mA$ lie in less than half of the complex plane ($\theta > \pi$).
    Our analysis relies on tools from numerical analysis, several subfields of complex analysis, and information-theoretic complexity, and the results are presented in a modular way to facilitate future use.

\end{abstract}

\section{Introduction}

The notion of \emph{worst-case analysis} sits at the heart of the theory of computation.
By certifying how an algorithm performs on \emph{every} input, worst-case analysis offers simple, reliable guidance to practitioners about which methods make the most efficient use of computational resources.

However, when studying problems involving \emph{numerical quantities}, the appropriate notion of a worst-case input can become unclear.
Consider, e.g., the task of computing a \((1\pm\eps)\)-approximation to \(\int_{0}^1 f(x) \, \d x\) from queries to a nonnegative function \(f:[0,1]\to\bbR_+\).
It is straightforward to see that no algorithm can accomplish this goal in the worst case with finitely many queries, since the mass of \(f\) can be concentrated on an arbitrarily small subset of the interval.
To remedy this, we have to assume that the function \(f\) is \emph{well-behaved}. 
For instance, we can assume that the derivative of \(f\) is bounded \cite{bakhvalov1959approximate,traub1988information}.

Different settings inevitably call for different complexity measures that quantify how ``well-behaved'' a problem is.
Signal processing methods may assume that data is generated from a signal with bounded maximum frequency \cite{shannon1949communication,tropp2009beyond,candes2014towards}.
Matrix completion methods often assume their data is coherent \cite{recht2011simpler,candes2012exact}. 
Eigenvalue algorithms regularly assume their input matrices have gaps between their eigenvalues \cite{kuczyski1992estimating,saad2011numerical,meyer2024unreasonable,chen2026does}.
Provided we have identified a descriptive measure of complexity, worst-case analysis helps us decide which algorithm for the problem is ``best''.
Contrapositively, if worst-case analysis fails to describe the behavior of algorithms in practice, then that indicates the correct complexity parameter(s) have yet to be identified.

Traditionally, TCS research on linear algebra has relied on coarse complexity measures for linear algebra problems.
Often these measures depend only on the singular values, like the condition number (ratio of largest and smallest singular values) or the minimum gap between singular values \cite{peng2024solving,meyer2024unreasonable,derezinski2026matrix}.
However, the understanding from the numerical analysis community is that, when working with \emph{nonsymmetric} matrices, the behavior of many algorithms is described by the positioning of eigenvalues \emph{on the complex plane} \cite{driscoll1998from,greenbaum1997iterative,trefethen2019approximation}.
(Recall that even a real nonsymmetric matrix generally has complex eigenvalues.)
We argue that, to have a useful TCS-style worst-case analysis, we need complexity measures that reflect this positioning.

The importance of eigenvalue positioning to complexity can be most vividly understood by studying the class of \emph{one-sided matrix--vector algorithms}, which access an input matrix \(\mA\in\bbR^{n\times n}\) through the forward matrix--vector product (matvec) \(\vx\mapsto\mA\vx\) but not the transposed product \(\vx\mapsto\mA^\top\vx\).
Such algorithms are widely used, and there are many reasons to study them; see below.
However, there exists a large theory--practice gap in the study of one-sided matrix--vector complexity.

Under the existing theory, which typically uses the condition number as the sole complexity parameter for solving linear systems, no one-sided matrix--vector algorithm can beat the trivial cost of $\Theta(n)$ matvecs for solving linear systems $\mA \vx = \vb$, even when \mA is perfectly conditioned \cite{derezinski2026matrix}.
Two-sided methods, which are also allowed to use the transposed product, can solve systems of bounded condition number by passing to the normal equations \(\mA^\top\mA\vx=\mA^\top\vb\), for which \(\mA^\top\mA\) always has positive \emph{real} eigenvalues.
Nevertheless, one-sided algorithms work very well in practice and often greatly outperform those two-sided methods.
So, we must conclude that new complexity parameters are needed to characterize the complexity of linear systems for one-sided methods.

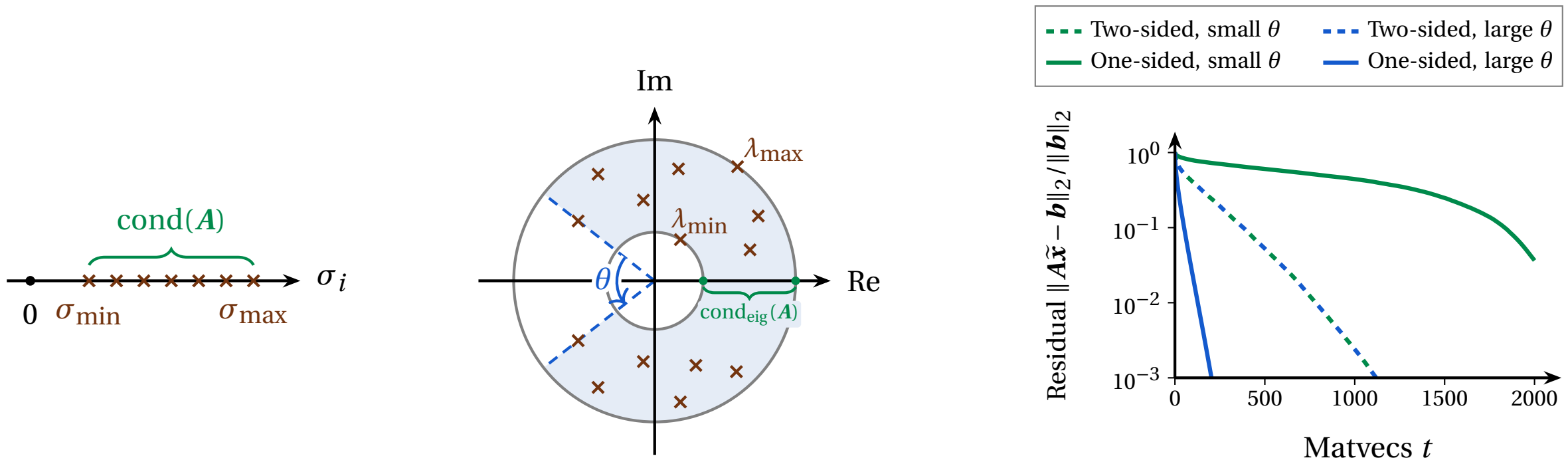
\begin{figure}[t]
    \centering
    \definecolor{EigBrown}{RGB}{115,52,15}
    \definecolor{SmallThetaGreen}{RGB}{0,138,75}
    \definecolor{LargeThetaBlue}{RGB}{20,90,205}

    \def\crossSize{0.072}
    \newcommand{\svdcross}[2]{\draw[EigBrown, thick]
            ($(#1,#2)+(-\crossSize,-\crossSize)$) -- ++({2*\crossSize},{2*\crossSize});
        \draw[EigBrown, thick]
            ($(#1,#2)+(-\crossSize,\crossSize)$) -- ++({2*\crossSize},{-2*\crossSize});}
    \newcommand{\eigcrosspolar}[2]{\draw[EigBrown, thick]
            ($({#1}:{#2})+(-\crossSize,-\crossSize)$) -- ++({2*\crossSize},{2*\crossSize});
        \draw[EigBrown, thick]
            ($({#1}:{#2})+(-\crossSize,\crossSize)$) -- ++({2*\crossSize},{-2*\crossSize});}

\begin{subfigure}[b]{0.25\textwidth}
        \centering
        \begin{tikzpicture}[
            scale=0.78,
            every node/.style={font=\footnotesize}
        ]
\path (-0.40,-2.40) rectangle (3.85,2.50);

\draw[-{Stealth[length=5pt]}, black, thick] (-0.30,0) -- (3.45,0)
                node[right=1pt, text=black] {$\sigma_i$};

\filldraw[black] (0,0) circle (1.5pt);
            \node[anchor=north, text=black] at (0,-0.14) {$0\vphantom{\sigma_{\min}}$};

\foreach \xpos in {0.75, 1.10, 1.45, 1.80, 2.15, 2.50, 2.85} {
                \svdcross{\xpos}{0}
            }

\node[anchor=north, text=EigBrown] at (0.75,-0.14) {$\sigma_{\min}$};
            \node[anchor=north, text=EigBrown] at (2.85,-0.14) {$\sigma_{\max}$};

\draw[SmallThetaGreen, thick, decorate, decoration={brace, amplitude=4.5pt}]
                (0.75,0.20) -- (2.85,0.20)
                node[midway, above=5.5pt, text=SmallThetaGreen] {$\cond(\mA)$};
        \end{tikzpicture}
        \caption{Singular values}
        \label{fig:intro-overview-svd}
    \end{subfigure}
    \hfill
\begin{subfigure}[b]{0.32\textwidth}
        \centering
        \def\Rmin{0.62}
        \def\Rmax{1.80}
        \def\gapHalf{38}
        \begin{tikzpicture}[
            scale=0.78,
            every node/.style={font=\footnotesize}
        ]
            \path (-2.45,-2.40) rectangle (2.65,2.50);

\fill[NavyBlue!9]
                ({180-\gapHalf}:\Rmax)
                arc[start angle={180-\gapHalf}, end angle={\gapHalf-180}, radius=\Rmax]
                -- ({\gapHalf-180}:\Rmin)
                arc[start angle={\gapHalf-180}, end angle={180-\gapHalf}, radius=\Rmin]
                -- cycle;

\draw[gray, thick] (0,0) circle (\Rmin);
            \draw[gray, thick] (0,0) circle (\Rmax);

\draw[LargeThetaBlue, thick, densely dashed] (0,0) -- ({180-\gapHalf}:\Rmax);
            \draw[LargeThetaBlue, thick, densely dashed] (0,0) -- ({180+\gapHalf}:\Rmax);

\draw[-{Stealth[length=5pt]}, black, thick] (-2.25,0) -- (2.25,0)
                node[right=1pt, text=black] {$\Re$};
            \draw[-{Stealth[length=5pt]}, black, thick] (0,-2.22) -- (0,2.22)
                node[above=1pt, text=black] {$\Im$};

\draw[white, line width=2.6pt, line cap=butt, shorten >=1.2pt]
                ({180-\gapHalf}:0.48) arc[start angle={180-\gapHalf}, end angle={180+\gapHalf}, radius=0.48];
            \draw[->, LargeThetaBlue, thick]
                ({180-\gapHalf}:0.48) arc[start angle={180-\gapHalf}, end angle={180+\gapHalf}, radius=0.48];
            \contourlength{1.2pt}
            \node[LargeThetaBlue, inner sep=0.5pt] at (180:0.65) {\contour{white}{$\theta$}};

\filldraw[SmallThetaGreen] (\Rmin,0) circle (1.3pt);
            \filldraw[SmallThetaGreen] (\Rmax,0) circle (1.3pt);
            \draw[SmallThetaGreen, thick, decorate, decoration={brace, mirror, amplitude=3.2pt}]
                (\Rmin,-0.08) -- (\Rmax,-0.08)
                node[midway, below=3.5pt, fill=NavyBlue!9, inner sep=0.6pt, rounded corners=1pt, text=SmallThetaGreen, font=\fontsize{6.3}{7.5}\selectfont] {$\scond(\mA)$};

\eigcrosspolar{58}{\Rmin}
            \node[EigBrown, font=\scriptsize, inner sep=1pt] at (53:0.96) {$\lambda_{\min}$};

            \eigcrosspolar{54}{\Rmax}
            \node[EigBrown, font=\scriptsize, above right=0.5pt, inner sep=1pt] at (54:\Rmax) {$\lambda_{\max}$};

\eigcrosspolar{180-\gapHalf}{1.24}
            \eigcrosspolar{180+\gapHalf}{1.24}

\eigcrosspolar{18}{1.28}
            \eigcrosspolar{32}{1.56}
            \eigcrosspolar{78}{1.46}
            \eigcrosspolar{98}{1.04}
            \eigcrosspolar{118}{1.54}
            \eigcrosspolar{-118}{1.54}
            \eigcrosspolar{-98}{1.04}
            \eigcrosspolar{-78}{1.58}
            \eigcrosspolar{-64}{1.20}
            \eigcrosspolar{-48}{1.56}
        \end{tikzpicture}
        \caption{Eigenvalues}
        \label{fig:intro-overview-eig}
    \end{subfigure}
    \hfill
\begin{subfigure}[b]{0.40\textwidth}
        \centering
        \begin{tikzpicture}[
            every node/.style={font=\footnotesize}
        ]
            \begin{semilogyaxis}[
                name=convaxis,
                width=3.85cm,
                height=2.45cm,
                scale only axis,
                axis x line=bottom,
                axis y line=left,
                axis on top,
                axis line style={-{Stealth[length=5pt]}, black, thick},
                xmin=0, xmax=2150,
                ymin=1e-3, ymax=1.9,
                unbounded coords=jump,
                xtick={0, 500, 1000, 1500, 2000},
                xticklabels={$0$, $500$, $1000$, $1500$, $2000$},
                ytick={1e-3, 1e-2, 1e-1, 1e0},
                yticklabels={$10^{-3}$, $10^{-2}$, $10^{-1}$, $10^{0}$},
                tick align=outside,
                tick pos=left,
                major tick length=2pt,
                xminorticks=false,
                yminorticks=false,
                tick style={black, thin},
                tick label style={font=\fontsize{7}{8.4}\selectfont, text=black, inner sep=1.5pt},
                xlabel={Matvecs $t$},
                ylabel={Residual $\norm{\mA\tilde{\vx} - \vb}_2 / \norm{\vb}_2$},
                xlabel style={text=black, font=\footnotesize, at={(axis description cs:0.5,-0.19)}, anchor=north},
                ylabel style={text=black, font=\scriptsize, at={(axis description cs:-0.24,0.5)}, anchor=south}
            ]
\addplot[SmallThetaGreen, very thick, dash pattern=on 2.2pt off 5.8pt, each nth point={5}] table[x=t, y=two_small, col sep=comma] {figures/gmres_vs_cg.csv};
                \addplot[LargeThetaBlue, very thick, dash pattern=on 2.2pt off 5.8pt, dash phase=4.0pt, each nth point={5}] table[x=t, y=two_large, col sep=comma] {figures/gmres_vs_cg.csv};
                \addplot[SmallThetaGreen, very thick, each nth point={5}] table[x=t, y=one_small, col sep=comma] {figures/gmres_vs_cg.csv};
                \addplot[LargeThetaBlue, very thick, each nth point={2}] table[x=t, y=one_large, col sep=comma] {figures/gmres_vs_cg.csv};
            \end{semilogyaxis}

\coordinate (plotwest) at (current bounding box.west);
            \coordinate (ploteast) at (current bounding box.east);
            \coordinate (legendnorth) at ([yshift={3pt+1em}]convaxis.north);
            \path let \p1 = (plotwest), \p2 = (ploteast), \p3 = (legendnorth), \n1 = {\x2 - \x1} in
                node[
                    draw=gray,
                    thin,
                    fill=white,
                    inner sep=2.5pt,
                    text width={\n1 - 5.4pt},
                    anchor=south west,
                    font=\fontsize{6.8}{8.2}\selectfont
                ] at (\x1,\y3) {\renewcommand{\arraystretch}{1.08}\setlength{\tabcolsep}{0pt}\begin{tabular*}{\linewidth}{@{\extracolsep{\fill}}ll@{}}
                        \tikz[baseline=-0.5ex]\draw[SmallThetaGreen, very thick, dash pattern=on 2.2pt off 1.8pt] (0pt,0pt)--(10pt,0pt); Two-sided, small $\theta$ &
                        \tikz[baseline=-0.5ex]\draw[LargeThetaBlue, very thick, dash pattern=on 2.2pt off 1.8pt] (0pt,0pt)--(10pt,0pt); Two-sided, large $\theta$ \\
                        \tikz[baseline=-0.5ex]\draw[SmallThetaGreen, very thick] (0pt,0pt)--(10pt,0pt); One-sided, small $\theta$ &
                        \tikz[baseline=-0.5ex]\draw[LargeThetaBlue, very thick] (0pt,0pt)--(10pt,0pt); One-sided, large $\theta$
                    \end{tabular*}};
        \end{tikzpicture}
        \caption{Solver convergence
}
        \label{fig:intro-overview-conv}
    \end{subfigure}

    \caption{
        Visualization of three aspects of matrices and their conditioning.
        \textbf{Subfigure (a)}:
The singular values of a matrix are always positive real numbers, and the condition number \(\cond(\mA)\) is the ratio of the largest to smallest singular value.
        \textbf{Subfigure (b)}:
For general matrices, the eigenvalues may lie anywhere on the complex plane.
        The spectral condition number \(\scond(\mA)\) is the ratio of the largest to smallest absolute value of eigenvalues.
        The angular gap \(\theta=\protect\measuredangle(\mA)\) is the angle of the largest circular sector on the complex plane that has no eigenvalues in it.
The blue shaded area, parameterized by the pair \((\scond(\mA),\protect\measuredangle(\mA))\), is called a \emph{polar rectangle}.
\textbf{Subfigure (c)}: We ran a one-sided method (GMRES) and a two-sided method (CGN) on two matrices, both of spectral condition number \(\kappa=200\).
        One matrix has a small angular gap of \(\pi/2\), and the other has a large gap \(3\pi/2\).
        Consistent with our theory, the matrix with a large gap converges faster with a one-sided method and the matrix with a smaller gap is faster with a two-sided method.
}
    \label{fig:intro-overview}
\end{figure}
 
In this article, we take linear system solving as our running example.
We demonstrate how to build an informative worst-case theory for the complexity of solving linear systems using one-sided matvec methods.
We do so by characterizing eigenvalue positioning with \emph{two} parameters:
the \emph{spectral condition number}
\begin{equation*}
    \scond(\mA)
\coloneqq \frac{\max_i |\lambda_i(\mA)|}{\min_i |\lambda_i(\mA)|}
\end{equation*}
and the \emph{angular eigenvalue gap} 
\[
    \measuredangle(\mA) 
\defeq \max\left\{
        ~
        \theta \in(0,2\pi] ~:~
        \text{there exists a sector of angle \(\theta\) containing no eigenvalues of \mA}
        ~
    \right\}.
\]
See \cref{fig:intro-overview} for a visualization.
By using these two parameters, we build a descriptive theory of one-sided matrix--vector algorithms.
Our main result can be summarized as follows.

\begin{theorem}[One-sided matrix--vector complexity of $\mA\vx = \vb$, informal]
    \label{thm:informal}
    Consider the class of normal matrices \mA with angular eigenvalue gap \(\measuredangle(\mA)\ge \theta\).
    \begin{enumerate}[label=(\alph*)]
        \item \label{item:informal-bounded} \textbf{Bounded spectral condition number.} Among such matrices with bounded spectral condition number \(\scond(\mA)\leq\kappa\), the one-sided matrix--vector complexity of solving linear systems to residual accuracy \(\varepsilon\) is \(t_\star = \Theta(\theta^{-2}\kappa^{\pi/\theta}\log(1/\eps))\).
        \item \label{item:informal-unbounded} \textbf{Unbounded spectral condition number.}
        For any such matrix, for any constant \(\theta\), the one-sided matrix--vector complexity of solving linear systems to backward accuracy \(\varepsilon\) is \(t_\star = \Theta_{\theta}(\varepsilon^{-\pi/\theta})\).
\end{enumerate}
\end{theorem}

This theorem contains two results, which use different error metrics (residual vs.\ backward error) and make different assumptions (bounded vs.\ unbounded spectral condition number).
In the first, we show that if the condition number is bounded, then the complexity depends logarithmically on the desired accuracy, with a prefactor $\theta^{-2}\kappa^{\pi/\theta}$ that depends on both the angular eigenvalue gap and the spectral condition number.
In the second result, we show that if the condition number is unbounded, then the complexity depends polynomially on the desired accuracy at a rate controlled by the angular eigenvalue gap.
Both results treat the class of complex normal matrices, consisting of all matrices with orthonormal eigenvectors.
For non-normal matrices, we obtain similar complexity upper bounds that contain a factor depending on the condition number of the eigenvector matrix.

\Cref{thm:informal} may be compared to the known complexity results for two-sided methods, which show that conjugate gradient on the normal equations (CGN) $\mat{A}^\top\mat{A}\vec{x} = \mat{A}^\top \vec{b}$ is optimal among the class of matrices with $\cond(\mA) \le \kappa$, with complexity $\Theta(\kappa\log(1/\eps))$ \cite{derezinski2026matrix,shamir2026simple}.
In particular, we obtain the following conclusion.
\actionbox{
    \vspace{-1.5em}
    \begin{center}
        The optimal one-sided solver (GMRES) achieves lower complexity than the ``optimal'' two-sided algorithm (CGN) if and only if the eigenvalues lie in at most \unline{half} of the complex plane ($\,\theta > \pi\!$).
\end{center}
}
\noindent This result provides a clear, complexity-theoretic explanation for the observed success of one-sided methods over two-sided methods in practice, and it provides a precise characterization of when and why one-sided methods are advantageous.

We note that messages somewhat similar to this idea have a history in numerical analysis.
When the eigenvalues fill at most half of the complex plane, Richardson iteration and several other Krylov methods are known to converge \cite{eiermann1989semiiterative,choi2015roots} and convergence bounds for GMRES are known for several eigenvalue geometries \cite{saad1986gmres,driscoll1998from,embree2022how}.
Further, many preconditioners for nonsymmetric problems are built with the intent of clustering eigenvalues onto as small a region on the complex plane as possible 
\cite{erlangga2006comparison,sidi2003zero,benzi2005numerical}.
Nonetheless, we are unaware of a reference in numerical analysis that characterizes the complexity of one-sided algorithms using the angular eigenvalue gap or a reference that clearly expresses the boxed maxim.
In particular, this paper is the first to establish \emph{necessary and sufficient conditions on eigenvalue positioning} for one-sided algorithms to be effective.

\subsection{Matrix--vector algorithms and complexity: One-sided and two-sided} \label{sec:intro-matvec}

\begin{table}[t]
    \centering
    \small
    \setlength{\tabcolsep}{4pt}
    \renewcommand{\arraystretch}{1.3}
    \begin{tabular}{@{}lcllccc@{}}
        \toprule
        & & & & \multicolumn{3}{c}{Relevant citations} \\
        \cmidrule(l){5-7}
        Matrix \(\mA\) & Access & Optimal matvecs & Achieved by & Alg. & Upper bd. & Lower bd. \\
        \midrule
        Psd
            & --
            & \(\Theta(\sqrt\kappa \log(1/\eps))\)
            & MINRES on \(\mA\vx=\vb\)
            & {\scriptsize\cite{paige1975solution}} & {\scriptsize \cite{meinardus1963uber,kaniel1966estimates}} & {\scriptsize \cite{nemirovskij1983problem,braverman2020gradient,derezinski2026matrix}} \\
        Symmetric
            & --
            & \(\Theta(\kappa \log(1/\eps))\)
            & CG on \(\mA^\top\mA\vx=\mA^\top\vb\)
            & {\scriptsize\cite{hestenes1952methods}} & {\scriptsize \cite{meinardus1963uber,kaniel1966estimates}} & {\scriptsize \cite{traub1988information,derezinski2026matrix}} \\
        General
            & two-sided
            & \(\Theta(\kappa \log(1/\eps))\)
            & CG on \(\mA^\top\mA\vx=\mA^\top\vb\)
            & {\scriptsize\cite{hestenes1952methods}} & {\scriptsize \cite{meinardus1963uber,kaniel1966estimates}} & {\scriptsize \cite{traub1988information,derezinski2026matrix}} \\
        General
            & one-sided
            & \(\Theta(n)\) \textcolor{gray}{ even with \(\kappa=1\)}
            & GMRES on \(\mA\vx=\vb\)
            & {\scriptsize\cite{saad1986gmres}} & {\scriptsize --} & {\scriptsize \cite{nachtigal1992how,derezinski2026matrix}} \\
            \midrule
        General
            & one-sided
            & $\Theta(\theta^{-2}\kappa^{\pi/\theta}\log(1/\varepsilon))$
            & GMRES on $\mA\vx = \vb$
            & {\scriptsize\cite{saad1986gmres}} & This paper & This paper
        \\
        \bottomrule
    \end{tabular}
    \caption{Worst-case matrix--vector complexity for \(\mA\vx=\vb\) to residual accuracy $\norm{\mA\tilde\vx - \vb}_2 \le \varepsilon \norm{\vb}_2$ for any normal matrix \mA with condition number \(\operatorname{cond}(\mA)\le\kappa\).
    On the last row, we additional assume that \(\measuredangle(\mA)\geq\theta\).
    Recent work \cite{derezinski2026matrix} provides the sharp lower bound for general randomized algorithms for each setting; see \cite{nemirovskij1983problem,traub1988information,nachtigal1992how,braverman2020gradient} for earlier work.
    For psd and symmetric systems, there is no difference between one-sided and two-sided methods.
    }
    \label{tab:matvec-complexity}
\end{table}

A quite substantial fraction of the fastest-known linear algebra algorithms are matrix--vector algorithms, which interact with the matrix \mA only via the matrix--vector product (matvec) operations $\vz \mapsto \mA\vz$ and $\vz \mapsto \mA^\top \vz$.
An algorithm is two-sided if it uses both primitives, and one-sided if it uses only the first primitive.
Examples of matrix--vector algorithms are numerous, including power and Krylov methods for eigenvalue problems \cite{lanczos1950iteration,kuczyski1992estimating,saad2011numerical}, the conjugate gradient algorithm and related methods for linear systems \cite{hestenes1952methods,greenbaum1997iterative,saad2003iterative,balabanov2025preconditioning}, stochastic trace estimation algorithms \cite{girard1987un,hutchinson1989stochastic,meyer2021hutch,persson2022improved,epperly2024xtrace}, and many more \cite{bakshi2022low,braverman2022sublinear,bakshi2023krylov,chen2024lanczos,chen2025near}.

The growing field of \emph{one-sided} or \emph{transpose-free} linear algebra \cite{boull2024operator,derezinski2026matrix,halikias2026transpose} seeks to understand the relative power of one- versus two-sided algorithms.
There are many motivations to understand the power of one-sided algorithms.
\begin{enumerate}[label=(\alph*)]
    \item \textbf{Transpose matvecs can be expensive or unavailable.} There are settings in which the transpose matvec is much more expensive than the forward matvec. For instance, on GPUs, the forward matvec may be cheaper than the transpose matvec for architectural reasons; and in large-scale machine learning applications, forward matvecs with the Jacobian of a neural network require a forward pass through the network, whereas transpose matvecs require backpropagation. In other applications, the transpose matvec may simply be unavailable. For instance, in operator learning, each matvec requires the solution of a PDE. The forward and transpose matvecs correspond to different PDEs, and one frequently has access to a solver only for one of them.
    \item \textbf{Primal vs.\ dual.} For the problem $\min_{\vx} \max_{\vy} \vy^\top \mA \vx + f(\vx) - g(\vy)$, forward matvecs access the gradient of the primal variable, and transpose matvecs access the dual. Therefore, one-sided algorithms are primal-only methods, and understanding their power is a natural question in optimization.
    \item \textbf{Explaining practice.} Even in settings where transpose matvecs are cheap and available, one-sided algorithms are often observed to be faster than the known two-sided algorithms in practice. Thus, understanding one-sided algorithms is important to understand algorithms in practice.
\end{enumerate}

The existing complexity theory for linear systems paints a bleak picture for one-sided algorithms; see \cref{tab:matvec-complexity} for a summary of existing results.
See \cite{halikias2026transpose} for further motivation on the study of one-sided algorithms.

\subsection{Why the angular gap helps: Polynomials on the complex plane} \label{sec:poly-complex-plane}

To understand the one-sided complexity of the linear system problem, it is helpful to consider the most widely used algorithm, known as the \emph{generalized minimum residual} (GMRES) method \cite{saad1986gmres}.
Ignoring matters of implementation, the method is simple:
First, it generates the Krylov subspace
\begin{equation} \label{eq:krylov}
    \mathcal{K}_t(\mA, \vb) \defeq \Span\{\vb, \mA\vb, \mA^2\vb, \dots, \mA^{t-1}\vb\}.
\end{equation}
Then, it outputs the vector \(\tilde\vx \in \mathcal{K}_t(\mA, \vb)\) within this space that minimizes the residual \(\norm{\mA\tilde\vx - \vb}_2\).
This strategy is natural: It repeatedly applies the matrix--vector oracle, then returns the best candidate within the span of the vectors generated.

Since the output of GMRES lies in the Krylov subspace, it takes the form $\tilde\vx = p(\mA)\vb$, where $p$ is a degree-$(t-1)$ polynomial.
In turn, the residual of GMRES is
\begin{equation*}
    \norm{\mA\tilde\vx - \vb}_2 = \norm{\vb - \mA p(\mA) \vb}_2 = \norm{(\mI - \mA p(\mA)) \vb}_2 = \norm{q(\mA)\vb}_2 \quad \text{where } q(z) = 1 - z p(z).
\end{equation*}
The polynomial $q$ has degree at most $t$, and it satisfies the condition $q(0) = 1$.
Hence, the GMRES algorithm identifies the polynomial that minimizes the norm of the vector $q(\mA)\vb$:
\begin{equation} \label{eq:gmres-optimality}
    \norm{\mA\tilde\vx - \vb}_2 = \min_{\vx \in \cK_t(\mA,\vb)} \norm{\mA\vx - \vb}_2 = \min_{\substack{\deg(q) \le t \\ q(0) = 1}} \norm{q(\mA)\vb}_2.
\end{equation}
Consequently, the GMRES algorithm succeeds if and only if there exists a polynomial for which $\norm{q(\mA)\vb}_2$ is small.
Since the eigenvalues of the matrix $q(\mA)$ are $q(\lambda_i(\mA))$, this amounts to showing that there exists a \emph{filter polynomial} $q(z)$ of degree \(t\) with $q(0) = 1$ while being small on the eigenvalues of \mA.
By trivial interpolation, this is always possible with a degree \(n\) polynomial.
The question is when such a polynomial \(q\) exists with \(t \ll n\).

With this polynomial framing in mind, it becomes clear why an angular gap is necessary for one-sided methods like GMRES to succeed.
Indeed, a straightforward application of the maximum modulus principle (\cref{fact:max-modulus}) shows that any polynomial satisfying $q(0) = 1$ cannot be uniformly small on a region that encircles the origin.

\begin{proposition}[Filter polynomials do not exist without an angular gap]
    Let \hangindent=-88pt \hangafter=-5\vadjust{\vbox to 0pt{\kern-1.05\baselineskip\kern0.15em\hbox to \textwidth{\hfill\begin{tikzpicture}[
    scale=0.58,
    every node/.style={font=\footnotesize},
    bluedotregion/.style={
        fill=NavyBlue!8,
        postaction={
            pattern={Dots[distance=4.2pt, radius=0.55pt]},
            pattern color=NavyBlue!65
        }
    }
]
\pgfmathsetmacro{\dotstep}{4.2 / (0.58 * 28.45274)}

\def\outerBlobPath{
        plot[smooth cycle, tension=0.75] coordinates {
            (1.08, 0.00)
            (0.76, 0.55)
            (0.36, 1.14)
            (-0.40, 1.18)
            (-1.10, 0.80)
            (-1.02, 0.00)
            (-0.86, -0.62)
            (-0.33, -1.02)
            (0.31, -0.96)
            (0.82, -0.60)
        }
    }

\def\innerBlobPath{
        (0.06, 0.03) ellipse [x radius=0.42, y radius=0.55]
    }

\path[bluedotregion, even odd rule] \outerBlobPath \innerBlobPath;

\fill[NavyBlue!8, rounded corners=1.2pt]
        ({-3.5*\dotstep}, {1.5*\dotstep}) rectangle ({-1.5*\dotstep}, {3.5*\dotstep});

\draw[->, gray] (-1.68,0) -- (1.75,0)
        node[right, text=black] {$\Re$};
    \draw[->, gray] (0,-1.68) -- (0,1.68)
        node[above, text=black] {$\Im$};

\draw[NavyBlue, very thick, dashed] \outerBlobPath;
    \draw[NavyBlue, very thick, dashed] \innerBlobPath;

\filldraw[black] (0,0) circle (1.5pt);
    \node[black, inner sep=0pt] at (0.20, 0.20) {$0$};
    \node[NavyBlue, inner sep=0pt] at ({-2.5*\dotstep}, {2.5*\dotstep}) {$\cS$};

\draw[gray!55, line width=0.4pt]
        ([shift={(-4pt,-3.5pt)}]current bounding box.south west)
        rectangle
        ([shift={(4pt,3.5pt)}]current bounding box.north east);
\end{tikzpicture}
 }\vss}} $\cS\subset\bbC$ be a set containing a simple closed curve enclosing the origin.
    Then any polynomial $q$ satisfying $q(0) = 1$ cannot be small on $\cS$: 
\begin{equation*}
        \max_{z \in \cS} |q(z)| \ge 1.
    \end{equation*}
\end{proposition}

That is, no filter polynomial exists on any set \cS that is allowed to encircle the origin, like the one shown to the right, and therefore GMRES can only succeed on all matrices \mA with eigenvalues in \cS if we take \(q\) to interpolate the eigenvalues of \mA exactly.
This recovers the trivial cost of \(t=n\) matvecs.
However, this point of failure vanishes if we insist upon an angular eigenvalue gap, as we formalize in the next section.

\subsection{Main result I: Bounded spectral condition number}

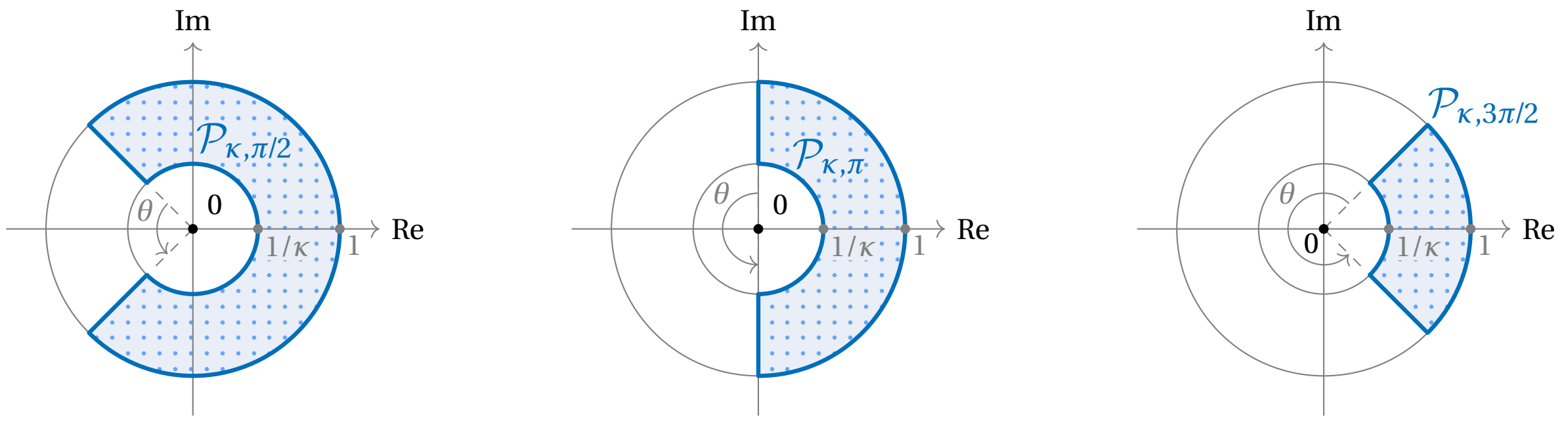
\begin{figure}[t]
    \centering
    \def\Rin{0.82}
    \def\Rout{1.85}
\pgfmathsetmacro{\dotstep}{4.5 / (0.80 * 28.45274)}

\begin{subfigure}[c]{0.31\textwidth}
        \centering
        \def\halfTheta{45}
        \begin{tikzpicture}[
            scale=0.80,
            every node/.style={font=\footnotesize},
            bluedotregion/.style={
                fill=NavyBlue!8,
                postaction={
                    pattern={Dots[distance=4.5pt, radius=0.60pt, xshift=1.6pt]},
                    pattern color=NavyBlue!65
                }
            }
        ]
            \coordinate (outerLower) at ({\halfTheta-180}:\Rout);
            \coordinate (outerUpper) at ({180-\halfTheta}:\Rout);
            \coordinate (innerUpper) at ({180-\halfTheta}:\Rin);
            \coordinate (innerLower) at ({\halfTheta-180}:\Rin);

\path[bluedotregion]
                (outerUpper)
                arc[start angle={180-\halfTheta}, end angle={\halfTheta-180}, radius=\Rout]
                -- (innerLower)
                arc[start angle={\halfTheta-180}, end angle={180-\halfTheta}, radius=\Rin]
                -- cycle;

\coordinate (PaPos) at ({3.3*\dotstep}, {5.5*\dotstep});
            \fill[NavyBlue!8, rounded corners=1.2pt]
                ({0.355*\dotstep}, {4.50*\dotstep}) rectangle ({3.355*\dotstep}, {6.70*\dotstep});
            \fill[NavyBlue!8, rounded corners=1.2pt]
                ({2.355*\dotstep}, {4.50*\dotstep}) rectangle ({6.355*\dotstep}, {5.70*\dotstep});
            \fill[NavyBlue!8, rounded corners=1.2pt]
                ({4.355*\dotstep}, {-2.00*\dotstep}) rectangle ({7.355*\dotstep}, {-0.08*\dotstep});

\draw[->, gray] (-2.35,0) -- (2.35,0)
                node[right, text=black] {$\Re$};
            \draw[->, gray] (0,-2.35) -- (0,2.35)
                node[above, text=black] {$\Im$};

\draw[gray] (0,0) circle (\Rin);
            \draw[gray] (0,0) circle (\Rout);

\draw[gray, dashed] (0,0) -- (innerUpper);
            \draw[gray, dashed] (0,0) -- (innerLower);
            \draw[->, gray]
                ({180-\halfTheta}:0.45) arc[start angle={180-\halfTheta}, end angle={180+\halfTheta}, radius=0.45];
            \node[gray] at (158:0.64) {$\theta$};

\draw[NavyBlue, very thick, line join=round]
                (outerUpper)
                arc[start angle={180-\halfTheta}, end angle={\halfTheta-180}, radius=\Rout]
                -- (innerLower)
                arc[start angle={\halfTheta-180}, end angle={180-\halfTheta}, radius=\Rin]
                -- cycle;

\node[NavyBlue, inner sep=0pt] at (PaPos) {\large $\cP_{\kappa,\pi\!/\!2}$};
            \filldraw[gray] (\Rin,0) circle (1.5pt);
            \filldraw[gray] (\Rout,0) circle (1.5pt);
            \node[gray, below right=1pt, inner sep=1pt] at (\Rin,0) {$1/\kappa$};
            \node[gray, below right=1pt, inner sep=1pt] at (\Rout,0) {$1$};
            \filldraw[black] (0,0) circle (1.5pt)
                node[above right=1pt] {$0$};
        \end{tikzpicture}
        \caption{
            \(\theta<\pi\): Hard for one-sided methods.
        }
        \label{fig:intro-annulus}
    \end{subfigure}
    \hfill
\begin{subfigure}[c]{0.33\textwidth}
        \centering
        \def\halfTheta{90}
        \begin{tikzpicture}[
            scale=0.80,
            every node/.style={font=\footnotesize},
            bluedotregion/.style={
                fill=NavyBlue!8,
                postaction={
                    pattern={Dots[distance=4.5pt, radius=0.60pt, xshift=1.6pt]},
                    pattern color=NavyBlue!65
                }
            }
        ]
            \coordinate (outerLower) at ({\halfTheta-180}:\Rout);
            \coordinate (outerUpper) at ({180-\halfTheta}:\Rout);
            \coordinate (innerUpper) at ({180-\halfTheta}:\Rin);
            \coordinate (innerLower) at ({\halfTheta-180}:\Rin);

\path[bluedotregion]
                (outerUpper)
                arc[start angle={180-\halfTheta}, end angle={\halfTheta-180}, radius=\Rout]
                -- (innerLower)
                arc[start angle={\halfTheta-180}, end angle={180-\halfTheta}, radius=\Rin]
                -- cycle;

\coordinate (PbPos) at ({4.55*\dotstep}, {4.5*\dotstep});
            \fill[NavyBlue!8, rounded corners=1.2pt]
                ({2.355*\dotstep}, {3.70*\dotstep}) rectangle ({5.355*\dotstep}, {5.70*\dotstep});
            \fill[NavyBlue!8, rounded corners=1.2pt]
                ({4.355*\dotstep}, {3.50*\dotstep}) rectangle ({6.355*\dotstep}, {4.70*\dotstep});
            \fill[NavyBlue!8, rounded corners=1.2pt]
                ({4.355*\dotstep}, {-2.00*\dotstep}) rectangle ({7.355*\dotstep}, {-0.08*\dotstep});

\draw[->, gray] (-2.35,0) -- (2.35,0)
                node[right, text=black] {$\Re$};
            \draw[->, gray] (0,-2.35) -- (0,2.35)
                node[above, text=black] {$\Im$};

\draw[gray] (0,0) circle (\Rin);
            \draw[gray] (0,0) circle (\Rout);

\draw[gray, dashed] (0,0) -- (innerUpper);
            \draw[gray, dashed] (0,0) -- (innerLower);
            \draw[->, gray]
                ({180-\halfTheta}:0.45) arc[start angle={180-\halfTheta}, end angle={180+\halfTheta}, radius=0.45];
            \node[gray] at (135:0.64) {$\theta$};

\draw[NavyBlue, very thick, line join=round]
                (outerUpper)
                arc[start angle={180-\halfTheta}, end angle={\halfTheta-180}, radius=\Rout]
                -- (innerLower)
                arc[start angle={\halfTheta-180}, end angle={180-\halfTheta}, radius=\Rin]
                -- cycle;

\node[NavyBlue, inner sep=0pt] at (PbPos) {\large $\cP_{\kappa,\pi}$};
            \filldraw[gray] (\Rin,0) circle (1.5pt);
            \filldraw[gray] (\Rout,0) circle (1.5pt);
            \node[gray, below right=1pt, inner sep=1pt] at (\Rin,0) {$1/\kappa$};
            \node[gray, below right=1pt, inner sep=1pt] at (\Rout,0) {$1$};
            \filldraw[black] (0,0) circle (1.5pt)
                node[above right=1pt] {$0$};
        \end{tikzpicture}
        \caption{
            \(\theta=\pi\): Tied with two-sided methods.
        }
        \label{fig:intro-polar-rect-panel}
    \end{subfigure}
    \hfill
\begin{subfigure}[c]{0.31\textwidth}
        \centering
        \def\halfTheta{135}
        \begin{tikzpicture}[
            scale=0.80,
            every node/.style={font=\footnotesize},
            bluedotregion/.style={
                fill=NavyBlue!8,
                postaction={
                    pattern={Dots[distance=4.5pt, radius=0.60pt, xshift=1.6pt]},
                    pattern color=NavyBlue!65
                }
            }
        ]
            \coordinate (outerLower) at ({\halfTheta-180}:\Rout);
            \coordinate (outerUpper) at ({180-\halfTheta}:\Rout);
            \coordinate (innerUpper) at ({180-\halfTheta}:\Rin);
            \coordinate (innerLower) at ({\halfTheta-180}:\Rin);

\path[bluedotregion]
                (outerUpper)
                arc[start angle={180-\halfTheta}, end angle={\halfTheta-180}, radius=\Rout]
                -- (innerLower)
                arc[start angle={\halfTheta-180}, end angle={180-\halfTheta}, radius=\Rin]
                -- cycle;

\fill[NavyBlue!8, rounded corners=1.2pt]
                ({4.355*\dotstep}, {-2.00*\dotstep}) rectangle ({7.355*\dotstep}, {-0.08*\dotstep});

\draw[->, gray] (-2.35,0) -- (2.35,0)
                node[right, text=black] {$\Re$};
            \draw[->, gray] (0,-2.35) -- (0,2.35)
                node[above, text=black] {$\Im$};

\draw[gray] (0,0) circle (\Rin);
            \draw[gray] (0,0) circle (\Rout);

\draw[gray, dashed] (0,0) -- (innerUpper);
            \draw[gray, dashed] (0,0) -- (innerLower);
            \draw[->, gray]
                ({180-\halfTheta}:0.45) arc[start angle={180-\halfTheta}, end angle={180+\halfTheta}, radius=0.45];
            \node[gray] at (135:0.64) {$\theta$};

\draw[NavyBlue, very thick, line join=round]
                (outerUpper)
                arc[start angle={180-\halfTheta}, end angle={\halfTheta-180}, radius=\Rout]
                -- (innerLower)
                arc[start angle={\halfTheta-180}, end angle={180-\halfTheta}, radius=\Rin]
                -- cycle;

\contourlength{1.5pt}
            \node[NavyBlue, inner sep=0pt] at (2.02,1.5) {\contour{white}{\large $\cP_{\kappa,3\pi\!/\!2}$}};
            \filldraw[gray] (\Rin,0) circle (1.5pt);
            \filldraw[gray] (\Rout,0) circle (1.5pt);
            \node[gray, below right=1pt, inner sep=1pt] at (\Rin,0) {$1/\kappa$};
            \node[gray, below right=1pt, inner sep=1pt] at (\Rout,0) {$1$};
            \filldraw[black] (0,0) circle (1.5pt)
                node[below left=0.5pt, inner sep=0.8pt] {$0$};
        \end{tikzpicture}
        \caption{
            \(\theta>\pi\): Fast for one-sided methods.
        }
        \label{fig:intro-psd}
    \end{subfigure}

    \caption{
        Visualization of the polar rectangle \(\cP_{\kappa,\theta}\) with different angle parameters \(\theta\).
        For any parameter \(\theta>0\), our main results \cref{thm:main-residual,thm:main-backward} show that one-sided methods converge to a solution using fewer than the trivial matvec complexity of \(\Omega(n)\) matvecs.
        Nonetheless, we can compare when one-sided methods are faster or slower than two-sided methods that solve the normal equations.
        For systems like \textbf{Subfigure (a)}, with a small angular gap, one-sided methods converge more slowly than methods solving the normal equations.
        At precisely \(\theta=\pi\), as shown in \textbf{Subfigure (b)}, one-sided methods are as fast as the two-sided methods.
        Once the angular gap gets larger, as shown in \textbf{Subfigure (c)}, one-sided methods provably outperform those methods based on the normal equations.
        This is consistent with the experimental plots in Subfigure (c) of \cref{fig:intro-overview}.
    }
    \label{fig:intro-polar-rect}
\end{figure}

Our first result treats matrices $\mA \in \bbC^{n\times n}$ that both have bounded spectral condition number $\scond(\mA) \le \kappa$ and non-trivial angular eigenvalue gap $\measuredangle(\mA) \ge \theta > 0$.
Any such matrix \mA can be scaled by a complex number so that the eigenvalues lie in the following set
\begin{equation*}
    \cP_{\kappa,\theta} \defeq \Bigl\{
        r \e^{\ri \phi}
        ~:~
        1/\kappa \le r \le 1, \: |\phi|\le \pi-\theta/2
    \Bigr\},
\end{equation*}
which we call a \emph{polar rectangle}; see \cref{fig:intro-polar-rect}.
Our main technical result establishes sharp upper and lower bounds on the degree needed to build a filter polynomial for this set.

\begin{theorem}[Filter polynomials on the polar rectangle]
    Fix \(\kappa \geq 600\), \(\theta\in(0,2\pi]\), and \(\eps\in(0,1)\).
    Let \(t_\star\) be the minimal degree $t$ for which there exists a degree-$t$ filter polynomial $q$ satisfying:
    \[
        q(0) = 1 \quad \text{and} \quad |q| \le \varepsilon \text{ on } \cP_{\kappa,\theta}.
    \]
    Then \(t_\star \asymp \theta^{-2} \kappa^{\pi/\theta}\log(2\pi/\eps)\).
\end{theorem}

As a consequence, we obtain a precise version of \cref{item:informal-bounded} of \cref{thm:informal}.

\begin{theorem}[Main result I: Bounded spectral condition number]\label{thm:main-residual}
    Consider a diagonalizable matrix $\mA = \mX\mLambda\mX^{-1}$ with spectral condition number at most $\kappa$ and angular eigenvalue gap at least $\theta$.
    Then the GMRES algorithm admits the complexity bound 
\begin{equation*}
        \norm{\mA\tilde\vx - \vb}_2 \le \varepsilon \norm{\vb}_2 \quad \text{after} \quad t = \left\lceil 1 + (\pi^3\e) \, \theta^{-2}\kappa^{\pi/\theta} \log \left( \frac{6\pi \cond(\mX)}{\varepsilon} \right)\right\rceil \text{ matvecs}.
    \end{equation*}
Conversely, any one-sided algorithm that produces such a $\tilde\vx$ with probability 2/3 for all normal matrices with this spectral condition number and angular eigenvalue gap must use \(t\gtrsim \theta^{-2}\kappa^{\pi/\theta}\log(1/\eps)\) matvecs.
\end{theorem}

Recall that a matrix is \emph{normal} if it has a set of orthonormal eigenvectors, which is equivalent to possessing an eigenvector matrix \mX of condition number one.
Therefore, this result shows that the matrix--vector complexity of solving $\mA\vx = \vb$ to $\varepsilon$-small residual error is $\Theta(\theta^{-2} \kappa^{\pi/\theta} \log(1/\varepsilon))$ for normal matrices.
For non-normal matrices, we obtain the same complexity result, up to a logarithmic factor in the eigenvector condition number.

It can be instructive to compare this complexity result for GMRES to other algorithms, focusing on the case where \mA is normal and \(\eps\) is constant for simplicity.
When $\theta = 2\pi$, the eigenvalues lie on a single ray, and the matrix is psd (up to scaling by a complex phase).
In this case, the complexity of GMRES is $\Theta(\sqrt{\kappa})$ matching 
the optimal complexity across all (possibly two-sided) matvec algorithms.
Next, recall that every symmetric matrix has real eigenvalues, and up to scaling therefore has eigenvalues in the real interval \(\cI_{\kappa} \defeq [-1,-1/\kappa]\cup[1/\kappa,1]\).
Up to another scaling, this is a subset of \(\cP_{\kappa,\pi}\).
Of course, not all matrices with angular eigenvalue gap \(\pi\) are real symmetric matrices.
Yet, the optimal complexities of one-sided methods on \(\cP_{\kappa,\pi}\) and of two-sided methods that solve symmetric linear systems are both \(\Theta(\kappa)\).
That is, we do not pay extra for having eigenvalues filling up half of the complex plane instead of just lying on the real axis.

The key threshold for comparing one-sided and two-sided methods is at \(\pi=\theta\).
When \(\pi<\theta\leq2\pi\), GMRES converges in fewer than \(\cO(\kappa)\) matvecs and is faster than any method which solves the normal equations.
On the other hand, when \(\theta<\pi\), one-sided methods provably require more matvecs than methods which solve the normal equations.

\subsection{Main result II: Unbounded spectral condition number}

For matrices \mA that have an unbounded or even infinite spectral condition number, there is no efficient matrix--vector algorithm, one- or two-sided, that can guarantee a small residual $\norm{\mA\tilde\vx - \vb}_2 \le \varepsilon \norm{\vb}_2$.
But there is a different natural concept of error---the \emph{backward error}---under which one can obtain non-trivial, $\kappa$-independent complexity results.

A vector \(\tilde\vx\) is called an \(\eps\)-backward error solution to \(\mA\vx=\vb\) if there exists a perturbation \(\mDelta\mA\) such that
\begin{equation} \label{eq:backward-error}
    (\mA + \mDelta\mA) \tilde\vx = \vb  \quad \text{and} \quad \norm{\mDelta\mA}_2 \le \varepsilon \norm{\mA}_2.
\end{equation}
The notion of backward error is sensible.
It reflects the fact that, when doing numerical computations involving matrices populated with data from the real world, the entries of the matrix \mA already come to us corrupted with some noise or measurement error, and we introduce further inaccuracy when we represent \mA using floating-point numbers.
Therefore, from the point of view of a practitioner, solving a system up to a suitably small perturbation to the input is (in a sense) the greatest possible accuracy an algorithm can achieve.
Backward error analysis dates back to the earliest days of computer science, featuring implicitly in works of von Neumann and Goldstine \cite{vonneumann1947numerical} and Turing \cite{turing1948rounding}, and Wilkinson received the Turing award in 1970 for his systematic development of backward error analysis.

Recent work of Derezi\'nski, Rebrova, and Nakatsukasa \cite{derezinski2026towards} exhibited a matrix--vector algorithm for solving a psd linear system to backward error \(\eps\) using $\order(\varepsilon^{-1/2})$ matvecs, independent of the condition number of \mA.
For general matrices, they produce a two-sided algorithm requiring $\order(\log(n) \cdot \varepsilon^{-1})$ matvecs.

Using our techniques, we develop condition number-independent results for one-sided linear system solving only under the assumption of an eigenvalue angle gap.

\begin{theorem}[Main result II: Unbounded spectral condition number]\label{thm:main-backward}
    Consider a diagonalizable matrix $\mA = \mX\mLambda\mX^{-1}$ with angular eigenvalue gap at least $\,\theta\!$.
    Then the GMBACK algorithm produces an $\varepsilon$-backward error solution after at most
    \begin{equation*}
        t \lesssim \theta^{-1} \left( \frac{2\cond(\mX)}{\varepsilon}\right)^{\pi / \theta} \text{ one-sided matrix--vector products}.
    \end{equation*}
Conversely, any one-sided algorithm that achieves such a guarantee with probability 2/3 for all normal matrices with this angular eigenvalue gap must use \(t\gtrsim \theta^{-1}\eps^{-\pi/\theta}\) matvecs.
\end{theorem}

Observe that, when $\theta = 2\pi$, the eigenvalues lie on the positive real line, and we recover the $\order(\varepsilon^{-1/2})$ complexity result by Derezi\'nski et al.
Generalizing the results from the psd case, we achieve backward error guarantees independently of the condition number of \mA.
Further, unlike the \(\cO(\log(n)\cdot\eps^{-1})\) cost of solving the normal equations, our guarantee has no dependence on the dimension \(n\).
So, as \(n\) grows large, \emph{for every fixed} \(\theta\) in the open interval \((0,2\pi)\), it is asymptotically faster to use a one-sided method than to use the method that is nominally optimal in the worst case.

Further, the work of Derezi\'nski, Mahoney, Meyer, and Musco \cite{derezinski2026from} shows that no \emph{deterministic} two-sided method can solve a general linear system with complexity independent of the condition number.
We show that, for every fixed \(\theta\) in the open interval \((0,2\pi)\), the deterministic GMBACK algorithm is fully able to avoid the dependence on the condition number.
Said another way, for deterministic methods, we have proven that we may choose to either run a two-sided method that is independent of \(\theta\) or a one-sided method that is independent of \(\kappa\).
However, it is not possible for deterministic methods to be fully independent of both.

\subsection{Technical overview}

To prove our main theorems (\cref{thm:main-residual,thm:main-backward}), we deploy and extend tools from numerical analysis \cite{greenbaum1997iterative,driscoll1998from}, potential theory \cite{axler2001harmonic,saff2010logarithmic}, complex analysis \cite{vuorinen2006conformal,beliaev2020conformal}, and information-theoretic matrix--vector complexity \cite{derezinski2026matrix,shamir2026simple}; see \cref{sec:related} for a discussion of prior work.
The analysis proceeds by first relating the matrix--vector complexity to 
existence guarantees of low-degree filter polynomials.
We connect these existence claims to a quantity from complex analysis called the \emph{Green's function}.
Finally, we use various tools from complex analysis to bound the Green's function.

For this introduction, we provide a brief overview of the key ideas in each step, with key theorems listed in \cref{tab:technical-overview}.
The arguments for the residual error and backward error are slightly different, so we first describe the argument for the residual error before describing the backward-error results.

\begin{table}[t]
    \centering
    \small
    \setlength{\tabcolsep}{4pt}
    \begin{tabular}{@{}llcb{2.6cm}b{2.8cm}b{2.5cm}@{}}
        \toprule
        Error metric & Bound & Main result & {From matvecs \newline to polynomials} & {Polynomials \newline to Green's function} &  {Bounding the \newline Green's function} \\
        \midrule
        Residual & Upper & \cref{thm:main-residual} & \cref{thm:gmres} & \cref{thm:poly-to-greens} & \cref{thm:harmonic-residual}\\
        Residual & Lower & \cref{thm:main-residual} & \cref{thm:lower-bound-forward} & \cref{thm:poly-to-greens} & \cref{thm:green-upper-polar-rect}\\[0.25em]
        Backward & Upper & \cref{thm:main-backward} & \cref{thm:gmberr} & \cref{thm:backward-green} & \cref{thm:harmonic-backward} \\
        Backward & Lower & \cref{thm:main-backward}\({}^\dagger\)  & -- & -- & -- \\
        \bottomrule
    \end{tabular}
    \caption{Overview of results. Our general pipeline is to reduce the question of matvec complexity to a question of the degree required for polynomial approximation. Then we relate the degree of approximation to the value of a specially chosen Green's function. Finally, we produce bounds on the Green's function to complete the main result. Note that any imported theorems are prior results from the literature.
    The dagger \({}^{\dagger}\) indicates that the lower bound in \cref{thm:main-backward} does not directly connect to a polynomial approximation problem and is instead proven by reduction using \cref{impthm:matvec-back-to-residual}.
    }
    \label{tab:technical-overview}
\end{table} 
\subsubsection{Step 1: Matrix--vector complexity and polynomials}

As described in \cref{sec:poly-complex-plane}, there are strong reasons to suspect that the complexity of solving $\mA\vx = \vb$ for a matrix with eigenvalues in a set \cS should be comparable to the minimum degree $t_\star$ needed to produce a \emph{filter polynomial} $q$ satisfying
\begin{equation} \label{eq:filter-poly-exist}
    q(0) = 1 \quad \text{and} \quad \norm{q}_{\cS} \defeq \max_{z \in \cS} |q(z)| \le \varepsilon.
\end{equation}
Indeed, this equivalence is true.
The first half of the connection follows from the following standard complexity bound for the GMRES algorithm.

\begin{fact}[Matrix--vector complexity and filter polynomials: Upper bound \protect{\cite[Prop.~4]{saad1986gmres}}]
    \label{thm:gmres}
    Let \mA be an invertible diagonalizable matrix with decomposition \(\mA=\mX\mLambda\mX^{-1}\), and suppose that the eigenvalues of \mA lie in a compact set \(\cS\subset\bbC\).
    Then if there exists a degree-$t$ polynomial satisfying \cref{eq:filter-poly-exist}, then the GMRES algorithm achieves the guarantee
\[
        \norm{\mA\widetilde\vx-\vb}_2 \leq \eps \cond(\mX) \norm{\vb}_2
    \]
    after $t$ one-sided matrix--vector products.
    If \mA is normal then $\cond(\mX) = 1$.
\end{fact}

Indeed, the proof is direct from \cref{eq:gmres-optimality}:
\begin{equation*}
    \norm{\mA\widetilde\vx-\vb}_2 \le \norm{q(\mA)\vb}_2 \le \norm{\mX}_2 \cdot \norm{q(\mLambda)}_2 \cdot \norm{\smash{\mX^{-1}}}_2 \cdot \norm{\vb}_2 \le \varepsilon\cond(\mX)\norm{\vb}_2.
\end{equation*}
Establishing the opposite direction requires an information-theoretic lower bound.
This is accomplished by the following new result, which we prove in \cref{sec:resid-lower-bound}.

\begin{theorem}[Matrix--vector complexity and filter polynomials: Lower bound] \label{thm:lower-bound-forward}
    Let \(\cS\subseteq\bbC\) be nonempty and compact, and let \(0 < \eps < 1\).
    Suppose that any polynomial \(p\) of degree at most \(t\) with \(p(0)=1\) has \(\norm{p}_{\cS} \geq \eps\).
    Then, any one-sided matvec algorithm that takes in a vector \vb and uses one-sided matvecs with a normal matrix \mA whose eigenvalues lie in \cS to produce a vector \(\widetilde\vx\) such that \(\norm{\mA\widetilde\vx-\vb}_2 \leq \eps\norm{\vb}_2\) with probability at least\ \(2/3\) must use at least \(\Omega(t)\) matvecs.
\end{theorem}

The proof follows the Krylov simulation argument developed by Chewi et al.\ \cite{chewi2024query} and refined by Derezi\'nski et al.\ \cite{derezinski2026matrix} and Shamir \cite{shamir2026simple}.
The proof of this result is modeled after the argument by Shamir, which treats only real symmetric matrices.
We extend Shamir's argument to the complex field, which requires results from complex polynomial approximation theory \cite{kroo2010strong}.

\subsubsection{Step 2: Polynomials and the Green's function}
\label{sec:intro-poly-to-greens}

\Cref{thm:gmres,thm:lower-bound-forward} reduce questions of existence or non-existence of algorithms to questions about the existence or non-existence of polynomials.
Fortunately, filter polynomials are widely studied in numerical analysis \cite{greenbaum1997iterative,driscoll1998from,saad2003iterative,saad2011numerical} and approximation theory \cite{saff2010logarithmic}, and there are established ideas to upper and lower bound the degree $t_\star$ needed to produce a filter polynomial satisfying \cref{eq:filter-poly-exist}.
In this article, we understand the degree $t_\star$ by drawing on conformal mappings.
Let us sketch out the basic idea; see \cite{driscoll1998from} for a different justification via potential theory.

Suppose that the eigenvalues of a nonsingular matrix \mA lie in a ``nice'' compact set \cS (specifically, a closed Jordan domain).
The region \cS may be complicated to analyze, making it difficult to explicitly construct a filter polynomial on \cS.
We instead turn to \emph{conformal maps}, one of the basic building blocks of complex analysis.
A conformal map \(\chi:\cQ_1 \to \cQ_2\) is a bijection between open sets \(\cQ_1\subseteq\bbC\) and \(\cQ_2\subseteq\bbC\) for which both \(\chi\) and \(\chi^{-1}\) are analytic.
One of the great observations of complex analysis is that conformal maps (approximately) preserve many properties of polynomials that we care about.
So, we can use conformal maps to reduce our difficult polynomial problem on \cS to an approximation problem on a much simpler domain.

We will use one of the most ubiquitous constructions of a conformal map.
By the Riemann mapping theorem, there exists a conformal map \(\chi_{\cS}\) from the outside of \cS to the outside of the unit disk \cite{beliaev2020conformal}.
That is, we take \(\cQ_1 = \bbC \setminus \cS\) and \(\cQ_2 = \bbC \setminus \overline\cD\), where \(\overline\cD = \{z\in\bbC~:~|z|\leq1\}\) is the closed unit disk; see \cref{fig:intro-conformal}.
\begin{figure}[t]
    \centering
    \begin{tikzpicture}[
        scale=0.72,
        every node/.style={font=\footnotesize},
        dotregion/.style={
            fill=PineGreen!8,
            postaction={
                pattern={Dots[distance=4.5pt, radius=0.60pt]},
                pattern color=PineGreen!65
            }
        }
    ]
\pgfmathsetmacro{\dotstep}{4.5 / (0.72 * 28.45274)}

        \def\kidneyPath{(0.65,0.45) .. controls (0.95,0.75) and (0.95,1.40) .. (0.45,1.65)
            .. controls (-0.10,1.90) and (-0.95,1.75) .. (-1.35,1.20)
            .. controls (-1.75,0.65) and (-1.45,0.10) .. (-1.30,-0.35)
            .. controls (-1.15,-0.80) and (-1.65,-1.20) .. (-1.25,-1.55)
            .. controls (-0.85,-1.90) and (-0.05,-1.75) .. (0.20,-1.30)
            .. controls (0.45,-0.85) and (-0.45,-0.55) .. (-0.40,-0.05)
            .. controls (-0.35,0.35) and (0.35,0.15) .. cycle}

\begin{scope}[shift={(0,0)}]
            \path[dotregion, even odd rule]
                (-2.35,-2.35) rectangle (2.35,2.35)
                \kidneyPath;

\fill[PineGreen!8, rounded corners=1.2pt]
                ({5.35*\dotstep}, {6.60*\dotstep}) rectangle ({7.45*\dotstep}, {9.40*\dotstep});
            \fill[PineGreen!8, rounded corners=1.2pt]
                ({7.00*\dotstep}, {6.60*\dotstep}) rectangle ({8.40*\dotstep}, {8.40*\dotstep});
            \fill[PineGreen!8, rounded corners=1.2pt]
                ({0.40*\dotstep}, {-1.60*\dotstep}) rectangle ({1.60*\dotstep}, {-0.40*\dotstep});

\draw[->, gray] (-2.60,0) -- (2.60,0)
                node[right, text=black] {$\Re$};
            \draw[->, gray] (0,-2.60) -- (0,2.60)
                node[above, text=black] {$\Im$};

\draw[NavyBlue, very thick] \kidneyPath;

\filldraw[black] (0,0) circle (1.5pt);
            \node[black, inner sep=0pt] at ({1.00*\dotstep}, {-1.00*\dotstep}) {$0$};
            \node[NavyBlue] at (-0.65,1.00) {\large $\cS$};
            \node[PineGreen!90!black, inner sep=0pt] at ({6.90*\dotstep}, {8.00*\dotstep}) {\large $\cQ_1$};
        \end{scope}

\begin{scope}[shift={({31*\dotstep},0)}]
            \def\Rdisk{1.05}
            \path[dotregion, even odd rule]
                (-2.35,-2.35) rectangle (2.35,2.35)
                (0,0) circle (\Rdisk);

\fill[PineGreen!8, rounded corners=1.2pt]
                ({5.35*\dotstep}, {6.60*\dotstep}) rectangle ({7.45*\dotstep}, {9.40*\dotstep});
            \fill[PineGreen!8, rounded corners=1.2pt]
                ({7.00*\dotstep}, {6.60*\dotstep}) rectangle ({8.40*\dotstep}, {8.40*\dotstep});
            \fill[PineGreen!8, rounded corners=1.2pt]
                ({5.55*\dotstep}, {3.35*\dotstep}) rectangle ({8.45*\dotstep}, {4.65*\dotstep});

\draw[->, gray] (-2.60,0) -- (2.60,0)
                node[right, text=black] {$\Re$};
            \draw[->, gray] (0,-2.60) -- (0,2.60)
                node[above, text=black] {$\Im$};
            \draw[gray] (\Rdisk,-0.10) -- (\Rdisk,0.10);
            \node[gray, below left=1pt, font=\scriptsize] at (\Rdisk,0) {$1$};

\draw[NavyBlue, very thick] (0,0) circle (\Rdisk);

\filldraw[black] ({7.00*\dotstep}, {3.00*\dotstep}) circle (1.5pt);
            \node[black, inner sep=0pt] at ({7.00*\dotstep}, {4.00*\dotstep}) {$w_\star$};
            \node[NavyBlue] at (-0.45,0.40) {\large $\overline\cD$};
            \node[PineGreen!90!black, inner sep=0pt] at ({6.90*\dotstep}, {8.00*\dotstep}) {\large $\cQ_2$};
        \end{scope}

\draw[->, >=Stealth, thick, black]
            (1.95,2.10) to[bend left=25] node[midway, above] {\large $\chi$} (4.85,2.10);
\end{tikzpicture}
    \caption{
        Visualization of the canonical conformal map used in \cref{sec:intro-poly-to-greens}.
        For any bounded simply connected region \cS, we define the complement region \(\cQ_1 = \bbC \setminus \cS\).
        Similarly, for the closed unit disk \(\overline\cD\), we have complement \(\cQ_2 = \bbC \setminus \overline\cD\).
        The Riemann mapping theorem guarantees the existence of a conformal map from \(\cQ_1\) to \(\cQ_2\).
        The point 0 gets mapped to the point \(w_\star \defeq \chi(0)\).
        A well-known fact is that the value of the Green's function at zero, \(G_{\cS}(0) = \log|w_\star|\), describes the minimal degree needed for there to exist a filter polynomial on \cS; see \cref{sec:intro-poly-to-greens}.
        As such, much of this article is devoted to bounding the Green's function for the polar rectangle.
}
    \label{fig:intro-conformal}
\end{figure}
 Then, instead of trying to construct a filter polynomial on \cS, we focus on constructing a filter polynomial in the transformed coordinates \(w = \chi_{\cS}(z)\).
We arrive at the simpler optimization problem
\begin{equation*}
    \min \norm{\hat{q}}_{\overline{\cD}} \quad \text{such that} \quad \hat{q}(w_\star) = 1
\end{equation*}
where \(w_\star \defeq \chi_{\cS}(0)\).
It can be shown using the maximum modulus principle (\cref{fact:max-modulus}) that $\hat{q}(w) = (w/w_\star)^t$ is the solution to this problem over all degree \(t\) polynomials.
Therefore, the exact minimum degree that suffices to have \(\norm{\hat{q}}_{\overline\cD} \leq \eps\) with \(\hat{q}(w_\star)=1\) is $\lceil (\log|\chi_{\cS}(0)|)^{-1} \log(1/\varepsilon) \rceil$.
This notation is unwieldy, so we introduce the logarithm of the absolute value of this conformal map, known as the \emph{Green's function}:
\begin{equation*}
    G_\cS(z) \defeq \log |\chi_{\cS}(z)| \quad \text{for } z \notin \cS.
\end{equation*}
Hence, we can instead write that the optimal degree for \(\hat q\) is $\lceil G_\cS(0)^{-1} \log(1/\varepsilon) \rceil$.
In words, in \(w\) coordinates, the \emph{degree required for a filter polynomial to exist is inversely proportional to the Green's function at zero}.
Unfortunately, the function $\hat{q}(w) = (\chi_{\cS}(z) / \chi_{\cS}(0))^t$ is not a degree-$t$ polynomial in $z$, but it is asymptotically close to one up to terms that vanish as $z\to \infty$ \cite{ellacott1983polynomial}.
This yields the maxim

\begin{equation} \label{eq:greens-degree-informal}
    t_\star \approx  G_\cS(0)^{-1} \log(1/\varepsilon).
\end{equation}

The informal relation \cref{eq:greens-degree-informal} is standard, and the lower bound $t_\star \ge  G_\cS(0)^{-1} \log(1/\varepsilon)$ is known as the Bernstein--Walsh theorem \cite{saff2010logarithmic}.
However, despite a broad literature on this topic, we were unable to find a reference for a comparable non-asymptotic upper bound, so we develop one in \cref{thm:poly-to-greens}.
For now, let us state a version specifically for the polar rectangle.

\begin{theorem}[Filter polynomials and the Green's function]\label{thm:poly-to-greens-intro}
    Let $G_{\cP_{\kappa,\theta}}$ be the Green's function of the polar rectangle $\cP_{\kappa,\theta}$, let $\varepsilon \in (0,1)$, and let  \begin{equation*}
        t_\star \coloneqq \min \{t ~:~ \text{there exists a degree-$t$ polynomial such that $q(0) = 1$ and $\norm{q}_{\cP_{\kappa,\theta}} \le \varepsilon$} \}.
    \end{equation*}
Then $G_{\cP_{\kappa,\theta}}(0)^{-1} \log(1/\varepsilon) \le t_\star \le 1 + G_{\cP_{\kappa,\theta}}(0)^{-1} \log(6\pi/\varepsilon)$.
\end{theorem}

Again, let us highlight that the lower bound is merely the Bernstein--Walsh theorem \cite{saff2010logarithmic}.
The upper bound is not much harder, relying on estimates for Faber polynomials \cite{ellacott1983polynomial}.
Our contribution in this section thus largely consists of identifying the correct ingredients from the literature and combining them in a straightforward way.

\subsubsection{Step 3: Bounding the Green's function}

By combining \cref{thm:gmres,thm:lower-bound-forward} with \cref{thm:poly-to-greens-intro}, we reduce the one-sided matrix--vector complexity of linear system solving to understanding the Green's function of the polar rectangle. 
Our main result shows matching lower and upper bounds of $\Theta(\theta^2 \, \kappa^{-\pi/\theta})$.

\begin{theorem}[Bounds on the Green's function for polar rectangle] \label{thm:greens-function-bound-residual}
    \label{thm:residual-greens}
    Fix \(\kappa \geq 600\) and \(\theta\in(0,2\pi]\).
    Then 
\begin{equation*}
    \frac{1}{\pi^3\e} \,\, \theta^2\kappa^{-\pi/\theta} \le G_{\cP_{\kappa,\theta}}(0) \le \frac{5}{2\pi^2} \,\, \theta^2 \kappa^{-\pi/\theta}.
\end{equation*}
\end{theorem}

These bounds are precise, with the numerical constants $(\pi^3\e)^{-1} \ge 0.01$ and $\tfrac{5}{2\pi^2} \le 0.26$ matching up to a factor of $26$.
The lower and upper bounds are separate results, and they are proven using different techniques.
The lower bound is proven in \cref{sec:harmonic} and uses comparison theorems from \emph{harmonic function theory}, reducing the bound to computations on an explicit function.
The upper bound is proven in \cref{sec:geometric}, and uses ideas from \emph{geometric function theory}, quantifying how ``difficult'' it is to separate $0$ from $\infty$ using curves supported on the complement domain \(\bbC \setminus \cP_{\kappa,\theta}\).
We have made an effort to make both proofs accessible to readers with no background in complex analysis, harmonic functions, or geometric function theory.

\subsubsection{Putting it all together: Proof of \cref{thm:main-residual}}

With all of these ingredients established, we can now prove \cref{thm:main-residual}.
First, \cref{thm:greens-function-bound-residual} gives matching upper and lower bounds on the Green's function for the polar rectangle, with \(G_{\cP_{\kappa,\theta}}(0) \asymp \theta^{2}\kappa^{-\pi/\theta}\).
Second, \cref{thm:poly-to-greens-intro} shows that the smallest degree \(t_\star\) that suffices for a filter polynomial to exist on the polar rectangle is \(t_\star \asymp \theta^{-2}\kappa^{\pi/\theta}\log(1/\eps)\).
Finally, \cref{thm:gmres} takes this upper bound on \(t_\star\) to show that GMRES converges with the specified number of matvecs, while \cref{thm:lower-bound-forward} shows that no one-sided method can succeed with fewer than \(\Omega(t_\star)\) matvecs.

\subsubsection{Unbounded spectral condition number results for backward error}\label{sec:intro-backward}

Our analysis for unbounded spectral condition numbers follows essentially the same path as our analysis in the bounded case, with a few twists.
Let \mA be a matrix with angular eigenvalue gap $\measuredangle(\mA) \ge \theta$.
By scaling by an appropriate complex number, we are free to assume the eigenvalues lie in the \emph{circular sector}
\begin{equation*}\label{eq:sector}
    \cT_\theta \defeq \Bigl\{
        r \e^{\ri \phi}
        ~:~
        0\le r \le 1, \: |\phi|\le \pi-\theta/2
    \Bigr\}.
\end{equation*}

We build on an analytical approach to bounding the backward error due to Derezi\'nski et al.\ \cite{derezinski2026towards}, who studied an algorithm called MINBERR that searches for the minimum backward-error solution to a psd linear system with a Krylov subspace.
For general (non-symmetric) systems, the appropriate one-sided algorithm is called GMBACK \cite{kasenally1995gmback,kasenally1997analysis}.
Our main result will be a complexity upper bound for GMBACK for matrices with an angular eigenvalue gap and a complexity lower bound showing that no one-sided algorithm can fundamentally improve on that complexity.

Following \cite{derezinski2026towards}, the appropriate question of polynomial analysis is as follows:
Given a matrix \mA with eigenvalues belonging to a set \cS, what polynomial degree $t_{\rm b}$ is required to attain the guarantee
\[
    \max_{z\in\cS} \left|z - \frac{1}{p(z)}\right| \leq \eps \max_{z\in\cS}|z| \quad \text{for some polynomial \(p\) of degree $t_{\rm b}$}?
\]
The importance of this quantity for the GMBACK algorithm is quantified by the following theorem.

\begin{theorem}[GMBACK and approximation of the inverse]
    \label{thm:gmberr}
    Let \(\mA=\mX\mLambda\mX^{-1}\) be an invertible diagonalizable matrix.
Suppose there exists a polynomial \(p\) of degree at most \(t_{\rm b}\) such that \(\max_i |\lambda_i - 1/p(\lambda_i)| \leq \eps \max_i |\lambda_i|\) for some $\eps \in (0,1)$.
    Then, for any vector \vb, running GMBACK with
    \(t_{\rm b}+1\) one-sided matvecs with \mA returns a vector \(\widetilde\vx\) such that 
    \[
        \widetilde\mA\widetilde\vx = \vb \text{ for some matrix \(\widetilde\mA\) with } \norm{\mA-\widetilde\mA}_2\leq\eps\cond(\mX)\norm{\mA}_2.
    \]
    In particular, if \mA is normal, then \(\cond(\mX)=1\), and we have \(\norm{\mA-\widetilde\mA}_2\leq\eps\norm{\mA}_2\).
\end{theorem}
We prove \cref{thm:gmberr} in \cref{sec:backward}.
The proof closely mirrors the analysis for MINBERR but is modified to handle the possible non-normality of \mA.

For a complexity lower bound, we appeal to reduction rather than a direct argument.
The argument uses the following theorem from a work of Derezi\'nski, Mahoney, Meyer, and Musco \cite{derezinski2026from}.

\begin{importedtheorem}[Lower bound: Reducing backward error to residual error \protect{\cite[Lem.~4.8]{derezinski2026from}}]
    \label{impthm:matvec-back-to-residual}
    Fix parameter \(\eps>0\).
    Let \mA have condition number at most \(\kappa\).
    For any vector \vb, suppose that \(\widetilde\vx\) solves the system \(\mA\vx=\vb\) to backward error at most \(\eps\).
    By computing a single one-sided matvec with \mA, we can construct a vector \(\widehat\vx\) with \(\norm{\mA\widehat\vx - \vb}_2 \leq \kappa \eps \norm{\vb}_2\).
\end{importedtheorem}

This result shows small backward error implies small residual.
Consequently, since solving to small residual is hard (\cref{thm:main-residual}), solving to small backward error is hard as well.
Using this reduction, we can directly prove the lower bound of our main result \cref{thm:main-backward}.

\begin{proof}[Proof of \cref{thm:main-backward} (lower bound)]
    Let \(\kappa \defeq \alpha/\eps\) for some \(\alpha\in(0,1)\) we will specify later.
    Suppose an algorithm uses \(t_{\rm b}\) one-sided matvecs to solve any linear system \(\mA\vx=\vb\) to backward error \(\eps\) with probability at least \(2/3\) for all normal matrices \mA whose eigenvalues lie in \(\cT_\theta\).
    Then, using only one additional matvec, we can recover a vector \(\widehat\vx\) that solves any linear system \(\mA\vx=\vb\) to relative residual \(\kappa\eps = \alpha\) for any normal matrix \mA whose eigenvalues lie in \(\cT_{\theta}\) with condition number at most \(\kappa\).
    Hence, the eigenvalues of \mA lie in the polar rectangle \(\cP_{\kappa,\theta}\).
    Therefore, by \cref{thm:main-residual}, we know that
    \[
        t_{\rm b} \gtrsim \theta^{-2} \left(\frac \alpha\eps\right)^{\pi/\theta} \log(1/\alpha).
    \]
    Maximizing over \(\alpha\) gives \(\alpha = \e^{-\theta/\pi}\).
    Plugging in and simplifying gives \(t_{\rm b} \gtrsim \theta^{-1} \eps^{-\pi/\theta}\).
\end{proof}

With the complexity lower bound handled by a separate argument, we are left with the upper bound.
Using a similar framework to the one we developed to analyze the residual error, we now need to prove the existence of a low-degree polynomial satisfying $\norm{z - 1/p(z)}_{\cT_\theta} \le \varepsilon$.
This result can also be attained using our basket of complex analysis tools: conformal mapping, Green's function, and comparison principles for harmonic functions.
The conclusion is as follows.

\begin{theorem}[Polynomial approximation: bounding backward error on the circular sector]
    \label{thm:poly-backward-upper}
    Fix \(\theta\in(0,2\pi]\).
    There exists a polynomial \(p\) of degree \(t_{\rm b} \lesssim \theta^{-1} (2/\eps)^{\pi/\theta}\) such that \(\norm{z - 1/p(z)}_{\cT_\theta} \leq \eps\).
\end{theorem}

We prove \cref{thm:poly-backward-upper} in \cref{sec:backward}.
Combining this result with \cref{thm:gmberr} proves the matvec upper bound for GMBACK in \cref{thm:main-backward}.

\subsection{Extension: Eigenvector approximation}

The results in this paper can also be used to analyze the Arnoldi algorithm for eigenvector approximation \cite{arnoldi1951principle,saad1980variations}.
In the Arnoldi algorithm, one begins from an initial vector \vb (often random) and builds an orthonormal basis for the Krylov subspace $\cK_t(\mA,\vb)$ (\cref{eq:krylov}), which then can be used to form estimates of eigenvalues or eigenvectors or to form low-rank approximations.

To study this problem, consider a diagonalizable matrix \(\mA = \mX\diag(\lambda_1,\ldots,\lambda_n)\mX^{-1}\) with eigenvalues \(\lambda_1,\ldots,\lambda_n\), and consider a particular eigenvalue \(\lambda_k\) with corresponding eigenvector \(\vx_k = \mX\ve_k\).
Suppose that \(\lambda_k\) is separated from the remaining eigenvalues by both an angular gap \(\theta\in(0,2\pi]\) and a relative distance \(g\in(0,1]\); see the inset diagram below.
That is, after shifting, rotating, and rescaling, the remaining eigenvalues lie in a polar rectangle \(\cP_{1/g,\theta}\).
Formally, there exist numbers $R > 0$, $g \in (0,1]$, and $\phi \in [0, 2\pi)$ such that
\begin{equation} \label{eq:eigenvalue-geometry}
    \lambda_j \in \lambda_k + R \e^{\ri\phi}\, \cP_{1/g,\theta} \qquad \text{for all } j\ne k.
\end{equation}

Under this setting, we have the following bound for the distance between an eigenvector and the Krylov subspace $\cK_t(\mA,\vb)$, which is an immediate corollary of the standard estimate \cite[Lem.~6.2]{saad2011numerical} and \cref{thm:poly-to-greens-intro,thm:greens-function-bound-residual}.

\medskip
\noindent
\begin{minipage}[c]{0.77\textwidth}
    \begin{corollary}[Recovering eigenvectors with an angular eigenvalue gap]
    \label{cor:eigvec-angular-gap}
    Assume the setting above and assume that \cref{eq:eigenvalue-geometry} holds with $g\le \tfrac{1}{600}$.
    Then 
    \begin{equation*} \label{eq:eigvec-approx}
        \norm{(\mI - \mPi_{\cK_t(\mA,\vb)}) \vx_k}_2
        \le
        6 \pi \exp\left( - \frac{(t-2)\,\theta^2 g^{\pi/\theta}}{\pi^3 \e} \right) \left(\frac{\norm{\mX^{-1} \vb}_1}{|(\mX^{-1} \vb)_k|} -1 \right),
    \end{equation*}
    where $\mPi_{\cK_t(\mA,\vb)}$ is the orthogonal projection onto $\cK_t(\mA,\vb)$.
    \end{corollary}
\end{minipage}\hfill
\begin{minipage}[c]{0.21\textwidth}
    \centering
    \definecolor{EigBrown}{RGB}{115,52,15}
    \definecolor{SmallThetaGreen}{RGB}{0,138,75}
    \definecolor{LargeThetaBlue}{RGB}{20,90,205}

    \def\crossSize{0.072}
    \newcommand{\eigcrosspolar}[2]{\draw[EigBrown, thick]
            ($({#1}:{#2})+(-\crossSize,-\crossSize)$) -- ++({2*\crossSize},{2*\crossSize});
        \draw[EigBrown, thick]
            ($({#1}:{#2})+(-\crossSize,\crossSize)$) -- ++({2*\crossSize},{-2*\crossSize});}

    \def\Rmin{0.92}    \def\Rmax{1.80}    \def\gapDir{140}   \def\gapHalf{50}   \pgfmathsetmacro{\gapLo}{\gapDir-\gapHalf}
    \pgfmathsetmacro{\gapHi}{\gapDir+\gapHalf}
    \pgfmathsetmacro{\regLo}{\gapHi-360}
    \begin{tikzpicture}[
        scale=0.72,
        every node/.style={font=\footnotesize}
    ]
\fill[NavyBlue!9]
            (\regLo:\Rmax) arc[start angle=\regLo, end angle=\gapLo, radius=\Rmax]
            -- (\gapLo:\Rmin) arc[start angle=\gapLo, end angle=\regLo, radius=\Rmin]
            -- cycle;

\draw[gray, thick] (0,0) circle (\Rmin);
        \draw[gray, thick] (0,0) circle (\Rmax);

\draw[LargeThetaBlue, thick, densely dashed] (0,0) -- (\gapLo:\Rmax);
        \draw[LargeThetaBlue, thick, densely dashed] (0,0) -- (\gapHi:\Rmax);

\draw[white, line width=2.6pt, line cap=butt, shorten >=1.2pt]
            (\gapLo:1.3) arc[start angle=\gapLo, end angle=\gapHi, radius=1.3];
        \draw[->, LargeThetaBlue, thick]
            (\gapLo:1.3) arc[start angle=\gapLo, end angle=\gapHi, radius=1.3];
        \contourlength{1.2pt}
        \node[LargeThetaBlue, inner sep=0.5pt] at (\gapDir:1.5) {\contour{white}{$\theta$}};

\filldraw[SmallThetaGreen] (\Rmin,0) circle (1.3pt);
        \filldraw[SmallThetaGreen] (\Rmax,0) circle (1.3pt);
        \draw[SmallThetaGreen, thick, decorate, decoration={brace, amplitude=3.2pt}]
            (0,0.08) -- (\Rmin,0.08)
            node[midway, above=2.5pt, inner sep=0.6pt, font=\scriptsize, text=SmallThetaGreen] {$gR$};
        \draw[SmallThetaGreen, thick, decorate, decoration={brace, mirror, amplitude=3.2pt}]
            (0,-0.08) -- (\Rmax,-0.08)
            node[pos=0.74, below=3.5pt, fill=NavyBlue!9, inner sep=0.6pt, rounded corners=1pt, text=SmallThetaGreen] {$R$};

\eigcrosspolar{-75}{\Rmin}
        \eigcrosspolar{32}{\Rmax}
        \node[EigBrown, font=\scriptsize, above right=0.5pt, inner sep=1pt] at (32:\Rmax) {$\lambda_j$};
        \foreach \a/\r in {-150/1.30, -125/1.20, -95/1.52, -60/1.28, -40/1.60,
                           20/1.30, 48/1.12, 62/1.50, 80/1.10, -108/1.08}{
            \eigcrosspolar{\a}{\r}
        }

\filldraw[black] (0,0) circle (1.5pt);
\node[black, above left=1.5pt, inner sep=1pt] at (0,0) {$\lambda_k$};

\draw[gray!55, line width=0.4pt]
            ([shift={(-4pt,-3.5pt)}]current bounding box.south west)
            rectangle
            ([shift={(4pt,3.5pt)}]current bounding box.north east);
    \end{tikzpicture}
 \end{minipage}
\medskip

When the vector \vb is chosen randomly, this bound shows that, $$t = \order\left(\theta^{-2} g^{-\pi/\theta} \log\left[\frac{\cond(\mX)n}{\varepsilon}\right]\right)$$ iterations suffice to ensure that the Krylov subspace $\cK_t(\mA,\vb)$ is $\varepsilon$-close to containing the eigenvector $\vx_k$ with high probability.
Bounds on the quantity $\norm{(\mI - \mPi_{\cK_t(\mA,\vb)}) \vx_k}_2$ are standard in the numerical analysis literature \cite{saad1980variations,jia2001analysis,bellalij2010further,saad2011numerical}.
Such a bound can be used to analyze the quality of low-rank approximations to a matrix \mA using the Krylov subspace $\cK_t(\mA,\vb)$ \cite{musco2015randomized,meyer2024unreasonable,tropp2023randomized,chen2026does} and to understand the quality of eigenvalue approximations obtained from the Arnoldi algorithm \cite{saad1980variations,jia2001analysis}.
We leave a more detailed discussion of these extensions to future work.

\subsection{Related work} \label{sec:related}

Understanding the complexity of Krylov linear solvers like GMRES has been a major research project in numerical analysis for decades, and eigenvalue positioning plays a core role in this theory; see \cite{embree2022how} for a survey, new results, and perspective.
The usage of conformal mappings in this theory is standard; see \cite{driscoll1998from} and the references therein.
By and large, these works focus on understanding the \emph{asymptotic} rate of convergence for matrices with eigenvalues in a region, and we were unable to find non-asymptotic convergence for GMRES from this literature.
More explicit bounds for a different class of semi-iterative methods were developed by Eiermann and collaborators \cite{eiermann1983construction,eiermann1985study,eiermann1989semiiterative,eiermann1989hybrid}; the techniques are similar to our \cref{thm:poly-to-greens}.
The bounds we study can be uninformative if the eigenvector matrix is ill-conditioned; in this case, bounds based on the pseudospectrum \cite{trefethen1990approximation} or numerical range \cite{eiermann1993fields,beckermann2005some,crouzeix2007numerical} can be more informative.

The relative power of one-sided methods (GMRES) and two-sided methods (CGN) has been discussed and debated by many authors; see, for instance, \cite{nachtigal1992how,greenbaum1997iterative,wathen2022some,epperly2026stable}.
Phenomena associated with slow GMRES convergence are discussed in \cite{nachtigal1992how,greenbaum1994max,greenbaum1996any}.

We believe that using the spectral condition number and the angular eigenvalue gap is a natural way to characterize the difficulty of a linear system of equations $\mA\vx = \vb$.
There is very limited prior work adopting this perspective.
Coleman and Myers \cite{coleman1995faber} produced an explicit representation for the conformal map $\chi$ in terms of integrals of radicals of rational functions and expressions for good filter polynomials.
As best we can tell, the basic complexity scaling $t_\star \asymp \theta^{-2} \kappa^{\pi/\theta}$ is not easily derivable from these expressions.
The paper \cite{modersitzki1995faber} studies filter polynomials on the polar rectangle and evaluates the integrals of \cite{coleman1995faber} to estimate the rate of convergence for specific values of $\kappa$ and $\theta$.
As best we can tell, the papers \cite{coleman1995faber,modersitzki1995faber} are the sole references to study convergence of linear algebra algorithms when the eigenvalues lie in a polar rectangle.
The paper \cite{koch2000conformal} studies convergence rates for linear systems where the eigenvalues lie in ``bratwurst-shaped'' regions.

 \section{Preliminaries}
\label{sec:prelims}

We write scalars using italics, vectors using bold lowercase, matrices in bold uppercase, and sets in calligraphic notation.
Universal constants may be upright non-italics.
The set of real numbers is \bbR and the set of complex numbers is \bbC.
The positive reals are \(\bbR_+\).
The extended complex plane is \(\widehat\bbC \defeq \bbC \cup \{\infty\}\).
The complex conjugate is denoted \(\overline z\), and the modulus is \(|z|\).

A set \cS has (topological) closure \(\overline\cS\) and boundary \(\partial\cS\).
A domain is a nonempty open connected set.
A curve is simple if it is not self-intersecting.
\cS is a Jordan domain if it is open, bounded, and if \(\partial\cS\) is a simple closed curve.
The open unit disk is \(\cD = \{z~:~|z|<1\}\).
The closed unit disk is \(\overline\cD\), and the unit circle is \(\partial\cD\).
For a function \(f\) defined on a set \cS, the supremum norm is \(\norm{f}_{\cS} = \sup_{z\in\cS}|f(z)|\).
A function is analytic at \(z\) if it has a power series that converges to the function in an open neighborhood of \(z\).
A conformal map is an analytic bijection between two domains.

The sets of real vectors and matrices are denoted \(\bbR^n\) and \(\bbR^{n \times d}\), with analogous notation for complex vectors and matrices.
The transpose is \(\mA^\top\) and the conjugate transpose is \(\mA^*\).
The identity matrix is \mI.
The Euclidean norm of a vector is \(\norm{\vx}_2\) and the matrix operator norm is \(\norm{\mA}_2\).

We use the standard big-Oh notations \(\cO(\cdot)\), \(\Omega(\cdot)\), \(\Theta(\cdot)\).
For two functions \(f\) and \(g\), we write \(f \lesssim g\) if \(f \leq \rC g\) for some universal constant \(\rC>0\).
We define \(f \gtrsim g\) analogously, and write \(f \asymp g\) if both \(f \lesssim g\) and \(g \lesssim f\).
All logarithms are natural.

 \section{Reduction: From polynomials to the Green's function}\label{sec:poly-to-greens}

This section is dedicated to developing a general version of \cref{thm:poly-to-greens-intro}, used in the analysis of both upper and lower bounds for the residual error.
This theorem connects polynomial approximation bounds for filter polynomials to the Green's function.
The mathematical arguments in this section are well known ideas in approximation theory, though existing bounds are often asymptotic.
We explain the argument here for two reasons:
for clarity of exposition for the TCS audience, and to explain where our non-asymptotic upper bound in \cref{thm:poly-to-greens-intro} originates from.

To state this result, we recall the Green's function from \cref{sec:intro-poly-to-greens}.
Let $\cS$ denote a bounded, closed, and simply connected subset of the complex plane.
For any such set, the Riemann mapping theorem (see \cite{beliaev2020conformal}) guarantees the existence of a conformal function 
\begin{equation} \label{eq:chi-def}
    \chi:\hat\bbC\setminus\cS\to \hat\bbC\setminus\overline\cD,
\end{equation}
which maps the outside of the region \cS to the outside of the unit disk \cD.
Here, $\hat\bbC \defeq \bbC \cup \{\infty\}$ denotes the extended complex plane.
In order to uniquely identify such a conformal map, we impose the normalization constraints \(\chi(\infty)=\infty\) and \(\lim_{z\to\infty} \chi(z)/z \in\bbR_+\), which are standard in complex analysis.
The \emph{Green's function} is
\[
    G_{\cS}(z) \defeq \log(|\chi(z)|),
\]
and recall that our goal is to understand its value at zero:
\[
    g \defeq G_{\cS}(0).
\]
To state the result, we need to define a measure of the geometric complexity of the spectrum \cS.
We present a simplified/specialized version of the general definition of boundary rotation.
For a general definition, see \cite[Pg.~45]{gaier1987lectures}.

\begin{definition}[Boundary rotation]
    Let \(\gamma = \gamma_1 \cup \cdots \cup \gamma_k\) be a continuous closed curve that is decomposable as a union of subcurves $\gamma_i$ that are either circular arcs or straight lines.
    Let \(\alpha_i\in[0,2\pi]\) be the interior angle between subcurves \(\gamma_{i-1}\) and \(\gamma_{i}\) at their intersection (with \(\gamma_0 \defeq \gamma_k\)), and let \(\beta_i\in[0,2\pi]\) be the angle of each arc, setting $\beta_i = 0$ for a line segment.
    Then, the boundary rotation of \(\gamma\) is
    \[
        V \defeq \sum_{i=1}^k (|\pi-\alpha_i| + \beta_i).
    \]
\end{definition}

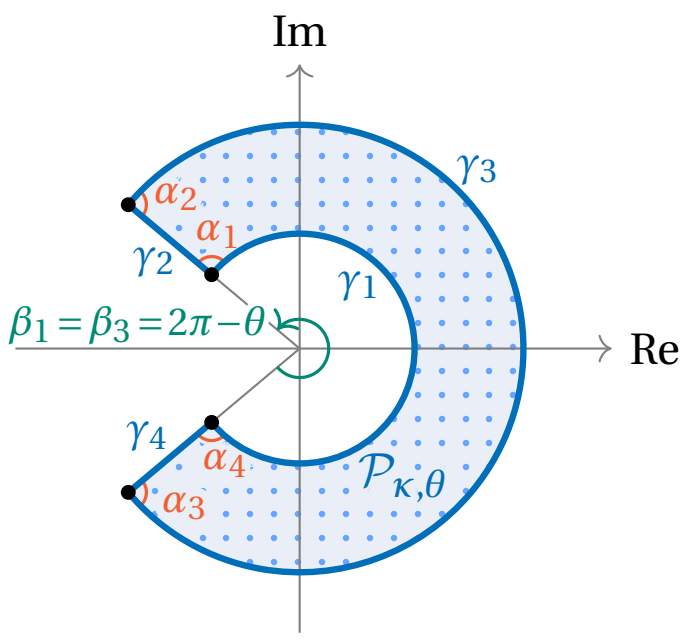
\begin{figure}[t]
    \centering
    \begin{tikzpicture}[
        scale=0.80,
        every node/.style={font=\footnotesize},
        bluedotregion/.style={
            fill=NavyBlue!8,
            postaction={
                pattern={Dots[distance=4.5pt, radius=0.60pt]},
                pattern color=NavyBlue!65
            }
        }
    ]
        \def\Rin{0.95}
        \def\Rout{1.85}
        \def\halfTheta{40}
        \def\rAlpha{0.15}
        \def\rAlphaMask{0.19}

        \pgfmathsetmacro{\dAngleIn}{asin(\rAlpha/(2*\Rin))}
        \pgfmathsetmacro{\dAngleOut}{asin(\rAlpha/(2*\Rout))}
        \pgfmathsetmacro{\dAngleInMask}{asin(\rAlphaMask/(2*\Rin))}
        \pgfmathsetmacro{\dAngleOutMask}{asin(\rAlphaMask/(2*\Rout))}

\begin{scope}[xscale=-1]
        \coordinate (innerUpper) at (\halfTheta:\Rin);
        \coordinate (outerUpper) at (\halfTheta:\Rout);
        \coordinate (outerLower) at ({-\halfTheta}:\Rout);
        \coordinate (innerLower) at ({-\halfTheta}:\Rin);

\path[bluedotregion]
            (outerUpper)
            arc[start angle=\halfTheta, end angle={360-\halfTheta}, radius=\Rout]
            -- (innerLower)
            arc[start angle=-\halfTheta, end angle={\halfTheta-360}, radius=\Rin]
            -- cycle;

\fill[NavyBlue!8]
            (innerUpper)
            -- ([shift={(\halfTheta:\rAlphaMask)}]innerUpper)
            arc[start angle=\halfTheta, end angle={\halfTheta+90+\dAngleInMask}, radius=\rAlphaMask]
            -- cycle;
        \fill[NavyBlue!8]
            (outerUpper)
            -- ([shift={({\halfTheta+90+\dAngleOutMask}:\rAlphaMask)}]outerUpper)
            arc[start angle={\halfTheta+90+\dAngleOutMask}, end angle={\halfTheta+180}, radius=\rAlphaMask]
            -- cycle;
        \fill[NavyBlue!8]
            (outerLower)
            -- ([shift={({-\halfTheta-180}:\rAlphaMask)}]outerLower)
            arc[start angle={-\halfTheta-180}, end angle={-\halfTheta-90-\dAngleOutMask}, radius=\rAlphaMask]
            -- cycle;
        \fill[NavyBlue!8]
            (innerLower)
            -- ([shift={({-\halfTheta-90-\dAngleInMask}:\rAlphaMask)}]innerLower)
            arc[start angle={-\halfTheta-90-\dAngleInMask}, end angle={-\halfTheta}, radius=\rAlphaMask]
            -- cycle;

\draw[gray] (0,0) -- (innerUpper);
        \draw[gray] (0,0) -- (innerLower);

\contourlength{1.0pt}
        \node[PineGreen, font=\scriptsize, inner sep=0pt, anchor=east]
            at (0.26,0.17) {\contour{white}{$\beta_1\!=\!\beta_3\!=\!2\pi\!-\!\theta$}};
        \node[NavyBlue, font=\scriptsize, inner sep=0pt]
            at (135:{\Rin-0.25}) {\contour{white}{$\gamma_1$}};
        \contourlength{1.6pt}
        \node[NavyBlue, inner sep=0pt]
            at (232:{0.5*(\Rin+\Rout)}) {\contour{NavyBlue!8}{$\cP_{\kappa,\theta}$}};
        \node[RedOrange, font=\scriptsize, inner sep=0pt]
            at ({\halfTheta+13.5}:{\Rin+0.18}) {\contour{NavyBlue!8}{$\alpha_1$}};
        \node[RedOrange, font=\scriptsize, inner sep=0pt]
            at ({\halfTheta+10.2}:{\Rout-0.26}) {\contour{NavyBlue!8}{$\alpha_2$}};
        \node[RedOrange, font=\scriptsize, inner sep=0pt]
            at ({-\halfTheta-14.5}:{\Rout-0.21}) {\contour{NavyBlue!8}{$\alpha_3$}};
        \node[RedOrange, font=\scriptsize, inner sep=0pt]
            at ({-\halfTheta-18}:{\Rin+0.21}) {\contour{NavyBlue!8}{$\alpha_4$}};

\node[NavyBlue, font=\scriptsize, inner sep=0pt]
            at (31:{0.5*(\Rin+\Rout)}) {$\gamma_2$};
        \node[NavyBlue, font=\scriptsize, inner sep=0pt]
            at (-30:{0.5*(\Rin+\Rout)+0.05}) {$\gamma_4$};

        \end{scope}

\draw[->, gray] (-2.35,0) -- (2.58,0)
            node[right, text=black] {$\Re$};
        \draw[->, gray] (0,-2.35) -- (0,2.35)
            node[above, text=black] {$\Im$};

        \begin{scope}[xscale=-1]

\draw[<-, PineGreen, semithick]
            (\halfTheta:0.24) arc[start angle=\halfTheta, end angle={360-\halfTheta}, radius=0.24];

\draw[RedOrange, semithick]
            ([shift={(\halfTheta:\rAlpha)}]innerUpper)
            arc[start angle=\halfTheta, end angle={\halfTheta+90+\dAngleIn}, radius=\rAlpha];

\draw[RedOrange, semithick]
            ([shift={({\halfTheta+90+\dAngleOut}:\rAlpha)}]outerUpper)
            arc[start angle={\halfTheta+90+\dAngleOut}, end angle={\halfTheta+180}, radius=\rAlpha];

\draw[RedOrange, semithick]
            ([shift={({-\halfTheta-180}:\rAlpha)}]outerLower)
            arc[start angle={-\halfTheta-180}, end angle={-\halfTheta-90-\dAngleOut}, radius=\rAlpha];

\draw[RedOrange, semithick]
            ([shift={({-\halfTheta-90-\dAngleIn}:\rAlpha)}]innerLower)
            arc[start angle={-\halfTheta-90-\dAngleIn}, end angle={-\halfTheta}, radius=\rAlpha];

\draw[NavyBlue, very thick, line join=round]
            (outerUpper)
            arc[start angle=\halfTheta, end angle={360-\halfTheta}, radius=\Rout]
            -- (innerLower)
            arc[start angle=-\halfTheta, end angle={\halfTheta-360}, radius=\Rin]
            -- cycle;

\filldraw[black] (innerUpper) circle (1.5pt);
        \filldraw[black] (outerUpper) circle (1.5pt);
        \filldraw[black] (outerLower) circle (1.5pt);
        \filldraw[black] (innerLower) circle (1.5pt);

\node[NavyBlue, font=\scriptsize] at (135:{\Rout+0.24}) {$\gamma_3$};
        \end{scope}
    \end{tikzpicture}
    \caption{
        Visualization of the boundary rotation calculation for the polar rectangle \(\cP_{\kappa,\theta}\).
        We decompose the boundary of \(\cP_{\kappa,\theta}\) into subcurves \(\gamma_1\), \(\gamma_2\), \(\gamma_3\), and \(\gamma_4\).
        The intersections between subcurves all have interior angle \(\alpha_1=\alpha_2=\alpha_3=\alpha_4=\pi/2\).
        Subcurves \(\gamma_2\) and \(\gamma_4\) are straight lines and therefore contribute \(\beta_2=\beta_4=0\).
        Subcurves \(\gamma_1\) and \(\gamma_3\) are circular arcs with angle \(\beta_1=\beta_3=2\pi-\theta\).
        As such, the total boundary rotation is \(V = 4|\pi-\pi/2| + 2(2\pi-\theta) = 6\pi-2\theta\).
    }
    \label{fig:bounded-rotation-polar-rectangle}
\end{figure}
 
As we illustrate in \cref{fig:bounded-rotation-polar-rectangle}, the polar rectangle has boundary rotation \(V = 6\pi - 2\theta\).
The following result shows that, to understand the degree of approximation needed to have small residual, 
it suffices to only examine the Green's function at zero.

\begin{theorem}[Green's function controls polynomial approximability]
    \label{thm:poly-to-greens}
    Fix \(\eps\in(0,1)\).
    Let \(\cS\subset\bbC\) be a compact and simply connected region where \(\partial\cS\) has boundary rotation \(V\), and let \(\chi\) be the normalized conformal map from \(\hat\bbC \setminus \cS\) to \(\hat\bbC \setminus \overline\cD\).
    Suppose that \(0\notin\cS\) and let \(g = G_{\cS}(0) = \log(|\chi(0)|)\) be the value of the Green's function at zero.
    Let \(t_\star\) be the smallest integer such that there exists a polynomial \(p\) of degree \(t_\star\) with $p(0)=1$ and \(\norm{p}_{\cS} \leq\eps\).
    Then,
    \[
        g^{-1}\log(1/\eps) \leq t_\star \leq 1 + g^{-1}\log(V/\eps).
    \]
\end{theorem}

The lower bound in \cref{thm:poly-to-greens} is an explicitly known result from potential theory called the Bernstein--Walsh theorem \cite[Lem.~3.7]{saff2010logarithmic}.
So, we focus on the upper bound of $t_\star$.
Results similar to this upper bound are explicitly known, like \cite[Thm.~2]{eiermann1989semiiterative} which instead assumes that \cS is a convex set.
For our result, we make use of the following known fact about Faber polynomials, whose definition will not be formally provided here.
We note that the exact Faber polynomials are actually known for the polar rectangle \cite{coleman1995faber}.
However, the expressions involved are sufficiently intricate that it is unclear how we can get asymptotic guarantees from them.
We will also utilize the maximum modulus principle, one of the most basic building blocks of complex analysis.

\begin{fact}[Maximum modulus principle] \label{fact:max-modulus}
    Let \(\cQ\subset\widehat\bbC\) be a nonempty open set.
    Let \(f:\cQ\to\bbC\) be analytic on \cQ, as well as continuous and bounded on its closure \(\overline\cQ\).
Over the domain \(\cQ\), the function \(z\mapsto|f(z)|\) is maximized on the boundary:
    \(
        \norm{f}_{\cQ} = \norm{f}_{\partial\cQ}.
    \)
\end{fact}

\begin{importedtheorem}[Faber polynomials and boundary rotation \protect{\cite[Lem.~2.1]{ellacott1983polynomial}}]
    \label{impthm:faber}
    Let \(\cS\subseteq\bbC\) be a closed set whose boundary is a simple closed curve with boundary rotation \(V\).
    Let \(\chi:\hat\bbC\setminus\cS\to\hat\bbC\setminus\overline\cD\) be the normalized conformal map from the outside of \cS to the outside of the unit disk.
    For all \(t \geq 1\) there exists a polynomial \(p\) of degree \(t\) such that
    \[
        \norm{p}_{\cS} \leq \frac{V}\pi
\quad\text{and}\quad
        \lim_{z \to \infty} p(z) - \chi(z)^t = 0.
    \]
\end{importedtheorem}

\begin{proof}[Proof of \cref{thm:poly-to-greens}]
    The lower bound side is exactly the Bernstein--Walsh theorem \cite[Lem.~3.7]{saff2010logarithmic}, so we prove the upper bound.

    Let \(\chi\) be the normalized conformal map from the outside of \cS to the outside of the unit disk, and let \(p\) be the polynomial constructed by \cref{impthm:faber}.
    Define \(f(z) \defeq p(z) - \chi(z)^t\), which is analytic on \(\cQ \defeq \widehat\bbC\setminus\cS\).
    Since \(f(\infty)=0\), we know that \(f\) is bounded on \cQ.
    By the maximum modulus principle, it holds that \(|f(0)| \leq \norm{f}_{\cQ} = \norm{f}_{\partial\cQ}\).

    Now we upper bound \(\norm{f}_{\partial\cQ}\).
    Since \(p\) is analytic on the entire complex plane, since \(\chi\) maps \(\partial\cS\) to the unit circle, and since \(\partial\cQ = \partial\cS\), we have
    \begin{equation*}
        \norm{f}_{\partial\cQ}
        = \norm{f}_{\partial\cS}
        \leq \norm{p}_{\partial\cS} + \norm{\chi}_{\partial\cS}^t
        = \norm{p}_{\cS} + 1
        \leq \tfrac V\pi + 1.
    \end{equation*}
    Therefore, we have \(|f(0)| = |p(0) - \chi(0)^t| \leq 1+\frac{V}{\pi}\).
    Since \(|\chi(0)| = \e^g\) and \(t \geq g^{-1} \log(1 + \frac V\pi + \frac V{\pi\eps})\),
    \begin{equation*}
        |p(0)|
        \geq \e^{gt} - 1 - \tfrac V\pi
        \geq \tfrac{V}{\pi\eps}.
    \end{equation*}
    We conclude by observing that the normalized polynomial \(\widehat p(z) = p(z) / p(0)\) has the desired properties \(\widehat p(0)=1\) and \(\norm{\widehat p}_{\cS} \leq \frac{\pi\eps}{V} \norm{p}_{\cS} \leq \eps\).
    To get the exact constants in the upper bound for \(t\), note that every set \cS has boundary rotation at least \(2\pi\), and therefore that \(1+\frac{V}{\pi}+\frac{V}{\pi\eps} \leq 2.5 \frac{V}{\pi\eps} \leq \frac{V}{\eps}\).
\end{proof}

 \section{Harmonic function theory: Lower bounding the Green's function}\label{sec:harmonic}

In this section we prove lower bounds on the Green's function, which in turn upper bounds the number of matvecs needed to minimize the residual.
The goal of this section is specifically to prove the lower bound \(G_{\cP_{\kappa,\theta}}(0) \gtrsim \theta^2 \kappa^{-\pi/\theta}\) from \cref{thm:greens-function-bound-residual}.
The corresponding upper bound on the Green's function is proven later, in \cref{sec:geometric}. 

We originally introduced the Green's function through conformal mapping. 
Under this interpretation, the Green's function $G_{\cS}(z) = \log(|\chi(z)|)$ measures how far the point $z$ is away from the boundary of the unit disk under the conformal mapping $\chi : \widehat{\bbC} \setminus \cS \to \widehat{\bbC} \setminus \overline{\cD}$.
An alternative perspective centers on viewing $G_{\cS}$ as a solution to \emph{Laplace's differential equation}
\begin{equation} \label{eq:harmonic}
    \frac{\partial^2 u}{\partial a^2}(a+
\ri b)+\frac{\partial^2 u}{\partial b^2}(a+
\ri b)=0
\end{equation}
on the region $\bbC \setminus \cS$.
Any function $u$ that satisfies \cref{eq:harmonic} is called a \emph{harmonic function}.
The class of harmonic functions is very rigid, enjoying many useful analytic properties and comparison principles.
By using results from the theory of harmonic functions, we will obtain the following lower bounds on Green's functions, as needed for the residual error analysis per \cref{thm:poly-to-greens}.

\begin{theorem}[Green's function lower bound for the polar rectangle]
    \label{thm:harmonic-residual}
    Let \(\kappa \geq 1\) and \(\theta\in(0,2\pi]\).
    Then, the Green's function for the polar rectangle \(\cP_{\kappa,\theta}\) satisfies \(G_{\cP_{\kappa,\theta}}(0) \geq (\pi^3\e)^{-1}\theta^2\kappa^{-\pi/\theta}\).
\end{theorem}

\subsection{Core results about harmonic functions and Green's functions}

We begin with the following list of standard properties about harmonic functions.

\begin{fact}[Harmonic functions]
    \label{fact:harmonic}
    Let \(\cQ\subseteq\bbC\) be an open set whose boundary is a finite union of simple curves. 
Suppose that $u,v : \overline{\cQ} \to \bbR$ are continuous functions that are harmonic on \cQ.
    Then the following hold:
    \begin{enumerate}[label=(\alph*)]
        \item \label{item:harmonic-boundary}
        \textbf{Boundary domination \textmd{\cite[Cor.~1.9]{axler2001harmonic}}.~} Assume \cQ is bounded. If \(u(z) \geq v(z)\) for all \(z\in\partial\cQ\), then \(u(z) \geq v(z)\) for all \(z\in\cQ\).
        \item \label{item:harmonic-mean-value}
        \textbf{Mean-value integral \textmd{\cite[Thm.~1.4]{axler2001harmonic}}.~} If \cQ contains the open disk \(\{z ~:~ |z - z_\star| < r\}\), then $u(z_\star)$ is the average of its values on the circle: \(u(z_\star) = \frac1{2\pi} \int_{0}^{2\pi} u(z_\star + r\e^{\ri\phi})\d\phi\).
\item \label{item:harmonic-analytic}
        \textbf{Components of an analytic function \textmd{\cite[p.~2]{axler2001harmonic}}.~} If \(f:\cQ\to\bbC\) is complex analytic on \(\cQ\), then \(w(z) \defeq \Re[f(z)]\) is harmonic on \(\cQ\).
\end{enumerate}
\end{fact}

We will also use the following two standard facts that relate these properties of harmonic functions to the Green's function:
the Green's function is harmonic, and Green's functions on nested domains satisfy a majorization relationship.

\begin{fact}[Green's functions]
    \label{fact:greens}
    Let \(\cS \subseteq \bbC\) be the closure of a Jordan domain, and let \(G_{\cS}(z) = \log|\chi(z)|\) denote the Green's function of \cS.
    Then the following properties hold:
    \begin{enumerate}[label=(\alph*)]
        \item \label{item:greens-harmonic}
        \textbf{Harmonicity \textmd{\cite[Sec.~2]{garnett2005harmonic}}.~} Define the open set \(\cQ = \bbC\setminus\cS\). Then \(G_{\cS}\) is harmonic on \(\cQ\), continuous on \(\overline{\cQ}\), and zero on the boundary \(\partial\cQ = \partial\cS\).
        \item \label{item:greens-monotone}
        \textbf{Domain monotonicity \textmd{\cite[p.~75]{garnett2005harmonic}}.~} If \(\cS'\supseteq \cS\) is the closure of a Jordan domain, then we have the monotonicity relation \(G_{\cS}(z) \geq G_{\cS'}(z)\) for all \(z\notin\cS'\).
    \end{enumerate}
\end{fact}

\subsection{Reduction to a wedge}

\begin{figure}[t]
    \centering
    \begin{tikzpicture}[
        scale=0.72,
        every node/.style={font=\footnotesize},
        dotregion/.style={
            fill=PineGreen!8,
            postaction={
                pattern={Dots[distance=4.5pt, radius=0.60pt]},
                pattern color=PineGreen!65
            }
        }
    ]
\def\Rin{0.82}
        \def\Rout{1.60}
        \def\Rbig{2.25}
        \def\halfTheta{38}
\def\bsep{0.058}

\pgfmathsetmacro{\qRin}{\Rin - \bsep}
            \pgfmathsetmacro{\qRout}{\Rout + \bsep}
            \pgfmathsetmacro{\mRin}{\Rin - 0.5*\bsep}
            \pgfmathsetmacro{\mRout}{\Rout + 0.5*\bsep}

            \pgfmathsetmacro{\pAngleInner}{\halfTheta + asin(\bsep/\Rin)}
            \pgfmathsetmacro{\pAngleOuter}{\halfTheta + asin(\bsep/\Rout)}
            \pgfmathsetmacro{\mAngleInner}{\halfTheta + asin(0.5*\bsep/\mRin)}
            \pgfmathsetmacro{\mAngleOuter}{\halfTheta + asin(0.5*\bsep/\mRout)}

            \pgfmathsetmacro{\rAlignIn}{sqrt(\Rin*\Rin - \bsep*\bsep)}
            \pgfmathsetmacro{\rAlignOut}{sqrt(\Rout*\Rout - \bsep*\bsep)}

        \begin{scope}[shift={(0,0)}, xscale=-1]
\coordinate (pOuterLower) at ({-\pAngleOuter}:\Rout);
            \coordinate (pOuterUpper) at (\pAngleOuter:\Rout);
            \coordinate (pInnerUpper) at (\pAngleInner:\Rin);
            \coordinate (pInnerLower) at ({-\pAngleInner}:\Rin);

\coordinate (mOuterLower) at ({-\mAngleOuter}:\mRout);
            \coordinate (mOuterUpper) at (\mAngleOuter:\mRout);
            \coordinate (mInnerUpper) at (\mAngleInner:\mRin);
            \coordinate (mInnerLower) at ({-\mAngleInner}:\mRin);

\coordinate (qRadInnerUpper) at (\halfTheta:\rAlignIn);
            \coordinate (qRadOuterUpper) at (\halfTheta:\rAlignOut);
            \coordinate (qRadInnerLower) at ({-\halfTheta}:\rAlignIn);
            \coordinate (qRadOuterLower) at ({-\halfTheta}:\rAlignOut);
            \coordinate (qArcOuterUpper) at (\pAngleOuter:\qRout);
            \coordinate (qArcInnerUpper) at (\pAngleInner:\qRin);

\path[dotregion, even odd rule]
                (-2.35,-2.35) rectangle (2.35,2.35)
                (mOuterUpper)
                arc[start angle=\mAngleOuter, end angle={360-\mAngleOuter}, radius=\mRout]
                -- (mInnerLower)
                arc[start angle=-\mAngleInner, end angle={\mAngleInner-360}, radius=\mRin]
                -- cycle;

\fill[NavyBlue!8]
                (mOuterUpper)
                arc[start angle=\mAngleOuter, end angle={360-\mAngleOuter}, radius=\mRout]
                -- (mInnerLower)
                arc[start angle=-\mAngleInner, end angle={\mAngleInner-360}, radius=\mRin]
                -- cycle;

\node[inner sep=0.8pt, fill=PineGreen!8, rounded corners=1.2pt]
                at (-1.65,1.75) {\phantom{\large $\cQ$}};

        \end{scope}
        \begin{scope}[shift={(0,0)}]
\draw[->, gray] (-2.60,0) -- (2.60,0)
                node[right, text=black] {$\Re$};
            \draw[->, gray] (0,-2.60) -- (0,2.60)
                node[above, text=black] {$\Im$};
        \end{scope}
        \begin{scope}[shift={(0,0)}, xscale=-1]

\draw[NavyBlue, very thick, dash pattern=on 3.60pt off 2.60pt]
                (pInnerUpper) -- (pOuterUpper);
            \draw[NavyBlue, very thick, dash pattern=on 3.60pt off 2.60pt]
                (pInnerLower) -- (pOuterLower);
            \draw[NavyBlue, very thick, dash pattern=on 3.60pt off 2.66pt]
                (pOuterUpper)
                arc[start angle=\pAngleOuter, end angle={360-\pAngleOuter}, radius=\Rout];
            \draw[NavyBlue, very thick, dash pattern=on 3.63pt off 2.81pt]
                (pInnerUpper)
                arc[start angle=\pAngleInner, end angle={360-\pAngleInner}, radius=\Rin];

\node[PineGreen!90!black] at (-1.65,1.75) {\large $\cQ$};
            \node[NavyBlue] at (246:{0.5*(\Rin+\Rout)}) {$\cP_{\kappa,\theta}$};
        \end{scope}

\pgfmathsetmacro{\pAngleInner}{\halfTheta + asin(\bsep/\Rin)}
            \pgfmathsetmacro{\pAngleOuter}{\halfTheta + asin(\bsep/\Rout)}
            \pgfmathsetmacro{\mAngleInner}{\halfTheta + asin(0.5*\bsep/\Rin)}
            \pgfmathsetmacro{\mAngleOuter}{\halfTheta + asin(0.5*\bsep/\Rout)}
            \pgfmathsetmacro{\rAlign}{sqrt(\Rin*\Rin - \bsep*\bsep)}

        \begin{scope}[shift={(7.4,0)}, xscale=-1]
            \coordinate (pOuterLower) at ({-\pAngleOuter}:\Rout);
            \coordinate (pOuterUpper) at (\pAngleOuter:\Rout);
            \coordinate (pInnerUpper) at (\pAngleInner:\Rin);
            \coordinate (pInnerLower) at ({-\pAngleInner}:\Rin);

            \coordinate (mOuterLower) at ({-\mAngleOuter}:\Rout);
            \coordinate (mOuterUpper) at (\mAngleOuter:\Rout);
            \coordinate (mInnerUpper) at (\mAngleInner:\Rin);
            \coordinate (mInnerLower) at ({-\mAngleInner}:\Rin);

            \coordinate (wAlignUpper) at (\halfTheta:\rAlign);
            \coordinate (wAlignLower) at ({-\halfTheta}:\rAlign);
            \coordinate (wOuterUpper) at (\halfTheta:\Rbig);
            \coordinate (wOuterLower) at ({-\halfTheta}:\Rbig);

\fill[NavyBlue!8]
                (mOuterUpper)
                arc[start angle=\mAngleOuter, end angle={360-\mAngleOuter}, radius=\Rout]
                -- (mInnerLower)
                arc[start angle=-\mAngleInner, end angle={\mAngleInner-360}, radius=\Rin]
                -- cycle;

\fill[PineGreen!8]
                (wAlignUpper) -- (mInnerUpper) -- (mOuterUpper) -- (\halfTheta:{sqrt(\Rout*\Rout - \bsep*\bsep)}) -- cycle;
            \fill[PineGreen!8]
                (wAlignLower) -- (mInnerLower) -- (mOuterLower) -- ({-\halfTheta}:{sqrt(\Rout*\Rout - \bsep*\bsep)}) -- cycle;

\path[dotregion]
                (0,0)
                -- (wOuterUpper)
                arc[start angle=\halfTheta, end angle=-\halfTheta, radius=\Rbig]
                -- cycle;

\node[inner sep=0.8pt, fill=PineGreen!8, rounded corners=1.2pt]
                at (-18:1.62) {\phantom{\large $\cW$}};

        \end{scope}
        \begin{scope}[shift={(7.4,0)}]
\draw[->, gray] (-2.60,0) -- (2.60,0)
                node[right, text=black] {$\Re$};
            \draw[->, gray] (0,-2.60) -- (0,2.60)
                node[above, text=black] {$\Im$};
        \end{scope}
        \begin{scope}[shift={(7.4,0)}, xscale=-1]

\draw[NavyBlue, very thick, dash pattern=on 3.6pt off 2.6pt]
                (pInnerUpper) -- (pOuterUpper);
            \draw[NavyBlue, very thick, dash pattern=on 3.6pt off 2.6pt]
                (pInnerLower) -- (pOuterLower);
            \draw[NavyBlue, very thick, dash pattern=on 3.6pt off 2.6pt]
                (pOuterUpper)
                arc[start angle=\pAngleOuter, end angle={360-\pAngleOuter}, radius=\Rout];
            \draw[NavyBlue, very thick, dash pattern=on 3.6pt off 2.6pt]
                (pInnerUpper)
                arc[start angle=\pAngleInner, end angle={360-\pAngleInner}, radius=\Rin];

\draw[PineGreen, very thick, dash pattern=on 3.38pt off 2.20pt]
                (0,0) -- (wAlignUpper);
            \draw[PineGreen, very thick, dash pattern=on 3.38pt off 2.20pt]
                (0,0) -- (wAlignLower);
            \draw[PineGreen, very thick, dash pattern=on 3.6pt off 2.6pt]
                (wAlignUpper) -- (wOuterUpper);
            \draw[PineGreen, very thick, dash pattern=on 3.6pt off 2.6pt]
                (wAlignLower) -- (wOuterLower);
            \draw[PineGreen, very thick, dash pattern=on 3.6pt off 2.6pt]
                (wOuterUpper)
                arc[start angle=\halfTheta, end angle=-\halfTheta, radius=\Rbig];

\node[NavyBlue] at (246:{0.5*(\Rin+\Rout)}) {$\cP_{\kappa,\theta}$};
            \node[PineGreen!90!black] at (-18:1.62) {\large $\cW$};
        \end{scope}

\draw[->, >=Stealth, thick, black]
            (2.3,1.24) to[bend left=32] (5.5,1.24);

\node[anchor=base] at (0,-3.20) {\small $G_{\cP_{\kappa,\theta}}$ harmonic on $\cQ$};
        \node[anchor=base] at (3.70,-3.20) {\large $\geq$};
        \node[anchor=base] at (7.40,-3.20) {\small $v$ harmonic on $\cW$};
    \end{tikzpicture}

    \caption{
        Visualization of the reduction to the wedge used in the harmonic function theory analysis.
        The first panel shows the polar rectangle \(\cP_{\kappa,\theta}\).
        Its complement is \cQ and is where the Green's function \(G_{\cP_{\kappa,\theta}}\) is guaranteed to be harmonic.
        Since this region is hard to analyze, we instead look at the wedge \(\cW\subseteq\cQ\).
        Since \(G_{\cP_{\kappa,\theta}}\) is harmonic on all of \cQ, it is also harmonic on \(\cW\).
        Since \cW is a simple shape, we can explicitly construct a function \(v\) that is harmonic on \cW.
        \cref{lem:greens-over-v} then shows that the Green's function is lower bounded by \(v\) on the wedge \cW.
}
    \label{fig:harmonic-reduction}
\end{figure}
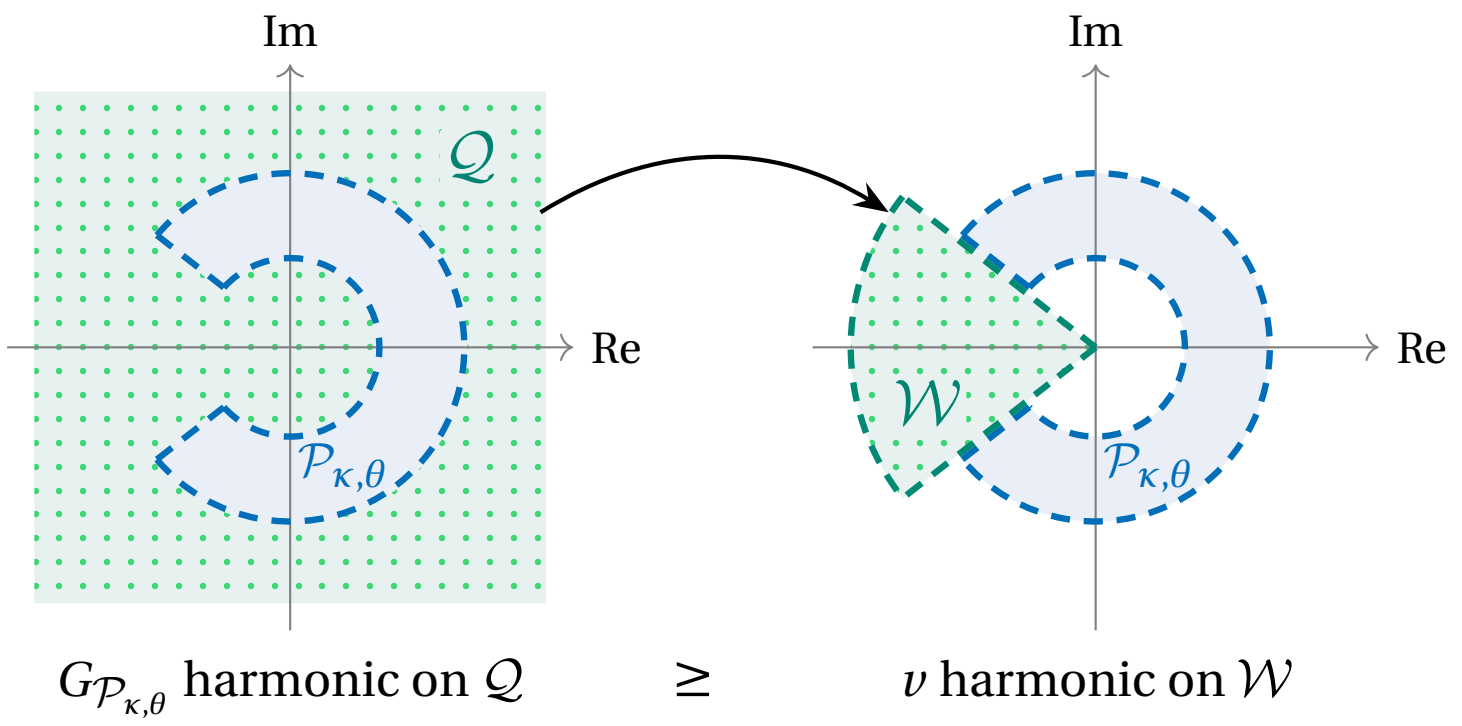
 
The Green's function is unwieldy because we do not know the exact expression for the conformal map \(\chi\) from \(\cQ = \widehat\bbC \setminus \cP_{\kappa,\theta}\) to \(\widehat\bbC \setminus\overline{\cD}\).
Instead, we can appeal to properties of harmonic functions to lower bound the Green's function.
The key idea is to find a simple subdomain \(\cW\subseteq\cQ\) and design an explicit, well-behaved harmonic function \(v\) on \cW.
Then by boundary domination, if we can show that the Green's function is greater than \(v\) on the boundary \(\partial\cW\), then we know that \(v\) is a lower bound for the Green's function on all of \cW.
This relationship shall be sharp enough to prove \cref{thm:harmonic-residual}.
The same idea will be used below in \cref{sec:backward} to prove a lower bound for the Green's function (\cref{thm:harmonic-backward}) for the circular sector, which is needed for the backward-error result.

In this section, we pick that subset \cW to be the wedge
\begin{equation}
    \label{eq:harmonic-wedge}
    \cW \defeq \bigl\{
        -r\e^{\ri \phi} ~:~ r\in(0,R),~ |\phi|<\theta/2
    \bigr\}
\end{equation}
for some parameter \(R \geq 1\) to be fixed later, and we take the function \(v\) to be
\begin{equation}
    \label{eq:harmonic-egg}
    v(-r\e^{\ri\phi}) \defeq (\log R) \left(\frac rR\right)^{\pi/\theta} \cos\left(\frac{\pi}{\theta}\phi\right).
\end{equation}
This wedge shape is chosen because it is geometrically natural.
The function \(v\), called the \emph{wedge minorant} \cite{ransford1995potential}, can be derived by solving Laplace's equation by separation of variables in polar coordinates subject to the boundary condition \(v(-r\e^{\ri \phi})=0\) when \(|\phi|=\theta/2\).
This choice of \cW and \(v\) shall suffice to lower bound the Green's function for \emph{both} the polar rectangle and the circular sector, meaning that this reduction is helpful for both the residual error and the backward error complexity results.

We will eventually want to show that the Green's function dominates \(v\) on the boundary of the wedge \cW.
To do so, we will need some elementary lower bounds on the value of the Green's function.
\begin{lemma}[Basic bounds on the Green's function]
    \label{lem:greens-circle-bound}
Let \(\cS\) be the closure of a Jordan domain with \(\cS \subseteq \overline\cD\).
    Then \(G_{\cS}(z) \geq 0\) for all \(z\notin\cS\) and \(G_{\cS}(z) \geq \log|z|\) for all \(|z|\geq1\).
\end{lemma}
\begin{proof}
Let \(\chi\) \hangindent=-88pt \hangafter=-6\vadjust{\vbox to 0pt{\kern-1.05\baselineskip\kern0.65em\hbox to \textwidth{\hfill\begin{tikzpicture}[
    scale=0.58,
    every node/.style={font=\footnotesize},
    dotregion/.style={
        fill=PineGreen!8,
        postaction={
            pattern={Dots[distance=4.2pt, radius=0.55pt]},
            pattern color=PineGreen!65
        }
    }
]
    \def\Rdisk{1.35}

\path[dotregion] (0,0) circle (\Rdisk);

\def\Spath{
        plot[smooth cycle, tension=0.75] coordinates {
            (0.65, 0.55)
            (-0.05, 0.98)
            (-0.82, 0.75)
            (-1.05, 0.00)
            (-0.78, -0.74)
            (-0.10, -0.82)
            (0.12, -0.22)
            (0.45, 0.08)
        }
    }
    \fill[NavyBlue!8] \Spath;

\node[inner sep=0.8pt, fill=PineGreen!8, rounded corners=1.2pt]
        at (0.55,-0.60) {\phantom{$\cD$}};

\draw[->, gray] (-1.68,0) -- (1.75,0)
        node[right, text=black] {$\Re$};
    \draw[->, gray] (0,-1.68) -- (0,1.68)
        node[above, text=black] {$\Im$};

\draw[PineGreen, very thick, dashed] (0,0) circle (\Rdisk);
    \draw[NavyBlue, very thick, dashed] \Spath;

\node[NavyBlue] at (-0.38,0.32) {$\cS$};
    \node[PineGreen!90!black] at (0.55,-0.60) {$\cD$};
    \node[gray, below right=2.5pt and 0.5pt, inner sep=0.5pt, font=\scriptsize] at (\Rdisk,0) {$1$};

\draw[gray!55, line width=0.4pt]
        ([shift={(-4pt,-3.5pt)}]current bounding box.south west)
        rectangle
        ([shift={(4pt,3.5pt)}]current bounding box.north east);
\end{tikzpicture}
 }\vss}} be the normalized conformal map from outside \cS to outside \(\overline\cD\).
    For the first bound, we have by definition that \(|\chi(z)| \geq 1\) for all \(z\notin\cS\), and therefore \(G_{\cS}(z) = \log|\chi(z)| \geq 0\) for all \(z\notin\cS\).
    For the second bound, by domain monotonicity (\cref{fact:greens}\ref{item:greens-monotone}), we have \(G_{\cS}(z) \geq G_{\overline\cD}(z)\) for \(z\notin\overline\cD\).
    The conformal map for the closed unit disk is \(\psi(z)=z\), so the corresponding Green's function is \(G_{\overline\cD}(z) = \log|z|\).
    Therefore, for any \(|z|\geq1\) we have \(G_{\cS}(z) \geq \log|z|\).
\end{proof}

Now, we are ready to show that the Green's function dominates \(v\) on the wedge.
This argument is just an application of boundary domination (\cref{fact:harmonic}\ref{item:harmonic-boundary}).

\begin{lemma}[The Green's function dominates \(v\)]
    \label{lem:greens-over-v}
    Let \(\cS\) be either the polar rectangle \(\cP_{\kappa,\theta}\) or the circular sector \(\cT_{\theta}\) for any \(\kappa \geq 1\) and \(\theta\in(0,2\pi]\).
    For any \(R \geq 1\), consider the wedge \cW and function \(v\) as defined in \cref{eq:harmonic-wedge,eq:harmonic-egg}.
    Then, \(G_{\cS}(z) \geq v(z)\) for all \(z\in\cW\).
\end{lemma}
\begin{proof}
    First, we show that \(v\) is harmonic on \cW.
    By Euler's formula, we recognize that \(\Re[(r\e^{\ri \phi})^\alpha] = r^\alpha \cos(\alpha\phi)\), so \(v\) is the real part of a complex analytic function:
    \[
        v(z) = \Re\left[(\log R) \left(-\tfrac{z}{R}\right)^{\pi/\theta}\right].
    \]
    Since the function on the inside is analytic on \cW, we know by \cref{fact:harmonic}\ref{item:harmonic-analytic} that \(v\) is harmonic on \(\cW\).
    By \cref{fact:greens}\ref{item:greens-harmonic}, we also have that \(G_{\cS}\) is harmonic on \cW.

    By boundary \hangindent=-74pt \hangafter=-4\vadjust{\vbox to 0pt{\kern-1.05\baselineskip\kern0.85em\hbox to \textwidth{\hfill\begin{tikzpicture}[
    scale=0.58,
    every node/.style={font=\footnotesize},
    dotregion/.style={
        fill=PineGreen!8,
        postaction={
            pattern={Dots[distance=4.2pt, radius=0.55pt]},
            pattern color=PineGreen!65
        }
    }
]
    \def\Rwedge{1.85}
    \def\halfTheta{38}

\begin{scope}[xscale=-1]
\path[dotregion]
        (0,0)
        -- (\halfTheta:\Rwedge)
        arc[start angle=\halfTheta, end angle=-\halfTheta, radius=\Rwedge]
        -- cycle;

\node[inner sep=0.6pt, fill=PineGreen!8, rounded corners=1pt]
        at (-19:0.92) {\phantom{\scriptsize $\theta$}};
    \node[inner sep=0.8pt, fill=PineGreen!8, rounded corners=1.2pt]
        at (-20:1.42) {\phantom{$\cW$}};

    \end{scope}

\draw[->, gray] (-2.22,0) -- (0.45,0)
        node[right, text=black] {$\Re$};
    \draw[->, gray] (0,-1.35) -- (0,1.35)
        node[above, text=black] {$\Im$};

    \begin{scope}[xscale=-1]

\draw[gray]
        ({-\halfTheta}:0.65) arc[start angle=-\halfTheta, end angle=\halfTheta, radius=0.65];
    \node[gray, font=\scriptsize] at (-19:0.92) {$\theta$};

\draw[black, very thick, line join=round]
        (0,0)
        -- (\halfTheta:\Rwedge)
        arc[start angle=\halfTheta, end angle=-\halfTheta, radius=\Rwedge]
        -- cycle;

\node[black, above right=0.5pt, inner sep=1pt] at (\halfTheta:{0.55*\Rwedge}) {$\gamma_1$};
    \node[black, below right=0.5pt, inner sep=1pt] at ({-\halfTheta}:{0.55*\Rwedge}) {$\gamma_2$};
    \node[black, above left=0.5pt, inner sep=1pt] at (20:\Rwedge) {$\gamma_3$};

\node[gray, above right=1.5pt, inner sep=0.5pt, font=\scriptsize] at (0,0) {$0$};
    \filldraw[gray] (\Rwedge,0) circle (1.5pt);
    \node[gray, below left=3.5pt and 0.5pt, inner sep=0.5pt, font=\scriptsize] at (\Rwedge,0) {$R$};

\filldraw[black] (0,0) circle (2.25pt);
    \filldraw[black] (\halfTheta:\Rwedge) circle (2.25pt);
    \filldraw[black] ({-\halfTheta}:\Rwedge) circle (2.25pt);

\node[PineGreen!90!black] at (-20:1.42) {$\cW$};
    \end{scope}

\draw[gray!55, line width=0.4pt]
        ([shift={(-4pt,-3.5pt)}]current bounding box.south west)
        rectangle
        ([shift={(4pt,3.5pt)}]current bounding box.north east);
\end{tikzpicture}
 }\vss}} domination (\cref{fact:harmonic}\ref{item:harmonic-boundary}), to complete the proof it suffices to show that \(v(z) \leq G_{\cS}(z)\) on the boundary \(\partial\cW\).
    We split the boundary \(\partial\cW\) into arcs \(\gamma_1\), \(\gamma_2\), and \(\gamma_3\) as shown in the diagram to the right.
    We have
    \begin{itemize}
        \item On \(\gamma_1 = \bigl\{-r\e^{- \ri \theta/2} ~:~ r\in[0,R]\bigr\}\) and \(\gamma_2 = \bigl\{-r\e^{\ri \theta/2} ~:~ r\in[0,R]\bigr\}\), we have \(v(z) = 0\).
        \item On \(\gamma_3 = \bigl\{-R\e^{\ri \phi} ~:~ |\phi|\leq\theta/2\bigr\}\), we have \(v(z) \leq \max_{|\phi|\leq\theta/2} v(-R\e^{\ri\phi}) = \log R\).
    \end{itemize}
    \hangindent=0pt
    By \cref{lem:greens-circle-bound}, since both the polar rectangle and the circular sector are subsets of the unit disk, we know that \(G_{\cS}(z) \geq 0\) on all of \(\partial\cW\), and that \(G_{\cS}(z) \geq \log R\) on \(\gamma_3\).
    Therefore, we have \(v(z) \leq G_{\cS}(z)\) on the entire boundary \(\partial\cW\), completing the proof.
\end{proof}

We can now prove the principal result of this section: a simple and clean lower bound on the Green's function for both the polar rectangle and circular sector.

\begin{theorem}[User-friendly lower bound on the Green's function]
    \label{thm:greens-to-wedge}
    Let \(\cS\) be either the polar rectangle \(\cP_{\kappa,\theta}\) or the circular sector \(\cT_{\theta}\) for any \(\kappa \geq 1\) and \(\theta\in(0,2\pi]\).
    The Green's function for \(\cS\) satisfies
\[
        G_{\cS}(-r\e^{\ri\phi})
\geq \frac\theta{\pi \e} r^{\pi/\theta} \cos\left(\frac{\pi}{\theta} \phi\right)
        \qquad
        \text{for }
        r\in(0,1]
        \text{ and }
        |\phi|\leq\frac\theta2.
    \]
\end{theorem}
\begin{proof}
    Invoking the prior lemma (\cref{lem:greens-over-v}), we have for any \(-r\e^{\ri \phi} \in \cW\) that
    \[
        G_{\cS}(-r\e^{\ri \phi})
\geq \bigl(\log R\bigr) \left(\frac{r}{R}\right)^{\pi/\theta} \cos\left(\frac{\pi}{\theta} \phi\right)
        = \bigl(R^{-\pi/\theta} \log R\bigr) r^{\pi/\theta} \cos\left(\frac{\pi}{\theta} \phi\right).
    \]
    Maximizing over \(R\) yields \(R = \e^{\theta/\pi}\) which recovers the stated bound.
    Since \(\e^{\theta/\pi} > \e^0 = 1\), any \(-r\e^{\ri \phi}\) with the stated parameters belongs to \cW.
\end{proof}

\subsection{Bounding the necessary quantities}

We are now very easily able to lower bound the quantities we are interested in.
We start with the lower bound on the Green's function needed for bounding residual error.
In this case, we need to lower bound \(G_{\cP_{\kappa,\theta}}(0)\).
If we were to naively plug in the bound from \cref{thm:greens-to-wedge}, we would only show that \(G_{\cP_{\kappa,\theta}}(0) \geq 0\), which is trivial.
To get a better guarantee, we instead turn to the mean value theorem, and lower bound \(G_{\cP_{\kappa,\theta}}(0)\) by integrating over the circle of radius \(\kappa^{-1}\), which includes much of the interior of \cW, giving us a nontrivial lower bound.
\begin{proof}[Proof of \cref{thm:harmonic-residual}]
    The circle \hangindent=-88pt \hangafter=-5\vadjust{\vbox to 0pt{\kern-1.05\baselineskip\kern0.65em\hbox to \textwidth{\hfill\begin{tikzpicture}[
    scale=0.58,
    every node/.style={font=\footnotesize},
]
    \def\Rcirc{0.90}
    \def\halfTheta{38}
\def\bsep{0.072}
    \pgfmathsetmacro{\Rblue}{\Rcirc + \bsep}
    \pgfmathsetmacro{\Rmid}{\Rcirc + 0.5*\bsep}
    \def\xmin{-1.92}
    \def\xmax{2.45}
    \def\ymin{-1.72}
    \def\ymax{2.32}

    \pgfmathsetmacro{\pAngleBlue}{\halfTheta + asin(\bsep/\Rblue)}
    \pgfmathsetmacro{\mAngleMid}{\halfTheta + asin(0.5*\bsep/\Rmid)}
    \pgfmathsetmacro{\tickAngle}{\halfTheta + 0.5*asin(\bsep/\Rblue)}
    \pgfmathsetmacro{\rOutStart}{\Rblue + 0.04}

\begin{scope}[xscale=-1]
\begin{scope}
        \clip (-\xmax,\ymin) rectangle (-\xmin,\ymax);

\fill[NavyBlue!8]
            (\mAngleMid:\Rmid)
            -- ++(\halfTheta:4.5)
            arc[start angle=\halfTheta, end angle={360-\halfTheta}, radius=5.4]
            -- ({-\mAngleMid}:\Rmid)
            arc[start angle={360-\mAngleMid}, end angle=\mAngleMid, radius=\Rmid]
            -- cycle;

\draw[NavyBlue, very thick, dash pattern=on 2.85pt off 2.10pt]
            ([shift={({\halfTheta+90}:\bsep)}]\halfTheta:\rOutStart) -- ++(\halfTheta:4.0);

\draw[NavyBlue, very thick, dash pattern=on 2.85pt off 2.10pt]
            ([shift={({-\halfTheta-90}:\bsep)}]{-\halfTheta}:\rOutStart) -- ++({-\halfTheta}:4.0);

\draw[NavyBlue, very thick, dash pattern=on 2.85pt off 2.10pt]
            (\pAngleBlue:\Rblue)
            arc[start angle=\pAngleBlue, end angle={360-\pAngleBlue}, radius=\Rblue];
    \end{scope}

    \end{scope}

\draw[->, gray] (-1.50,0) -- (1.42,0)
        node[right, text=black] {$\Re$};
    \draw[->, gray] (0,-1.50) -- (0,1.50)
        node[above, text=black] {$\Im$};

    \begin{scope}[xscale=-1]

\draw[BrickRed, very thick]
        (\tickAngle:\Rcirc) arc[start angle=\tickAngle, end angle={360-\tickAngle}, radius=\Rcirc];
    \draw[PineGreen, very thick]
        ({-\tickAngle}:\Rcirc) arc[start angle=-\tickAngle, end angle=\tickAngle, radius=\Rcirc];

\draw[BrickRed, line width=1.0pt]
        ([shift={({\tickAngle+90}:0.030)}]\tickAngle:{\Rcirc-0.12})
        -- ([shift={({\tickAngle+90}:0.030)}]\tickAngle:{\Rblue+0.05});
    \draw[PineGreen, line width=1.0pt]
        ([shift={({\tickAngle-90}:0.030)}]\tickAngle:{\Rcirc-0.12})
        -- ([shift={({\tickAngle-90}:0.030)}]\tickAngle:{\Rblue+0.05});

    \draw[BrickRed, line width=1.0pt]
        ([shift={({-\tickAngle-90}:0.030)}]{-\tickAngle}:{\Rcirc-0.12})
        -- ([shift={({-\tickAngle-90}:0.030)}]{-\tickAngle}:{\Rblue+0.05});
    \draw[PineGreen, line width=1.0pt]
        ([shift={({-\tickAngle+90}:0.030)}]{-\tickAngle}:{\Rcirc-0.12})
        -- ([shift={({-\tickAngle+90}:0.030)}]{-\tickAngle}:{\Rblue+0.05});

\node[NavyBlue] at (-1.65,-1.04) {$\cP_{\kappa,\theta}$};

\filldraw[gray] (\Rcirc,0) circle (1.5pt);
    \node[gray, above left=1.7pt and 0pt, inner sep=0.5pt, font=\scriptsize] at (\Rcirc,0) {$\kappa^{-1}$};

\filldraw[gray] (0,0) circle (1.5pt);
    \node[gray, above right=1.7pt and 0pt, inner sep=0.5pt, font=\scriptsize] at (0,0) {$0$};
    \end{scope}

\draw[gray!55, line width=0.4pt] (\xmin,\ymin) rectangle (\xmax,\ymax);
\end{tikzpicture}
 }\vss}} \(\{z~:~|z|=\kappa^{-1}\}\) can be decomposed into two parts, as shown in the diagram on the right: a green curve covered by \cref{thm:greens-to-wedge} and a red curve on the boundary of \(\cP_{\kappa,\theta}\).
    Recall from \cref{fact:greens}\ref{item:greens-harmonic} that \(G_{\cP_{\kappa,\theta}}\) is zero on the boundary of \(\cP_{\kappa,\theta}\).
    So, by the mean value theorem (\cref{fact:harmonic}\ref{item:harmonic-mean-value}) and the harmonic lower bound (\cref{thm:greens-to-wedge}),
    \[
        G_{\cP_{\kappa,\theta}}(0)
        = \frac1{2\pi} \int_{0}^{2\pi} G_{\cP_{\kappa,\theta}}(-\kappa^{-1}\e^{\ri \phi}) \d\phi
        \geq \frac1{2\pi} \int_{-\theta/2}^{\theta/2} \frac{\theta}{\pi\e} \kappa^{-\pi/\theta} \cos\left(\frac\pi\theta\phi\right) \d\phi
= \frac{1}{\pi^3\e} \theta^2 \kappa^{-\pi/\theta}. \hspace*{88pt}\qedhere 
    \]  
\end{proof}

 \section{Geometric function theory: Upper bounding the Green's function}\label{sec:geometric}

In this section, we introduce and use tools from the field of geometric function theory to upper bound the Green's function as it appears in \cref{sec:poly-to-greens}. An upper bound on the Green's function gives a lower bound on the degree \(t_\star\) needed for solving linear systems to small residual, and the opposing lower bound on the Green's function is found in \cref{sec:harmonic}.
The use of tools from geometric function theory and, more specifically, moduli of families of curves, has been systematically exploited by Andrievskii and collaborators in a series of works spanning decades \cite{andrievskii1995conformal,andrievskii1994maximal,andrievskii2004conformal,andrievskii2004green,andrievskii2006constructive,andrievskii2006optimal,andrievskii2007remez}.
Here, we adapt this general strategy to build precise tools for understanding the Green's function of the polar rectangle.

This section has three parts.
In \cref{sec:geometric-setup}, we introduce an ensemble of definitions and facts needed to work with geometric function theory, culminating in the definition of the modulus of a family of cross cuts that separate 0 from \(\infty\).
In \cref{sec:geometric-greens}, we prove that a lower bound on the modulus suffices to prove an upper bound on the Green's function.
In \cref{sec:geometric-polar-rect}, we bound said modulus needed to control residual error, proving the following:
\begin{theorem}[Green's function upper bound for the polar rectangle]
    \label{thm:green-upper-polar-rect}    
    Let \(\kappa \geq 536\) and \(\theta\in(0,2\pi]\).
    Then, the Green's function for the polar rectangle satisfies \(G_{\cP_{\kappa,\theta}}(0) \leq (5/2\pi^2) \, \theta^2 \kappa^{-\pi/\theta}\).
\end{theorem}

\subsection{Geometric function theory: Setup}
\label{sec:geometric-setup}

The fundamental building blocks in our discussion of geometric function theory are \emph{curves}, defined in the standard way as a continuous function $\gamma : (0,1) \to \widehat\bbC$.
A curve $\gamma$ is \emph{closed} if its endpoints are equal ($\lim_{t\to 0}\gamma(t) = \lim_{t\to 1}\gamma(t)$) and \emph{simple} if it never intersects itself ($\gamma(t) \ne \gamma(s)$ for all $t\ne s$).
For an absolutely continuous curve, its \emph{arclength} is $\int_0^1 |\gamma'(t)| \, \d t$, and the curve is \emph{rectifiable} if its arclength is finite.
We will abuse notation to let \(\gamma(t)\) denote the parameterized curve while also letting \(\gamma\subset\bbC\) denote the set of points traced by the curve.

\begin{figure}[t]
    \centering
    \def\blobPath{(1.05,0.15)
        .. controls (1.06,0.35) and (1.03,0.50) .. (0.98,0.65)
        .. controls (0.88,0.98) and (0.68,1.18) .. (0.45,1.28)
        .. controls (0.25,1.37) and (0.05,1.43) .. (-0.15,1.45)
        .. controls (-0.85,1.52) and (-1.35,1.05) .. (-1.25,0.45)
        .. controls (-1.15,-0.05) and (-1.65,-0.45) .. (-1.45,-1.00)
        .. controls (-1.25,-1.55) and (-0.45,-1.45) .. (-0.05,-1.15)
        .. controls (0.35,-0.85) and (0.55,-1.55) .. (1.10,-1.35)
        .. controls (1.65,-1.15) and (1.70,-0.35) .. cycle}

\begin{subfigure}[c]{0.47\textwidth}
        \centering
        \begin{tikzpicture}[
            scale=0.80,
            every node/.style={font=\footnotesize},
            dotregion/.style={
                fill=PineGreen!8,
                postaction={
                    pattern={Dots[distance=4.5pt, radius=0.60pt]},
                    pattern color=PineGreen!65
                }
            }
        ]
\path[dotregion] \blobPath;

\draw[->, gray] (-2.55,0) -- (2.55,0)
                node[right, text=black] {$\Re$};
            \draw[->, gray] (0,-2.35) -- (0,2.35)
                node[above, text=black] {$\Im$};

\draw[PineGreen, very thick, dashed] \blobPath;

\draw[BrickRed, very thick]
                (1.05,0.15)
                .. controls (0.32,0.12) and (0.08,-0.32) .. (-0.05,-1.15);
            \node[BrickRed, above left=-0.5pt] at (0.38,-0.22) {$\gamma$};

\filldraw[black] (0.764,1.051) circle (1.5pt)
                node[below left=1pt] {$a$};
            \filldraw[black] (1.12,-0.65) circle (1.5pt)
                node[left=2pt] {$b$};

\node[PineGreen!90!black] at (-0.62,0.58) {\Large $\cQ$};
        \end{tikzpicture}
        \caption{Example of a separating cross cut \(\gamma\in\Gamma_{\cQ}(a,b)\). Here, the shaded region \(\cQ\) lies inside of the dashed boundary, so \(\gamma\) also lies inside the boundary.}
        \label{fig:crosscut-interior}
    \end{subfigure}
    \hfill
\begin{subfigure}[c]{0.47\textwidth}
        \centering
        \begin{tikzpicture}[
            scale=0.80,
            every node/.style={font=\footnotesize},
            dotregion/.style={
                fill=PineGreen!8,
                postaction={
                    pattern={Dots[distance=4.5pt, radius=0.60pt]},
                    pattern color=PineGreen!65
                }
            }
        ]
\path[dotregion, even odd rule]
                (-2.35,-2.15) rectangle (2.35,2.15)
                \blobPath;

\draw[->, gray] (-2.55,0) -- (2.55,0)
                node[right, text=black] {$\Re$};
            \draw[->, gray] (0,-2.35) -- (0,2.35)
                node[above, text=black] {$\Im$};

\draw[PineGreen, very thick, dashed] \blobPath;

\draw[black, very thick, {Bar[width=3.5pt]}-{Bar[width=3.5pt]}]
                (0.98,0.65)
                .. controls (0.88,0.98) and (0.68,1.18) .. (0.45,1.28);
            \node[black, below left=1pt] at (0.76,1.05) {$\cA$};

\draw[BrickRed, very thick]
                (1.05,0.15)
                .. controls (1.85,0.45) and (1.75,1.55) .. (0.95,1.85)
                .. controls (0.35,2.08) and (-0.05,1.80) .. (-0.15,1.45);
            \node[BrickRed, above right=1pt] at (1.35,1.45) {$\gamma$};

\node[PineGreen!90!black] at (-1.65,1.55) {\Large $\cQ$};
        \end{tikzpicture}
        \caption{Example of a separating cross-cut \(\gamma\in\Gamma_{\cQ}(\cA,\infty)\). Here, \(\gamma\) ensures that every path from \(\cA\) to \(\infty\) on the shaded area \(\cQ\) must intersect \(\gamma\).}
        \label{fig:crosscut-exterior}
    \end{subfigure}

    \caption{
        Examples of separating cross-cuts on various domains \(\cQ\).
        It is possible for the domain \(\cQ\) to be either bounded, as in \cref{fig:crosscut-interior}, or to be unbounded, as in \cref{fig:crosscut-exterior}.
    }
    \label{fig:crosscut}
\end{figure}
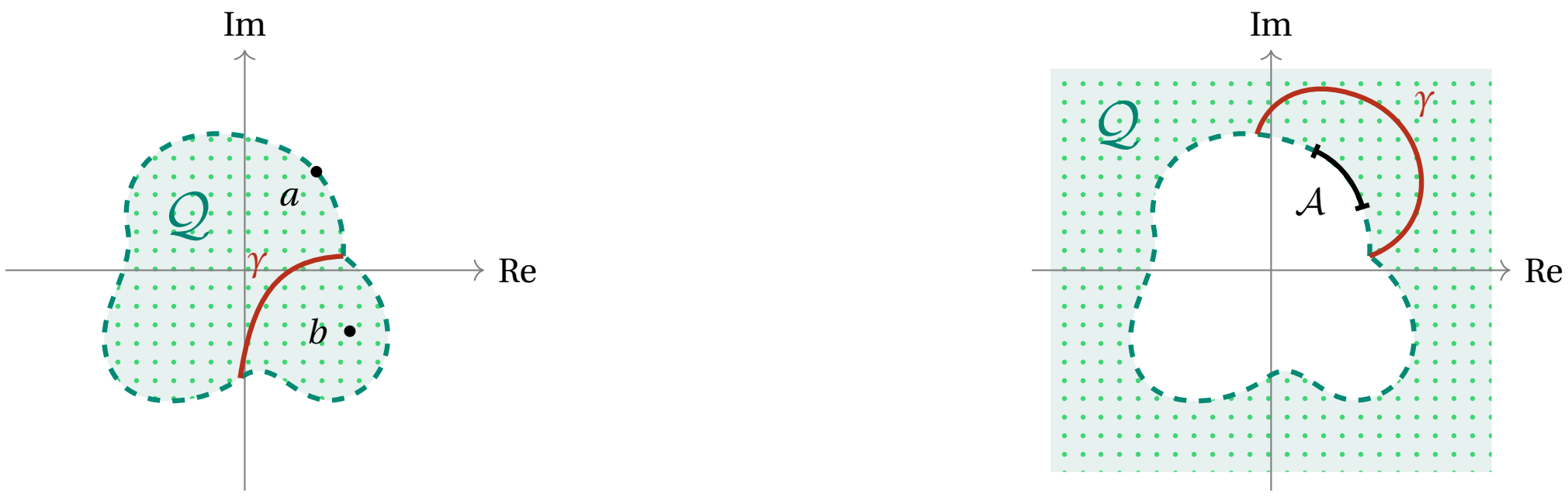

Our study will focus on a particular type of curve known as a \emph{cross cut}.
\begin{definition}[Cross cut]
    Let $\cS\subseteq \hat{\bbC}$ be open and simply connected.
    A curve \(\gamma\) in \cS is said to be a \emph{cross cut} of \cS if \(\gamma\) is simple, rectifiable, and has distinct endpoints lying on the boundary $\partial \cS$.
    A cross cut partitions $\overline\cS \setminus \overline\gamma$ into two regions.
    The cross cut $\gamma$ is said to \emph{separate} sets \cA and \cB if they lie in separate regions, and we let
\begin{equation*}
        \Gamma_{\cS}(\cA,\cB) \defeq \{ \gamma ~:~ \gamma \text{ separates \cA from \cB in \cS} \}.
    \end{equation*}
The collection \(\Gamma_{\cS}(\cA,\cB)\) is called the \emph{family of cross cuts} that separate \cA and \cB.
\end{definition}
We can apply the notion of separation also to points $a,b \in \overline\cS$ by identifying them with the singleton sets $\{a\},\{b\}$.
In particular, we can describe the set $\Gamma_\cS(a,b)$ or $\Gamma_\cS(a,\cB)$ of curves separating two points or a point from a set.
The notions of cross cuts and separating cross cuts are illustrated in \cref{fig:crosscut}.

We shall be interested in quantifying \emph{how difficult it is to separate \cA from \cB}.
To this end, for any (Borel-measurable) function $\rho : \cS \to \bbR_+$ and any rectifiable curve $\gamma$, we define the $\rho$-arclength as 
\begin{equation*}
    \int_\gamma \rho (z) \, \d s(z) = \int_0^1 \rho(\gamma(t)) |\gamma'(t)| \, \d t.
\end{equation*}
Here, \(s\) denotes the arclength measure on the complex plane.
The function $\rho$ is often called a \emph{metric}.

\begin{definition}[Modulus of a set of curves]
    Let $\cS \subseteq \hat{\bbC}$ be simply connected, and let $\Gamma$ be a set of rectifiable curves in \cS.
    We say a metric $\rho : \cS \to \bbR_+$ is \emph{admissible} for \(\Gamma\) if each $\gamma \in \Gamma$ has $\rho$-arclength of at least one.
    The \emph{modulus} of $\Gamma$ is the smallest $\rho^2$-area of \cS under any admissible metric
\begin{equation*}
        \rM_{\cS}(\Gamma) \defeq \inf \left\{ \iint_{\cS} |\rho(z)|^2 \, \d A(z) ~:~ \rho \text{ is admissible for } \Gamma \right\}.
    \end{equation*}
Here, $A$ denotes the area measure for the complex plane.
\end{definition}
We use the shorthands
\[
    \rM_{\cS}(a,b) \defeq \rM_{\cS}(\Gamma_{\cS}(a,b)),
\quad
    \rM_{\cS}(\cA,\cB) \defeq \rM_{\cS}(\Gamma_{\cS}(\cA,\cB)), \quad \text{etc.}
\]
for moduli of families of cross cuts, which shall be the essential quantities in our analysis.
For the reader interested in learning more about the origins and principles of this quantity, we suggest reading Chapter 4 of the excellent and approachable book by Beliaev \cite{beliaev2020conformal}.

As one would expect for any natural concept in complex analysis, the set of cross cuts and the modulus interact nicely with conformal mappings.

\begin{importedtheorem}[Conformal maps respect the modulus of cross cuts, \protect{\cite[Thms.~2.24 and 4.2]{beliaev2020conformal}}]
    \label{impthm:conformal-mods-cuts}
    Let \(\cQ,\cQ'\subseteq\widehat\bbC\) be domains whose boundaries are simple closed curves.
    Let \(\chi : \cQ \to \cQ'\) be a conformal bijection.
    For any disjoint sets \(\cA,\cB\subseteq\overline{\cQ}\), we have \(\rM_{\cQ}(\cA,\cB) = \rM_{\cQ'}(\chi(\cA),\chi(\cB))\). 
\end{importedtheorem}

\subsection{Relating the Green's function and the modulus}
\label{sec:geometric-greens}

In order to use the modulus, we need to know about the relationship between the modulus and the Green's function.
We prove the following bound.

\begin{theorem}[Modulus bounds the Green's function]
    \label{thm:modulus-bounds-greens}
    Let \cS be the closure of a Jordan domain with \(0 \notin\cS\).
    Let \(\cQ \defeq \widehat\bbC \setminus \cS\) and assume that \(\rM_{\cQ}(0,\infty) \geq 1\).
    Then, \(G_{\cS}(0) \leq 10 \e^{-\pi \rM_{\cQ}(0,\infty)}\).
\end{theorem}

Results similar to \cref{thm:modulus-bounds-greens} are implicit in prior works \cite{andrievskii2004green,beliaev2020conformal}, but we are unaware of a reference for this result.
To prove the bound, we start by importing a modulus bound on a commonly studied shape called the \emph{Gr\"otzsch ring}.
Here, we look at the set of curves on the unit disk that \emph{connect} the unit circle to an interval inside the circle.

\medskip
\noindent
\begin{minipage}[c]{0.75\textwidth}
    \begin{importedtheorem}[Modulus of connecting cuts on the disk, \protect{\cite[Eqn.~2.4]{betsakos1998harmonic} and \cite[Eqns.~5.57, 5.58]{vuorinen2006conformal}}]
        \label{impthm:modulus-connecting-grotzsch}
        Let \(\cD = \{z~:~|z|<1\}\) be the unit disk, and let \(w\in(0,1)\).
        Let \(\Gamma_w\) be the set of rectifiable curves with one endpoint on \(\partial\cD\) and the other endpoint on the interval \([0,w]\).
        Then,
        \[
\rM_{\cD}(\Gamma_w) \leq \frac{4}{\pi}\log\left(4 + \frac{8}{w^{-1}-1}\right).
        \]
    \end{importedtheorem}
\end{minipage}\hfill
\begin{minipage}[c]{0.23\textwidth}
    \vspace{-0.5em}
    \centering
    \begin{tikzpicture}[
    scale=0.58,
    every node/.style={font=\footnotesize},
    dotregion/.style={
        fill=PineGreen!8,
        postaction={
            pattern={Dots[distance=4.2pt, radius=0.55pt]},
            pattern color=PineGreen!65
        }
    }
]
    \def\Rdisk{1.35}

\path[dotregion] (0,0) circle (\Rdisk);

\draw[->, gray] (-1.68,0) -- (1.75,0)
        node[right, text=black] {$\Re$};
    \draw[->, gray] (0,-1.68) -- (0,1.68)
        node[above, text=black] {$\Im$};

\draw[PineGreen, very thick, dashed] (0,0) circle (\Rdisk);

\draw[BrickRed, very thick]
        (124:\Rdisk)
        .. controls (-0.42,0.72) and (0.30,0.62) .. (0.38,0);
    \node[BrickRed, inner sep=1pt] at (-0.88,1.34) {$\gamma$};

\draw[black, very thick] (0,0) -- (0.82,0);
    \filldraw[black] (0,0) circle (1.5pt);
    \filldraw[black] (0.82,0) circle (1.5pt);
    \node[anchor=base, inner sep=0pt, font=\scriptsize] at (-0.20,0.20) {$0$};
    \node[anchor=base, inner sep=0pt, font=\scriptsize] at (0.82,0.20) {$w$};
    \node[PineGreen!90!black] at (-0.64,-0.64) {$\cD$};

\draw[gray!55, line width=0.4pt]
        ([shift={(-4pt,-3.5pt)}]current bounding box.south west)
        rectangle
        ([shift={(4pt,3.5pt)}]current bounding box.north east);
\end{tikzpicture}

     \par\vspace{0.5em}
\end{minipage}
\medskip

The result in \cref{impthm:modulus-connecting-grotzsch} uses connecting curves instead of cross curves.
To make this result easier for us to use in the context of cross cuts, we perform a simple reduction.

\medskip
\noindent
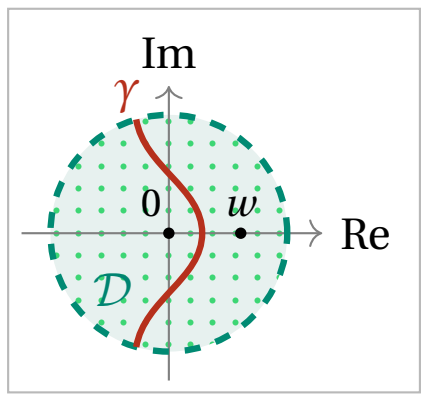
\begin{minipage}[c]{0.23\textwidth}
    \vspace{0.5em}
    \centering
    \begin{tikzpicture}[
    scale=0.58,
    every node/.style={font=\footnotesize},
    dotregion/.style={
        fill=PineGreen!8,
        postaction={
            pattern={Dots[distance=4.2pt, radius=0.55pt]},
            pattern color=PineGreen!65
        }
    }
]
    \def\Rdisk{1.35}

\path[dotregion] (0,0) circle (\Rdisk);

\draw[->, gray] (-1.68,0) -- (1.75,0)
        node[right, text=black] {$\Re$};
    \draw[->, gray] (0,-1.68) -- (0,1.68)
        node[above, text=black] {$\Im$};

\draw[PineGreen, very thick, dashed] (0,0) circle (\Rdisk);

\draw[BrickRed, very thick]
        (106:\Rdisk)
        .. controls (-0.28,0.75) and (0.38,0.52) .. (0.38,0)
        .. controls (0.38,-0.52) and (-0.28,-0.75) .. (-106:\Rdisk);
    \node[BrickRed, inner sep=1pt] at (-0.48,1.56) {$\gamma$};

\filldraw[black] (0,0) circle (1.5pt);
    \filldraw[black] (0.82,0) circle (1.5pt);
    \node[anchor=base, inner sep=0pt, font=\scriptsize] at (-0.20,0.20) {$0$};
    \node[anchor=base, inner sep=0pt, font=\scriptsize] at (0.82,0.20) {$w$};
    \node[PineGreen!90!black] at (-0.64,-0.64) {$\cD$};

\draw[gray!55, line width=0.4pt]
        ([shift={(-4pt,-3.5pt)}]current bounding box.south west)
        rectangle
        ([shift={(4pt,3.5pt)}]current bounding box.north east);
\end{tikzpicture}

     \par\vspace{-0.5em}
\end{minipage}
\hfill
\begin{minipage}[c]{0.75\textwidth}
    \begin{lemma}[Modulus of cross cuts on the disk]
        \label{lem:modulus-cutting-grotzsch}
        Let \(\cD = \{z~:~|z|<1\}\) be the unit disk, and let \(w\in(0,1)\).
        Then, the modulus of cross cuts separating \(0\) and \(w\) on the disk satisfies
        \[
\rM_{\cD}(0,w) \leq \frac{1}{\pi}\log\left(4 + \frac{8}{w^{-1}-1}\right).
        \]
    \end{lemma}
\end{minipage}\medskip

\begin{proof}
Let \(\gamma\) \hangindent=-88pt \hangafter=-5 be any curve in \(\Gamma_{\cD}(0,w)\).\vadjust{\vbox to 0pt{\kern-1.05\baselineskip\kern0.65em\hbox to \textwidth{\hfill\begin{tikzpicture}[
    scale=0.58,
    every node/.style={font=\footnotesize},
    dotregion/.style={
        fill=PineGreen!8,
        postaction={
            pattern={Dots[distance=4.2pt, radius=0.55pt]},
            pattern color=PineGreen!65
        }
    }
]
    \def\Rdisk{1.35}

\path[dotregion] (0,0) circle (\Rdisk);

\draw[->, gray] (-1.68,0) -- (1.75,0)
        node[right, text=black] {$\Re$};
    \draw[->, gray] (0,-1.68) -- (0,1.68)
        node[above, text=black] {$\Im$};

\draw[PineGreen, very thick, dashed] (0,0) circle (\Rdisk);

\draw[gray!85!black, line width=1.6pt] (0,0) -- (0.86,0);

\draw[orange!80!Goldenrod, very thick]
        (0.24,0)
        .. controls (0.28,-0.34) and (0.40,-0.34) .. (0.44,0)
        .. controls (0.48,0.34) and (0.60,0.34) .. (0.64,0);

\draw[BrickRed, very thick]
        (114:\Rdisk)
        .. controls (-0.28,0.72) and (0.20,0.34) .. (0.24,0);
    \node[BrickRed, inner sep=1pt] at (-0.82,1.48) {$\gamma_1$};

    \draw[BrickRed, very thick]
        (0.64,0)
        .. controls (0.70,-0.50) and (0.66,-0.90) .. (-64:\Rdisk);
    \node[BrickRed, inner sep=1pt] at (0.96,-1.44) {$\gamma_2$};

\filldraw[black] (0,0) circle (1.5pt);
    \filldraw[black] (0.86,0) circle (1.5pt);
    \node[anchor=base, inner sep=0pt, font=\scriptsize] at (-0.20,0.20) {$0$};
    \node[anchor=base, inner sep=0pt, font=\scriptsize] at (0.86,0.20) {$w$};
    \node[PineGreen!90!black] at (-0.64,-0.64) {$\cD$};

\draw[gray!55, line width=0.4pt]
        ([shift={(-4pt,-3.5pt)}]current bounding box.south west)
        rectangle
        ([shift={(4pt,3.5pt)}]current bounding box.north east);
\end{tikzpicture}
 }\vss}}
    Since \(\gamma\) separates 0 and \(w\), it must intersect the interval \([0,w]\) at least once.
    As shown in the diagram to the right, we let \(\gamma_1\) be the subarc of \(\gamma\) before its first intersection with the interval and let \(\gamma_2\) be the subarc of \(\gamma\) after its last intersection with the interval.
    Since both \(\gamma_1\) and \(\gamma_2\) have one endpoint on \(\partial\cD\) and one endpoint on the interval \([0,w]\), we have that \(\gamma_1,\gamma_2\in\Gamma_w\).

Now let \(\rho\) \hangindent=-88pt \hangafter=-1 be any metric admissible on \(\Gamma_w\).
    We know that \(\gamma_1\) and \(\gamma_2\) have \(\rho\)-arclength at least one.
    So, we know that \(\gamma\) has \(\rho\)-arclength at least two.
    As such, the metric \(\widehat\rho(z) \defeq \frac12\rho(z)\) is admissible for \(\Gamma_{\cD}(0,w)\).
    So, we have
    \[
        \rM_{\cD}(0,w)
        \leq \iint_{\cD} |\widehat\rho(z)|^2 \, \d A(z)
        = \frac14 \iint_{\cD} |\rho(z)|^2 \, \d A(z).
\]
    Taking the infimum over all \(\rho\) admissible for \(\Gamma_w\) gives \(\rM_{\cD}(0,w) \leq \frac14 \rM_{\cD}(\Gamma_w)\).
    Applying \cref{impthm:modulus-connecting-grotzsch} completes the theorem.
\end{proof}

We have now set ourselves up to prove the modulus-to-Green's relation in \cref{thm:modulus-bounds-greens}.

\begin{figure}[t]
    \centering
    \begin{tikzpicture}[
        scale=0.72,
        every node/.style={font=\footnotesize},
        curve label/.style={font=\scriptsize, inner sep=1pt},
        dotregion/.style={
            fill=PineGreen!8,
            postaction={
                pattern={Dots[distance=4.5pt, radius=0.60pt]},
                pattern color=PineGreen!65
            }
        }
    ]
        \def\kidneyPath{(0.65,0.45) .. controls (0.95,0.75) and (0.95,1.40) .. (0.45,1.65)
            .. controls (-0.10,1.90) and (-0.95,1.75) .. (-1.35,1.20)
            .. controls (-1.75,0.65) and (-1.45,0.10) .. (-1.30,-0.35)
            .. controls (-1.15,-0.80) and (-1.65,-1.20) .. (-1.25,-1.55)
            .. controls (-0.85,-1.90) and (-0.05,-1.75) .. (0.20,-1.30)
            .. controls (0.45,-0.85) and (-0.45,-0.55) .. (-0.40,-0.05)
            .. controls (-0.35,0.35) and (0.35,0.15) .. cycle}

\begin{scope}[shift={(0,0)}]
            \path[dotregion, even odd rule]
                (-2.35,-2.35) rectangle (2.35,2.35)
                \kidneyPath;

\draw[->, gray] (-2.60,0) -- (2.60,0)
                node[right, text=black] {$\Re$};
            \draw[->, gray] (0,-2.60) -- (0,2.60)
                node[above, text=black] {$\Im$};

\draw[NavyBlue, very thick, dashed] \kidneyPath;

\draw[BrickRed, very thick]
                (0.65,0.45)
                .. controls (1.50,0.35) and (1.45,-1.90) .. (0.30,-2.00)
                .. controls (-0.20,-2.05) and (-0.42,-1.88) .. (-0.47,-1.73);
            \node[BrickRed, below right=1pt] at (1.02,-1.50) {$\gamma$};

\filldraw[black] (0,0) circle (1.5pt)
                node[below right=1pt] {$0$};
            \node[NavyBlue] at (-0.65,1.00) {\large $\cS$};
            \node[PineGreen!90!black] at (1.45,1.75) {\large $\cQ$};
        \end{scope}

\begin{scope}[shift={(6.8,0)}]
            \def\Rdisk{1.05}
            \path[dotregion, even odd rule]
                (-2.35,-2.35) rectangle (2.55,2.35)
                (0,0) circle (\Rdisk);

\draw[->, gray] (-2.60,0) -- (2.80,0)
                node[right, text=black] {$\Re$};
            \draw[->, gray] (0,-2.60) -- (0,2.60)
                node[above, text=black] {$\Im$};
            \draw[gray] (\Rdisk,-0.10) -- (\Rdisk,0.10);
            \node[gray, below left=1pt, font=\scriptsize] at (\Rdisk,0) {$1$};

\draw[NavyBlue, very thick, dashed] (0,0) circle (\Rdisk);

\draw[BrickRed, very thick]
                (45:\Rdisk)
                .. controls (1.65,1.15) and (2.35,0.55) .. (2.35,0)
                .. controls (2.35,-0.55) and (1.65,-1.15) .. (-45:\Rdisk);
            \node[BrickRed, below right=1pt] at (1.85,-0.75) {$\gamma$};

\filldraw[black] (1.60,0) circle (1.5pt)
                node[above=2pt] {$\e^g$};
            \node[NavyBlue] at (-0.45,0.40) {\large $\cD$};
        \end{scope}

\begin{scope}[shift={(13.8,0)}]
            \def\Rdisk{1.05}
            \path[dotregion]
                (0,0) circle (\Rdisk);

\draw[->, gray] (-2.60,0) -- (2.60,0)
                node[right, text=black] {$\Re$};
            \draw[->, gray] (0,-2.60) -- (0,2.60)
                node[above, text=black] {$\Im$};
            \node[gray, below right=1pt, font=\scriptsize] at (\Rdisk,0) {$1$};

\draw[NavyBlue, very thick, dashed] (0,0) circle (\Rdisk);

\draw[BrickRed, very thick]
                (104:\Rdisk)
                .. controls (-0.22,0.58) and (0.30,0.40) .. (0.30,0)
                .. controls (0.30,-0.40) and (-0.22,-0.58) .. (-104:\Rdisk);
            \node[BrickRed, inner sep=1pt] at (-0.36,1.24) {$\gamma$};

\filldraw[black] (0,0) circle (1.5pt)
                node[above left=0.5pt, inner sep=1pt] {$0$};
            \filldraw[black] (0.58,0) circle (1.5pt)
                node[above right=0.5pt, font=\scriptsize, inner sep=1pt] {$w=\e^{-g}$};
            \node[PineGreen!90!black] at (-0.52,-0.45) {\large $\cD$};
        \end{scope}

\draw[->, >=Stealth, thick, black]
            (1.95,2.10) to[bend left=25] node[midway, above] {\large $\widehat{\chi}$} (4.85,2.10);
        \draw[->, >=Stealth, thick, black]
            (8.85,2.10) to[bend left=25] node[midway, above] {\large $\psi$} (11.75,2.10);
    \end{tikzpicture}
    \caption{
        Visualization of the sequence of transformations performed in the proof of \cref{thm:modulus-bounds-greens}.
        \textbf{First panel:~}
        We start with an arbitrary Jordan domain \cS that does not contain the point 0, and focus on cross cuts separating \(0\) from \(\infty\) on \(\cQ = \widehat\bbC\setminus\cS\).
        \textbf{Second panel:~}
        The conformal map \(\widehat\chi\) maps \(\cQ\) to the outside of the unit disk \cD.
        By \cref{eq:chi-def}, we know that \(\widehat\chi(0) = \e^{g}\) where \(g = G_{\cS}(0)\) is the Green's function at 0.
        This map preserves \(\widehat\chi(\infty)=\infty\), so we are interested in cross cuts that separate \(\e^g\) from \(\infty\).
        \textbf{Third panel:~}
        The conformal map \(\psi\) maps the outside of the unit disk to the inside of the unit disk.
        We now want curves \(\gamma\) that separate \(w \defeq \psi(\e^{g}) = \e^{-g}\) and \(\psi(\infty)=0\).
}
    \label{fig:modulus-greens}
\end{figure}
 
\begin{proof}[Proof of \cref{thm:modulus-bounds-greens}.]
    This proof works by chaining together a sequence of conformal maps.
    We visualize these steps in \cref{fig:modulus-greens}.

    Our first conformal transformation maps from \(\cQ\defeq\widehat{\bbC} \setminus \cS\) to the outside $\widehat\bbC\setminus\overline\cD$ of the unit disk, as shown in the first step of \cref{fig:modulus-greens}.
    Recall that exactly such a map $\chi$ was introduced above in \cref{eq:chi-def} to define the Green's function.
    In particular, $|\chi(0)| = \e^g$ is the exponential of the Green's function at zero $g \coloneqq G_{\cS}(0)$.
    To simplify bookkeeping, it will be helpful to multiply $\chi$ by a phase $\e^{\ri\phi}$ for $\phi \in \bbR$ so that
\begin{equation*}
        \hat{\chi}(z) \defeq \e^{\ri\phi}\chi(z)
        \quad\text{satisfies}\quad
        \hat{\chi}(0) = \e^g.
    \end{equation*}
That is, we have simply rotated the conformal map so that it sends zero to a point on the real line, exactly as shown in \cref{fig:modulus-greens}.
    Next, we apply the conformal map \(\psi(z) \defeq 1/z\), which maps the outside of the unit circle to the inside of the unit circle, as shown in the second step of \cref{fig:modulus-greens}.

    The composition $\psi \circ \hat{\chi}$ conformally maps the open exterior $\widehat{\bbC} \setminus \cS$ of the original region to the unit disk \cD.
    The points \(0\) and \(\infty\) are sent to
    \[
        \psi(\widehat\chi(0)) = \psi(\e^g) = \e^{-g}
        \quad\text{and}\quad
        \psi(\widehat\chi(\infty)) = \psi(\infty) = 0.
    \]
By the conformal invariance of the modulus (\cref{impthm:conformal-mods-cuts}) and our bound on the cross cuts in the unit disk (\cref{lem:modulus-cutting-grotzsch}), we have
    \[
        \rM_{\cQ}(0,\infty)
        = \rM_{\widehat\bbC \setminus \overline\cD}(\e^g,\infty)
        = \rM_{\cD}(\e^{-g},0)
\leq \frac{1}{\pi} \log\left(4 + \frac{8}{\e^g-1}\right)
        \leq \frac1\pi \log\left(4 + \frac{8}g\right),
    \]
    where the last inequality uses \(\e^g - 1 \geq g\).
    Equivalently, we have \(\e^{\pi\rM_{\cQ}(0,\infty)} \leq 4 + 8/g\).
    Since \(\rM_{\cQ}(0,\infty) \geq 1\), we have \(\e^{\pi\rM_{\cQ}(0,\infty)} - 4 \geq 4/5 \cdot \e^{\pi\rM_{\cQ}(0,\infty)}\).
    Hence,
    \[
        \frac45 \e^{\pi\rM_{\cQ}(0,\infty)} \leq \e^{\pi\rM_{\cQ}(0,\infty)} - 4 \leq \frac 8g. 
    \]
    Rearranging the outermost terms recovers \(g \leq 10 \e^{-\pi\rM_{\cQ}(0,\infty)}\), completing the proof.
\end{proof}

\subsection{The modulus for the polar rectangle}
\label{sec:geometric-polar-rect}

In this section, we finally see the full power of the modulus of cross cuts, with it allowing us to effortlessly bound the Green's function for a region like $\cP_{\kappa,\theta}$ by carving its complement $\cQ \defeq \widehat\bbC \setminus \cP_{\kappa,\theta}$ into geometrically simple subregions.
We can then compute the corresponding modulus of cross cuts on each subregion to bound the modulus on \(\cQ\).
The fundamental result of this section is the following.
\begin{theorem}[Bounding the modulus for the polar rectangle]
    \label{thm:polar-rect-modulus}
    For any $\kappa \ge 1$ and $\theta \in (0,2\pi]$, it holds that 
    \[
        \rM_{\cQ}(0,\infty)
        \geq \frac{\log \kappa}{\theta} + \frac2\pi \log\left(\frac{2\pi}{\theta}\right).
    \]
\end{theorem}

Analogously to the harmonic function theory section, we will geometrically split up the complement space \(\cQ \defeq \widehat\bbC \setminus \cP_{\kappa,\theta}\) into smaller geometric regions.
The following fact, sometimes also called ``the parallel rule'', shows one way this can be done for the modulus.

\begin{lemma}[Superadditivity of the modulus; see also \protect{\cite[Prop.~4.7]{beliaev2020conformal}}]
    \label{lem:parallel-rule}
    Let \(\cQ\subseteq\widehat\bbC\) be a domain and let \(\Gamma\) be a set of curves on \cQ.
    Let \(\cQ_1,\ldots,\cQ_k\subseteq\cQ\) be disjoint subsets of \cQ.
    Let \(\Gamma_1,\ldots,\Gamma_k\subseteq\Gamma\) such that \(\Gamma_i\) contains only curves supported on \(\cQ_i\).
    Then, \(\rM_{\cQ}(\Gamma) \geq \sum_{i=1}^k \rM_{\cQ_i}(\Gamma_i)\).
\end{lemma}
\begin{proof}
    Let \(\rho\) be any admissible metric for \(\Gamma\).
    Since every \(\Gamma_i\) is a subset of \(\Gamma\), we know that \(\rho\) is admissible for \(\Gamma_i\), and therefore that \(\iint_{\cQ_i} |\rho(z)|^2 \, \d A(z) \geq \rM_{\cQ_i}(\Gamma_i)\).
    Since the domains \(\cQ_i\) are disjoint, we have
    \[
        \iint_{\cQ} |\rho(z)|^2 \, \d A(z)
        \geq \sum\limits_{i=1}^k \iint_{\cQ_i} |\rho(z)|^2 \, \d A(z)
        \geq \sum\limits_{i=1}^k \rM_{\cQ_i}(\Gamma_i).
    \]
    Taking the infimum over all admissible \(\rho\) completes the proof.
\end{proof}

This proof gives us permission to consider all cross cuts of \(\cQ\) that separate 0 and infinity, split \(\cQ\) into three subregions, and consider corresponding crosscuts on those subregions.
In our analysis, we will see two simpler modulus problems arise.

\medskip
\noindent
\begin{minipage}[c]{0.75\textwidth}
    \begin{importedtheorem}[Modulus of cross cuts across the rectangle, \protect{\cite[Pg.~137]{beliaev2020conformal}}]
        \label{lem:rectangle-modulus}
        Consider the rectangle \cR of width \(w\) and height \(h\):
        \[
            \cR \defeq \bigl\{x+\ri y ~:~ x\in(0, w), y\in(0,h)\bigr\}.
        \]
Let \(\cA = \{\ri y~:~y\in[0,h]\}\) and \(\cB = \{w+\ri y~:~y\in[0,h]\}\) be the left and right sides of the rectangle.
        Then, \(\rM_{\cR}(\cA,\cB) = \frac{w}{h}\).
    \end{importedtheorem}
\end{minipage}\hfill
\begin{minipage}[c]{0.23\textwidth}
    \centering
    \begin{tikzpicture}[
    scale=0.62,
    every node/.style={font=\footnotesize},
    dotregion/.style={
        fill=PineGreen!8,
        postaction={
            pattern={Dots[distance=4.2pt, radius=0.55pt]},
            pattern color=PineGreen!65
        }
    }
]
    \def\Wrect{2.2}
    \def\Hrect{1.35}

\path[dotregion] (0,0) rectangle (\Wrect,\Hrect);

\draw[->, gray] (-0.38,0) -- (2.72,0)
        node[right, text=black] {$\Re$};
    \draw[->, gray] (0,-0.38) -- (0,1.82)
        node[above, text=black] {$\Im$};

\draw[PineGreen, very thick, dashed] (0,\Hrect) -- (\Wrect,\Hrect);
    \draw[PineGreen, very thick, dashed] (0,0) -- (\Wrect,0);

\draw[black, very thick] (0,0) -- (0,\Hrect)
        node[midway, left=1.5pt, text=black] {$\cA$};
    \draw[black, very thick] (\Wrect,0) -- (\Wrect,\Hrect)
        node[midway, right=1.5pt, text=black] {$\cB$};

\draw[BrickRed, very thick]
        (0.95,\Hrect)
        .. controls (1.10,0.85) and (0.78,0.50) .. (0.92,0);
    \node[BrickRed, left=0.5pt] at (0.90,0.62) {$\gamma$};

\node[PineGreen!90!black] at (1.58,0.68) {$\cR$};
    \filldraw[black] (\Wrect,0) circle (1.4pt);
    \filldraw[black] (\Wrect,\Hrect) circle (1.4pt)
        node[above right=-1pt, font=\scriptsize] {$w+\ri h$};

\draw[gray!55, line width=0.4pt]
        ([shift={(-4pt,-3.5pt)}]current bounding box.south west)
        rectangle
        ([shift={(4pt,3.5pt)}]current bounding box.north east);
\end{tikzpicture}

 \end{minipage}
\medskip

\noindent
\begin{minipage}[c]{0.23\textwidth}
    \centering
    \begin{tikzpicture}[
    scale=0.58,
    every node/.style={font=\footnotesize},
    dotregion/.style={
        fill=PineGreen!8,
        postaction={
            pattern={Dots[distance=4.2pt, radius=0.55pt]},
            pattern color=PineGreen!65
        }
    }
]
    \def\Rdisk{1.35}
    \def\halfTheta{38}

\begin{scope}[xscale=-1]
\path[dotregion] (0,0) circle (\Rdisk);

    \end{scope}

\draw[->, gray] (-1.68,0) -- (1.75,0)
        node[right, text=black] {$\Re$};
    \draw[->, gray] (0,-1.68) -- (0,1.68)
        node[above, text=black] {$\Im$};

    \begin{scope}[xscale=-1]

\draw[gray] (0,0) -- (\halfTheta:\Rdisk);
    \draw[gray] (0,0) -- ({-\halfTheta}:\Rdisk);
    \draw[gray]
        ({-\halfTheta}:0.62) arc[start angle=-\halfTheta, end angle=\halfTheta, radius=0.62];
    \node[font=\scriptsize] at (16:0.86) {$\theta$};

\draw[PineGreen, very thick, dashed]
        (\halfTheta:\Rdisk) arc[start angle=\halfTheta, end angle={360-\halfTheta}, radius=\Rdisk];
    \draw[black, very thick, {Bar[width=3.5pt]}-{Bar[width=3.5pt]}]
        ({-\halfTheta}:\Rdisk) arc[start angle=-\halfTheta, end angle=\halfTheta, radius=\Rdisk];
    \node[black, above left=0pt] at (22:\Rdisk) {$\cC_\theta$};

\draw[BrickRed, very thick]
        (108:\Rdisk)
        .. controls (0.15,0.88) and (0.35,0.38) .. (0.35,0)
        .. controls (0.35,-0.38) and (0.15,-0.88) .. (-108:\Rdisk);
    \node[BrickRed, right=1pt] at (-0.12,0.88) {$\gamma$};

\filldraw[black] (0,0) circle (1.5pt)
        node[above right=0.5pt, font=\scriptsize] {$0$};
    \node[PineGreen!90!black] at (-0.75,-0.68) {$\cD$};
    \end{scope}

\draw[gray!55, line width=0.4pt]
        ([shift={(-4pt,-3.5pt)}]current bounding box.south west)
        rectangle
        ([shift={(4pt,3.5pt)}]current bounding box.north east);
\end{tikzpicture}
 \end{minipage}
\hfill
\begin{minipage}[c]{0.75\textwidth}
    \begin{importedtheorem}[Modulus of cross cuts escaping the circle, \protect{\cite[Thm.~4.15 and Def.~2.45]{beliaev2020conformal}}]
        \label{impthm:circular-escape}
        Let \(\cD \defeq \{z~:~|z|<1\}\) be the unit disk, and
        define the circular arc \(\cC_\theta \defeq \{-\e^{\ri \phi} ~:~ |\phi| < \theta/2\}\).
        Then \(\rM_{\cD}(\Gamma_{\cD}(0,\cC_\theta)) \geq \frac1\pi \log(2\pi/\theta)\).
    \end{importedtheorem}
\end{minipage}\medskip

We are now ready to prove \cref{thm:polar-rect-modulus}, the lower bound on the modulus of cross cuts that separate 0 and \(\infty\) on \(\cQ\).

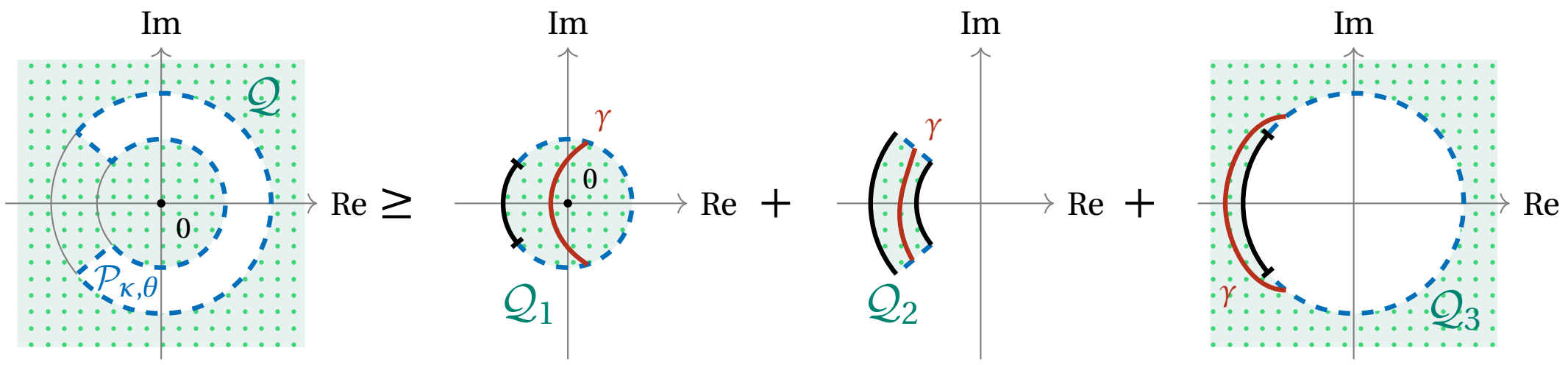
\begin{figure}[t]
    \centering
    \begin{tikzpicture}[
        scale=0.55,
        every node/.style={font=\footnotesize},
        dotregion/.style={
            fill=PineGreen!8,
            postaction={
                pattern={Dots[distance=4.2pt, radius=0.58pt]},
                pattern color=PineGreen!65
            }
        }
    ]
\def\Rin{1.05}
        \def\Rout{1.80}
        \def\halfTheta{40}

\begin{scope}[shift={(0,0)}, xscale=-1]
            \coordinate (outerLower) at ({-\halfTheta}:\Rout);
            \coordinate (outerUpper) at (\halfTheta:\Rout);
            \coordinate (innerUpper) at (\halfTheta:\Rin);
            \coordinate (innerLower) at ({-\halfTheta}:\Rin);

            \path[dotregion, even odd rule]
                (-2.35,-2.35) rectangle (2.35,2.35)
                (outerUpper)
                arc[start angle=\halfTheta, end angle={360-\halfTheta}, radius=\Rout]
                -- (innerLower)
                arc[start angle=-\halfTheta, end angle={\halfTheta-360}, radius=\Rin]
                -- cycle;

        \end{scope}
        \begin{scope}[shift={(0,0)}]
\draw[->, gray] (-2.55,0) -- (2.55,0)
                node[right, text=black] {$\Re$};
            \draw[->, gray] (0,-2.55) -- (0,2.55)
                node[above, text=black] {$\Im$};
        \end{scope}
        \begin{scope}[shift={(0,0)}, xscale=-1]

\draw[gray]
                (innerLower) arc[start angle=-\halfTheta, end angle=\halfTheta, radius=\Rin];
            \draw[gray]
                (outerLower) arc[start angle=-\halfTheta, end angle=\halfTheta, radius=\Rout];

\draw[NavyBlue, very thick, dashed, line join=round]
                (outerUpper)
                arc[start angle=\halfTheta, end angle={360-\halfTheta}, radius=\Rout]
                -- (innerLower)
                arc[start angle=-\halfTheta, end angle={\halfTheta-360}, radius=\Rin]
                -- cycle;

            \filldraw[black] (0,0) circle (1.6pt)
                node[below right=0.5pt, font=\scriptsize] {$0$};
            \node[NavyBlue] at (293:1.41) {$\cP_{\kappa,\theta}$};
            \node[PineGreen!90!black] at (-1.65,1.75) {\large $\cQ$};
        \end{scope}

\node at (3.85,0) {\Large $\geq$};

\begin{scope}[shift={(6.65,0)}, xscale=-1]
            \path[dotregion] (0,0) circle (\Rin);

        \end{scope}
        \begin{scope}[shift={(6.65,0)}]
\draw[->, gray] (-1.85,0) -- (1.95,0)
                node[right, text=black] {$\Re$};
            \draw[->, gray] (0,-2.55) -- (0,2.55)
                node[above, text=black] {$\Im$};
        \end{scope}
        \begin{scope}[shift={(6.65,0)}, xscale=-1]

\draw[NavyBlue, very thick, dashed]
                (\halfTheta:\Rin) arc[start angle=\halfTheta, end angle={360-\halfTheta}, radius=\Rin];
            \draw[black, very thick, {Bar[width=3.5pt]}-{Bar[width=3.5pt]}]
                ({-\halfTheta}:\Rin) arc[start angle=-\halfTheta, end angle=\halfTheta, radius=\Rin];

\draw[BrickRed, very thick]
                (108:\Rin)
                .. controls (0.12,0.72) and (0.28,0.32) .. (0.28,0)
                .. controls (0.28,-0.32) and (0.12,-0.72) .. (-108:\Rin);
            \node[BrickRed, above right=-1.2pt, font=\scriptsize] at (105:\Rin) {$\gamma$};

            \filldraw[black] (0,0) circle (1.6pt)
                node[above right=0.5pt, font=\scriptsize] {$0$};
            \node[PineGreen!90!black] at (0.65,-1.65) {\large $\cQ_1$};
        \end{scope}

\node at (10.05,0) {\Large $+$};

\begin{scope}[shift={(13.40,0)}, xscale=-1]
            \coordinate (wInnerLower) at ({-\halfTheta}:\Rin);
            \coordinate (wInnerUpper) at (\halfTheta:\Rin);
            \coordinate (wOuterUpper) at (\halfTheta:\Rout);
            \coordinate (wOuterLower) at ({-\halfTheta}:\Rout);

            \path[dotregion]
                (wInnerUpper) -- (wOuterUpper)
                arc[start angle=\halfTheta, end angle=-\halfTheta, radius=\Rout]
                -- (wInnerLower)
                arc[start angle=-\halfTheta, end angle=\halfTheta, radius=\Rin]
                -- cycle;

        \end{scope}
        \begin{scope}[shift={(13.40,0)}]
\draw[->, gray] (-2.35,0) -- (1.20,0)
                node[right, text=black] {$\Re$};
            \draw[->, gray] (0,-2.55) -- (0,2.55)
                node[above, text=black] {$\Im$};
        \end{scope}
        \begin{scope}[shift={(13.40,0)}, xscale=-1]

\draw[NavyBlue, very thick, dashed] (wInnerUpper) -- (wOuterUpper);
            \draw[NavyBlue, very thick, dashed] (wInnerLower) -- (wOuterLower);
            \draw[black, very thick]
                (wInnerLower) arc[start angle=-\halfTheta, end angle=\halfTheta, radius=\Rin];
            \draw[black, very thick]
                (wOuterLower) arc[start angle=-\halfTheta, end angle=\halfTheta, radius=\Rout];

\draw[BrickRed, very thick]
                (\halfTheta:1.40)
                .. controls (1.25,0.32) and (1.52,-0.22) .. ({-\halfTheta}:1.45);
            \node[BrickRed, above right=-1.5pt, font=\scriptsize] at (\halfTheta:1.38) {$\gamma$};

            \node[PineGreen!90!black] at (1.45,-1.65) {\large $\cQ_2$};
        \end{scope}

\node at (16.00,0) {\Large $+$};

\begin{scope}[shift={(19.50,0)}, xscale=-1]
            \path[dotregion, even odd rule]
                (-2.35,-2.35) rectangle (2.35,2.35)
                (0,0) circle (\Rout);

        \end{scope}
        \begin{scope}[shift={(19.50,0)}]
\draw[->, gray] (-2.55,0) -- (2.55,0)
                node[right, text=black] {$\Re$};
            \draw[->, gray] (0,-2.55) -- (0,2.55)
                node[above, text=black] {$\Im$};
        \end{scope}
        \begin{scope}[shift={(19.50,0)}, xscale=-1]

\draw[NavyBlue, very thick, dashed]
                (\halfTheta:\Rout) arc[start angle=\halfTheta, end angle={360-\halfTheta}, radius=\Rout];
            \draw[black, very thick, {Bar[width=3.5pt]}-{Bar[width=3.5pt]}]
                ({-\halfTheta}:\Rout) arc[start angle=-\halfTheta, end angle=\halfTheta, radius=\Rout];

\draw[BrickRed, very thick]
                (52:\Rout)
                .. controls (1.72,1.42) and (2.10,0.60) .. (2.10,0)
                .. controls (2.10,-0.60) and (1.72,-1.42) .. (-52:\Rout);
            \node[BrickRed, below left=-0.5pt, font=\scriptsize] at (1.72,-1.15) {$\gamma$};

            \node[PineGreen!90!black] at (-1.65,-1.75) {\large $\cQ_3$};
        \end{scope}
    \end{tikzpicture}
    \caption{
        Visualization of the decomposition used in the proof of \cref{thm:polar-rect-modulus}.
        We decompose \(\cQ = \widehat\bbC\setminus\cP_{\kappa,\theta}\) into regions \(\cQ_1\), \(\cQ_2\), and \(\cQ_3\).
        For each region \(\cQ_i\), we look at the subset of cross cuts \(\Gamma_i\subset\Gamma_{\cQ}(0,\infty)\) that lie entirely on \(\cQ_i\), and visualize an example \(\gamma\in\Gamma_i\) above.
        We have \(\rM_{\cQ}(0,\infty) \geq \rM_{\cQ_1}(\Gamma_1) + \rM_{\cQ_2}(\Gamma_2) + \rM_{\cQ_3}(\Gamma_3)\).
        For \(\cQ_1\), we look at cross cuts which separate 0 from the black circular arc, recovering \cref{impthm:circular-escape}.
        For \(\cQ_2\), we look at cross cuts which separate the left and right sides of a channel.
        Up to a conformal map, this is equivalent to the rectangle studied in \cref{lem:rectangle-modulus}.
        For \(\cQ_3\), we look at cross cuts separating \(\infty\) from another circular arc.
        Up to a conformal map, this is equivalent to \cref{impthm:circular-escape}.
}
    \label{fig:modulus-decomp}
\end{figure}
 
\begin{proof}[Proof of \cref{thm:polar-rect-modulus}]
    We begin by decomposing \(\cQ\) into three regions as shown in \cref{fig:modulus-decomp}.
    We will apply \cref{lem:parallel-rule} to these regions.
    Let \(\Gamma_1\), \(\Gamma_2\), and \(\Gamma_3\) be the subsets of \(\Gamma = \Gamma_{\cQ}(0,\infty)\) that respectively lie on the regions \(\cQ_1\), \(\cQ_2\), and \(\cQ_3\):
    \[
        \Gamma_i = \bigl\{
            \gamma\in\Gamma ~:~ \gamma \subset \cQ_i
        \bigr\}.
    \]
    We can apply the superadditivity rule \cref{lem:parallel-rule} to say
    \begin{equation}
        \label{eq:polar-mod-decomp}
        \rM_{\cQ}(\Gamma) \geq \rM_{\cQ_1}(\Gamma_1) + \rM_{\cQ_2}(\Gamma_2) + \rM_{\cQ_3}(\Gamma_3).
    \end{equation}
    Our goal now shifts to lower bound these three separate moduli.
    We proceed casewise.

    \textbf{Region 1.~}
    Up to scaling by a factor of \(\kappa\), the region \(\cQ_1=\{z~:~|z|<\kappa^{-1}\}\) is the unit disk \(\cD\).
    So, up to scaling, \(\Gamma_1\) is the set of cross cuts of \(\cD\) that separate 0 from \(\cC_\theta\), where \(\cC_\theta\) is the set defined in \cref{impthm:circular-escape}.
    That is, \(\rM_{\cQ_1}(\Gamma_1) = \rM_{\cD}(\Gamma_{\cD}(0,\cC_\theta)) \geq \frac1\pi \log(2\pi/\theta)\).

    \textbf{Region 2.~}
    The next region is \(\cQ_2=\{-r\e^{\ri \phi}~:~r\in(\kappa^{-1},1),\, \phi\in(-\theta/2,\theta/2)\}\).
    The set of cross cuts \(\Gamma_2\subseteq\Gamma\) is the set of curves on \(\cQ_2\) that separate the inner radial arc \(\{-\kappa^{-1} \e^{\ri \phi} ~:~ \phi\in(-\theta/2,\theta/2)\}\) from the outer radial arc \(\{-\e^{\ri \phi} ~:~ \phi\in(-\theta/2,\theta/2)\}\).
    The map \(\psi(z) = \log(-\kappa z)+\ri \, \theta/2\) is conformal on \(\cQ_2\) and we can recognize
    \[
        \psi(\cQ_2) = \bigl\{ x+\ri\phi ~:~ x\in(0,\log\kappa),\, \phi\in(0,\theta) \bigr\}
    \]
    as a rectangle with width \(\log\kappa\) and height \(\theta\).
    Similarly, the image of the cross cuts under \(\psi\) is the set of all curves that connect the bottom of the rectangle and the top of the rectangle.
    By the conformal invariance of the modulus of cross cuts (\cref{impthm:conformal-mods-cuts}) and the modulus of cuts across the rectangle (\cref{lem:rectangle-modulus}) we therefore have
    \(
        \rM_{\cQ_2}(\Gamma_2)
        = \rM_{\psi(\cQ_2)}(\psi(\Gamma_2))
        = \frac{\log\kappa}{\theta}
    \).

    \textbf{Region 3.~}
    The third region \(\cQ_3=\{z~:~|z|>1\}\) consists of the outside of the circle of radius \(1\), and the set \(\Gamma_3\) is the set of cross cuts on \(\cQ_3\) that separate the radial arc \(\{-\e^{\ri \phi}~:~|\phi|<\theta/2\}\) from \(\infty\).
Note that the map \(\psi(z) = 1/z\) conformally maps \(\cQ_3\) to the unit disk \cD and maps \(\Gamma_3\) to the exact set of cross cuts used in \cref{impthm:circular-escape}.
    So, by the conformal invariance (\cref{impthm:conformal-mods-cuts}) and the modulus bound for circular escape (\cref{impthm:circular-escape}), we find that \(\rM_{\cQ_3}(\Gamma_3) = \rM_{\psi(\cQ_3)}(\psi(\Gamma_3)) \geq \frac1\pi \log(2\pi/\theta)\).

    Combining these three bounds with \cref{eq:polar-mod-decomp} completes the proof.
\end{proof}

We have at last assembled the ingredients needed to prove \cref{thm:green-upper-polar-rect}, the upper bound on the Green's function needed to control the residual error.

\begin{proof}[Proof of \cref{thm:green-upper-polar-rect}.]
For \(\theta < 2\pi\), the combination of \cref{thm:modulus-bounds-greens} and \cref{thm:polar-rect-modulus} tells us that
\[
    G_{\cP_{\kappa,\theta}}(0)
    \leq 10 \cdot \e^{-\frac{\pi}{\theta}\log\kappa - 2\log(1/\theta) - 2\log
    (2\pi)}
    = \tfrac{5}{2\pi^2} \cdot \theta^2 \kappa^{-\pi/\theta}
\]
so long as \(\rM_{\cQ}(0,\infty) \geq 1\).
For any \(\theta\in(0,2\pi)\), this is guaranteed to hold by \cref{thm:polar-rect-modulus} if \(\kappa \geq 536\).
To recover the result for \(\theta=2\pi\), take the limit as \(\theta\to2\pi\).
\end{proof}

 \section{Matrix--vector lower bounds and complex polynomial approximation}\label{sec:resid-lower-bound}

In this section we prove \cref{thm:lower-bound-forward}, which shows that the complexity of finding a small-residual solution to a linear system with eigenvalues in a set \cS must exceed the degree $t$ required to create a filter polynomial on \cS with norm $\varepsilon$.
For the polar rectangle, lower bounds on the degree are provided by \cref{thm:greens-function-bound-residual,thm:poly-to-greens-intro}.

\restatetheorem{thm:lower-bound-forward}

To get from polynomial approximation to complexity, we can employ the \emph{Krylov simulation argument} for randomized algorithms.
This argument was originally developed by Narayanan and coauthors~\cite{chewi2024query,bakshi2023krylov}, extended and simplified by Derezi\'nski et al.~\cite{derezinski2026matrix}, and further simplified by Shamir~\cite{shamir2026simple}.
The core of the argument is that a general algorithm which computes \(t\) matvecs is no more powerful than a randomized block Krylov method which has block size \(t\) and depth \(t\).
The existing results all treat real symmetric matrices and therefore only consider real eigenvalues.
We generalize the arguments from Shamir's approach \cite{shamir2026simple} to normal matrices with complex eigenvalues.
Our proof very closely tracks that of Shamir; we include the full argument for completeness.

\subsection{Setup}

We begin by presenting Shamir's version of the Krylov simulation argument, adapted to complex matrices.
Consider a linear algebra problem involving a matrix $\mA \in \bbC^{n\times n}$ and a vector $\vb \in \bbC^n$.
We are interested in proving complexity lower bounds for the solution of the problem \(\mA\vx=\vb\) by one-sided matrix--vector algorithms, which are allowed to be randomized and adaptive.

By Yao's minimax principle, it is sufficient to prove a lower bound against (adaptive) deterministic one-sided algorithms for a single hard random distribution of inputs.
In this argument, we construct the hard instance as $\mA = \mU\mLambda\mU^*$, where $\mLambda \in \bbC^{n\times n}$ is a diagonal matrix that is known to the algorithm and $\mU$ is a Haar-random unitary matrix.
We shall prove lower bounds against algorithms that access the more powerful oracle
\begin{equation} \label{eq:shamir}
    \vz \mapsto (\mA\vz, \mU^*\vz).
\end{equation}

\subsection{The Krylov simulation lemma}\label{sub:haarhard}

The core result in Shamir's version of the Krylov simulation argument is the following lemma.
    
\begin{lemma}[Krylov simulation \protect{\cite[Lemma 4]{shamir2026simple}}]\label{lem:gaussianquery}
    Let $\mLambda \in \bbC^{n\times n}$ be diagonal and fixed, let $\mU \in \bbC^{n\times n}$ be a Haar-random unitary matrix, and let $\vb \in \bbC^n$ be a standard complex Gaussian vector.
    Consider an adaptive deterministic algorithm that, on input $(\mLambda,\vb)$, makes $t-1$ queries to the oracle given by \cref{eq:shamir} for \(\mA=\mU\mLambda\mU^*\) and outputs a vector $\tilde\vx$.
    Then there exist standard complex Gaussian vectors $\vg_0,\ldots,\vg_t$, independent of one another, such that $\mU^*\vb = \vg_0$ and
    \[
        \mU^*\mA\tilde\vx \in \Span \big\{\mLambda^j \vg_i ~:~ 1\le j\leq t, 0 \leq i \leq t \big\}.
    \]
\end{lemma}
\noindent 
The proof of \cite[Lem.~4]{shamir2026simple} carries over to the complex case unmodified, so we omit the proof.

\subsection{Choosing the set of eigenvalues}\label{sub:weightedapprox}

To complete the proof, we must populate the diagonal matrix \(\mLambda\in\bbC^{n \times n}\) with entries from \cS that are challenging to make small by a filter polynomial.
The following optimality result for filter polynomials is useful towards this end.

\begin{importedtheorem}[Filter polynomials: Optimality conditions, \protect{\cite[Thm.~10.3]{kroo2010strong}}] \label{impthm:filter-optimality}
    Let $\cS \subseteq \bbC$ be nonempty and compact, and consider the optimization problem
\begin{equation*}
    \min_{\deg(q)\le t} \norm{q}_{\cS} \quad \text{such that} \quad q(0) = 1.
    \end{equation*}
A polynomial $q$ of degree at most $t$ with $q(0) = 1$ is an optimal solution to this problem if and only if there are $1 \le k \le 2t+1$ locations $\lambda_1,\ldots,\lambda_k \in \cS$ and convex combination coefficients $w_1,\ldots,w_k$ such that
\begin{equation} \label{eq:filter-optimality-1}
        |q(\lambda_i)| = \norm{q}_{\cS} \quad \text{for each } i=1,\ldots,k
    \end{equation}
and
\begin{equation} \label{eq:filter-optimality-2}
        \sum_{i=1}^k w_i \, \overline{p(\lambda_i)} \, q(\lambda_i) = 0 \quad \text{for every polynomial $p$ with $\deg(p) \le t$ and $p(0) = 0$}.
    \end{equation}
\end{importedtheorem}

This result is a specific case of a more general theorem; see \cite[Thm.~10.3]{kroo2010strong} for the general result.
Using \cref{impthm:filter-optimality}, we obtain a continuous-to-discrete conversion principle, which is a direct generalization of \cite[Lem.~2]{shamir2026simple}. 

\begin{lemma}[From continuous to discrete]\label{lem:polypoints}
    Let \cS be nonempty and compact, and let $t\ge 0$.
    Then there exist $k\le 2t+1$ points $\lambda_1,\ldots,\lambda_k \in \cS$ and convex combination coefficients $w_1,\ldots,w_k$ such that
\begin{equation} \label{eq:polypoints}
        \min \bigl\{ \norm{q}_{\cS}^2 ~:~ \deg(q) \le t, \: q(0) = 1 \bigr\} = \min \biggl\{ \sum_{i=1}^k w_i |q(\lambda_i)|^2 ~:~ \deg(q) \le t, \: q(0) = 1 \biggr\}.
    \end{equation}
\end{lemma}

\begin{proof}
    Let $\{\lambda_i\}_{i=1}^k \subseteq \cS$ be the points provided by \cref{impthm:filter-optimality}, and let $\{w_i\}_{i=1}^k$ be the corresponding weights.
    The expression $\sum_{i=1}^k w_i |q(\lambda_i)|^2$ is always smaller than $\norm{q}_{\cS}^2$, so the left-hand side of \cref{eq:polypoints} is always smaller than the right-hand side.
    We establish the reverse inequality.

    Let $q_\star$ be the optimal solution to the left-hand side of \cref{eq:polypoints}, and consider any admissible polynomial $q_\star + p$ for a degree-$t$ polynomial $p$ with $p(0) = 0$.
    Then 
\begin{multline*}
        \sum_{i=1}^k w_i |q_\star(\lambda_i) + p(\lambda_i)|^2 = \sum_{i=1}^k w_i \overline{[q_\star(\lambda_i) + p(\lambda_i)]}[q_\star(\lambda_i) + p(\lambda_i)] \\ = \sum_{i=1}^k w_i |q_\star(\lambda_i)|^2 + \cancelto{0}{2 \Re \left[\sum_{i=1}^k w_i \overline{p(\lambda_i)}q_\star(\lambda_i)\right]} + \sum_{i=1}^k w_i |p(\lambda_i)|^2 \ge \sum_{i=1}^k w_i |q_\star(\lambda_i)|^2 = \norm{q_\star}_{\cS}^2.
    \end{multline*}
The middle term vanished by \cref{eq:filter-optimality-2}.
    The final equality comes from \cref{eq:filter-optimality-1}.
\end{proof}

\subsection{Proof of \texorpdfstring{\cref{thm:lower-bound-forward}}{Theorem \ref*{thm:lower-bound-forward}}} 
\label{sub:lowerboundconstruct}

We are now ready to prove \cref{thm:lower-bound-forward}.
We note some basic concentration bounds we shall use, which can easily be proven using, e.g., \cite[Lem.~1]{laurent2000adaptive}:

\begin{fact}[Norm of a Gaussian vector]\label{fact:normofgauss}
    Let $\vg\in\bbC^n$ be a standard complex Gaussian vector with $n \gtrsim 1$.
    Then \[\mathbb{P}\left\{ \|\vg\|_2^2 \geq 2n \right\} \leq \frac{1}{n^2}
    \quad\text{ and }\quad
    \mathbb{P}\left\{ \|\vg\|_2^2 \leq n/2 \right\} \leq \frac{1}{n^2}.\] 
\end{fact}

\begin{proof}[Proof of \cref{thm:lower-bound-forward}]

    Invoking \cref{lem:polypoints}, we know that every polynomial \(q\) of degree at most \(t\) and \(q(0)=1\) must have \(\sum_{j=1}^k w_j |q(\lambda_j)|^2 \geq \eps^2\) for some nonnegative weights \(w_1,\ldots,w_k\) and nodes \(\lambda_1,\ldots,\lambda_k\in\cS\), where \(k \leq 2t+1\).
    For a universal constant \(\rC \geq 1\), define the constants \(N = \rC^2 t^2\) and the multiplicities
    \[
        m_j \defeq \lceil N w_j + \rC t \rceil.
    \]
    Let \(n = \sum_{j=1}^k m_j\) and construct the diagonal matrix \(\mLambda\in\bbC^{n \times n}\) to have eigenvalues \(\lambda_1,\ldots,\lambda_k\) with multiplicities \(m_1,\ldots,m_k\), respectively.

    We construct \(\mA=\mU\mLambda\mU^*\) where \mU is a Haar-random unitary matrix, and let \(\vb\in\bbC^n\) be a standard complex Gaussian vector.
    Consider any fixed deterministic one-sided matvec algorithm that computes \(t-1\) matvecs with \mA and returns a vector \(\tilde\vx\).
    By \cref{lem:gaussianquery}, we know there exist standard complex Gaussian vectors \(\vg_0,\ldots,\vg_t\) such that \(\mU^*\vb = \vg_0\) and
    \[
        \mLambda\mU^*\tilde\vx = \sum_{i=0}^t p_i(\mLambda)\vg_i
        \quad
        \text{for some }
        \deg(p_i) \leq t
        \text{ and }
        p_i(0)=0.
    \]
    The constraint that \(p_i(0)=0\) is a consequence of the exponent on \mLambda always being at least 1 in \cref{lem:gaussianquery}.
    
    Now let \(\mPi_j\) be the projection onto the eigenspace of \mLambda associated with eigenvalue \(\lambda_j\).
    Define the sub-vectors
    \(\vg_{i,j} \defeq \mPi_j \vg_i\), and let \(\mPi'_j\) be the orthogonal projection \textit{away} from \(\Span\{\vg_{1,j},\ldots,\vg_{t,j}\}\).
    Further define \(\vh_j \defeq \mPi'_j \vg_{0,j}\).
By applying a union bound on \cref{fact:normofgauss}, we know that
    \[
        \norm{\vg_0}_2^2 \lesssim n
        \quad\text{and}\quad
        \norm{\vh_j}_2^2 \gtrsim m_j
        \quad
        \text{for all } j\in\{1,\ldots,k\}
    \]
    with probability at least \(2/3\).
    We now bound the residual.
    Expand
    \[
        \norm{\mA\tilde\vx-\vb}_2^2
        = \norm{\mLambda\mU^*\tilde\vx - \vg_0}_2^2
        = \norm*{(p_0(\mLambda)-\mI)\vg_0 + \sum_{i=1}^t p_i(\mLambda)\vg_i}_2^2
        = \sum\limits_{j=1}^k \norm*{(p_0(\lambda_j)-1)\vg_{0,j} + \sum_{i=1}^t p_i(\lambda_j)\vg_{i,j}}_2^2.
    \]
    Project the vector inside of the norm of each summand using \(\mPi_j'\).
    Since orthogonal projections reduce norms, and using that \(m_j \geq N w_j\) and the fact that \(q(z) \defeq 1-p_0(z)\) has \(q(0)=1\) and degree at most \(t\),
    \[
        \norm{\mA\tilde\vx-\vb}_2^2
        \geq \sum\limits_{j=1}^k \norm{(p_0(\lambda_j)-1)\vh_{j}}_2^2
        \gtrsim \sum\limits_{j=1}^k m_j |q(\lambda_j)|^2
        \geq N \sum\limits_{j=1}^k w_j |q(\lambda_j)|^2
        \geq \eps^2 N.
    \]
    We conclude by observing that \(\norm{\vb}_2^2 \lesssim n \lesssim N\), and therefore that \(\norm{\mA\tilde\vx-\vb}_2 \gtrsim \eps \norm{\vb}_2\).
\end{proof}

 \section{Unbounded spectral condition numbers: The GMBACK algorithm}\label{sec:backward}

In this section, we prove \cref{thm:gmberr}, our complexity result for the GMBACK algorithm \cite{kasenally1995gmback,kasenally1997analysis} for solving a linear system with unbounded spectral condition numbers to small backward error.
Conceptually, the GMBACK algorithm generates the Krylov subspace $\cK_t(\mA,\vb)$, as defined in \cref{eq:krylov}, and outputs the vector $\tilde\vx \in \cK_t(\mA,\vb)$ achieving the smallest backward error, as defined in \cref{eq:backward-error}.
We defer to the original papers \cite{kasenally1995gmback,kasenally1997analysis} for the details of the algorithm and its implementation.
Other Krylov subspace algorithms that seek to minimize the backward error or a related quantity include MINPERT \cite{kasenally1997analysis}, CGBACK \cite{gratton2010minimizing}, and LSMB \cite{hallman2018lsmb}.

Our first main theorem reduces the analysis of GMBACK to a question of polynomial approximation.
The result and its analysis closely mirrors the analysis of the MINBERR algorithm for real symmetric matrices \cite[Sec.~4]{derezinski2026towards}.

\restatetheorem{thm:gmberr}

\begin{proof}
    Let \(p\) be the polynomial guaranteed to exist by the premise.
    Since \(p\) has degree at most \(t_{\rm b}\), the vector \(\vy = p(\mA)\vb \in\cK_{t_{\rm b}+1}(\mA,\vb)\) lives in the degree-$(t_{\rm b}+1)$ Krylov subspace.
    By the hypothesis on $p$, it holds that \(p(\lambda_i) \neq 0\) for all \(i\), so the matrix \(p(\mA)\) is invertible.

    Let \(\widehat\mA = p(\mA)^{-1}\), so that \(\widehat\mA\vy = \vb\).
    We bound the backward error of \(\vy\) as
    \[
        \norm{\widehat\mA - \mA}_2
        = \norm{\mX(p(\mLambda)^{-1} - \mLambda)\mX^{-1}}_2
        \leq \cond(\mX) \max_{i} \left|\lambda_i - \frac1{p(\lambda_i)}\right|
        \leq \eps\cond(\mX)\norm{\mA}_2.
    \]
Thus, $\vy$ is an $\eps\cond(\mX)$-backward error solution to $\mA\vx = \vb$.
    Consequently, since GMBACK outputs the solution in \(\cK_{t_{\rm b}+1}(\mA,\vb)\) that achieves the smallest backward error, GMBACK must return an $\eps\cond(\mX)$-backward error solution as well, completing the proof.
\end{proof}

With this theorem in hand, the analysis of GMBACK reduces to demonstrating the existence of a polynomial $p$ with the required properties when the angular eigenvalue gap is $\measuredangle(\mA) \ge \theta$.
Equivalently, by scaling by an appropriate complex number, we can assume that the eigenvalues lie in the \sector $\cT_\theta$, defined above in \cref{sec:intro-backward}; see Panel~(b) of \cref{fig:univalent-proof}.
This is the content of \cref{thm:poly-backward-upper}, which we shall spend the rest of the section proving.

\restatetheorem{thm:poly-backward-upper}

The proof will consist of two parts.
In \cref{sec:backward-poly-to-greens}, we relate polynomial approximation to the Green's function.
In \cref{sec:backward-green}, we bound the Green's function.

\subsection{Approximation of the inverse via the Green's function}\label{sec:backward-poly-to-greens}

To analyze GMBACK for a matrix \mA, we must exhibit a low-degree polynomial $p$ for which $1/p(\lambda) \approx \lambda$ for each eigenvalue of \mA. 
As before, we let \cS denote a set of eigenvalues, which we assume for this section to be compact and simply connected.
Additionally, since we are interested in the unbounded spectral condition number case, we assume that $0 \in \cS$ lies in the set of admissible eigenvalues.

A technical challenge is that, since $0$ lies in the set of admissible eigenvalues, the Green's function satisfies $G_{\cS}(0) = 0$.
Consequently, we cannot directly use the Green's function at zero to bound the polynomial approximation error.
Fortunately, the value of the Green's function \emph{just outside} the set of admissible eigenvalues will provide us the information we need. 
We have the following result.

\begin{theorem}[Green's function controls polynomial approximation, backward variant]
    \label{thm:backward-green}
    Let \(\cS\subseteq\bbC\) be a compact and simply connected region, and let \(R = \max_{z\in\cS}|z|\).
    Suppose further that \cS contains at least two distinct elements, that \(0\in\cS\), and that \(\,0\!\) lies on the boundary of \cS.
    Let \(G_{\cS}(z)\) be the Green's function of \cS.
    For any \(\eps \in (0,1)\), there exists a polynomial \(p\) of degree \(t_{\rm b}\) such that
    \[
        \lrnorm{z - \frac{1}{p(z)}}_{\cS} \leq \eps\norm{z}_{\cS}
        \quad\text{and}\quad
        t_{\rm b} \lesssim 1 + \frac1{\max\limits_{\substack{z\notin\cS:\\|z|\leq \eps R/2}} G_{\cS}(z)}.
    \]
\end{theorem}
\noindent In particular, to prove \cref{thm:poly-backward-upper}, it shall suffice to show that there exists a single point \(z\) with \(|z|\leq\eps/2\) such that \(G_{\cT_\theta}(z) \gtrsim \theta \eps^{\pi/\theta}\).

The proof will first construct a polynomial approximation $p$ to the function $(z-w)^{-1}$ for some $w$ that is close to, but not contained in, the set \cS.
Then, we will show this polynomial satisfies $1/p(z) \approx z$ on \cS.
The core lemma is as follows.

\begin{lemma}[Polynomial approximation to \((z-w)^{-1}\)]
    \label{cor:approx-1/z-w}
    Let \(\cS\subseteq\bbC\) be a compact, simply connected set, and assume \(w\notin\cS\).
    There exists a polynomial \(p\) of degree \(t \lesssim 1 + \frac{1}{G_{\cS}(w)}\) such that
    \begin{equation}
        \label{eq:univalent-approx-cor}
        \biggl|
            p(z) - \frac{1}{z-w}
        \biggr|
        \leq \frac19 \cdot \min_{v \in \cS} \biggl|\frac{w-v}{z-v}\biggr| \cdot \biggl|\frac{1}{z-w}\biggr|
        \qquad
        \text{for all }
        z\in\cS.
    \end{equation}
\end{lemma}

We shall prove this lemma by appealing to the theory of \emph{univalent functions}.
Univalent functions are very structured, and they possess remarkably strong approximation theoretic properties.
Our proof follows an established approach, but our instantiation of it appears to be novel.
For earlier work, see the paper \cite{abdullayev2020exact}, which uses univalent function theory to analyze the extremal properties of Faber polynomials; see also \cite[Sec.~2]{mattei2021explicit}.
We will interact with this theory only through the following result.

\begin{importedtheorem}[Uniform approximation of univalent functions, \protect{\cite[Lems.~4 and 6]{andrievskii1994maximal}}]
    \label{impthm:univalent-approx}
    Consider a family of univalent functions $\{f_z : \cD \to \bbC\}$ parametrized by a number $z \in \cS$ taking values in a set $\cS \subseteq \bbC$.
    Moreover, assume $f_z(0) = 0$ for each $z \in \cS$.
    Then, for every \(\zeta\in\cD\), there exist \(m \lesssim (1-|\zeta|)^{-1}\) coefficients \(\beta_1,\ldots,\beta_{m}\in\bbC\) such that the following bound holds for all $z \in \cS$:
    \begin{equation}
        \label{eq:univalent-approx}
        \left|
            \sum_{j=1}^{m} \beta_j a_j(z) - f_z(\zeta)
        \right|
        \leq \frac19 \min_{s \notin f_z(\cD)} |f_z(\zeta) - s|,
    \end{equation}
where $a_j(z)$ denote the coefficients of the power series expansion $f_z(u) =\sum_{j=1}^\infty a_j(z) u^j$ for each $z \in \cS$.
\end{importedtheorem}

The conclusion of this result is strong.
For given $m$ and $\zeta$, the same set of coefficients \(\beta_j\) achieves the approximation guarantee \cref{eq:univalent-approx} simultaneously for \emph{every} univalent function in the family.

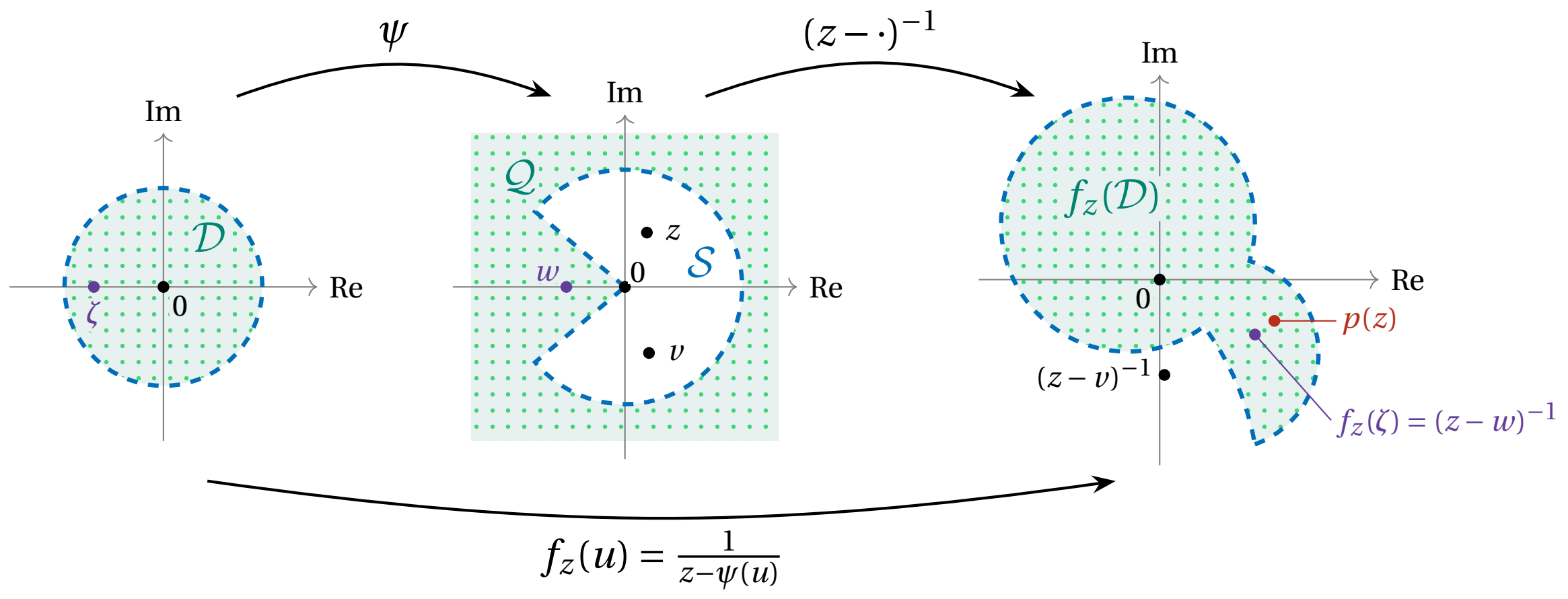
\begin{figure}[t]
    \centering
\begin{tikzpicture}[
        scale=0.72,
        every node/.style={font=\footnotesize},
        dotregion/.style={
            fill=PineGreen!8,
            postaction={
                pattern={Dots[distance=4.5pt, radius=0.60pt]},
                pattern color=PineGreen!65
            }
        },
        backing/.style={fill=PineGreen!8, inner sep=0.8pt, rounded corners=1.2pt},
        axis/.style={->, gray},
        pt/.style={black},
        map/.style={->, >=Stealth, thick, black},
    ]
        \colorlet{zetacolor}{RoyalPurple}  \def\Rdisk{1.35}  \def\Rsec{1.6}    

\begin{scope}
            \path[dotregion] (0,0) circle (\Rdisk);
            \draw[axis] (-2.1,0) -- (2.1,0) node[right, text=black] {$\Re$};
            \draw[axis] (0,-2.1) -- (0,2.1) node[above, text=black] {$\Im$};
            \draw[NavyBlue, very thick, dashed] (0,0) circle (\Rdisk);

            \filldraw[pt] (0,0) circle (2pt);
            \node[backing, below right=2pt] at (0,0) {$0$};
            \coordinate (zetaA) at (-0.95,0);
            \filldraw[zetacolor] (zetaA) circle (2pt);
            \node[backing, text=zetacolor, below=3pt] at (zetaA) {$\zeta$};
            \node[backing, text=PineGreen!90!black] at (0.62,0.68) {\large $\cD$};
        \end{scope}

\begin{scope}[shift={(6.3,0)}]
            \path[dotregion, even odd rule]
                (-2.1,-2.1) rectangle (2.1,2.1)
                (0,0) -- (140:\Rsec) arc[start angle=140, end angle=-140, radius=\Rsec] -- cycle;
            \draw[axis] (-2.35,0) -- (2.35,0) node[right, text=black] {$\Re$};
            \draw[axis] (0,-2.35) -- (0,2.35) node[above, text=black] {$\Im$};
            \draw[NavyBlue, very thick, dashed]
                (0,0) -- (140:\Rsec) arc[start angle=140, end angle=-140, radius=\Rsec] -- cycle;

            \filldraw[pt] (0,0) circle (2pt);
            \node[above right=1pt, inner sep=0.5pt, fill=white] at (0,0) {$0$};
            \coordinate (wB) at ({-0.5*\Rsec},0);
            \filldraw[zetacolor] (wB) circle (2pt);
            \node[backing, text=zetacolor, above left=1pt] at (wB) {$w$};
            \filldraw[pt] (68:{0.5*\Rsec}) circle (2pt);
            \node[right=1.5pt] at (68:{0.5*\Rsec}) {$z$};
            \filldraw[pt] (-70:{0.6*\Rsec}) circle (2pt);
            \node[right=1.5pt] at (-70:{0.6*\Rsec}) {$v$};
            \node[text=NavyBlue] at (18:1.1) {\large $\cS$};
            \node[backing, text=PineGreen!90!black] at (-1.45,1.5) {\large $\cQ$};
        \end{scope}

\begin{scope}[shift={(13.6,0.1)}, scale=1.3] \path[dotregion] plot coordinates {(1.000,-1.732) (1.123,-1.678) (1.240,-1.606) (1.346,-1.520) (1.440,-1.420) (1.518,-1.310) (1.580,-1.193) (1.624,-1.072) (1.653,-0.952) (1.667,-0.834) (1.667,-0.721) (1.655,-0.615) (1.634,-0.517) (1.605,-0.426) (1.571,-0.344) (1.532,-0.270) (1.491,-0.204) (1.447,-0.145) (1.403,-0.094) (1.359,-0.048) (1.315,-0.008) (1.271,0.027) (1.229,0.058) (1.188,0.085) (1.149,0.108) (1.111,0.128) (1.074,0.146) (1.040,0.161) (1.006,0.174) (0.974,0.185) (0.944,0.195) (0.954,0.232) (0.964,0.271) (0.973,0.312) (0.981,0.354) (0.988,0.399) (0.994,0.446) (0.997,0.495) (1.000,0.547) (1.000,0.601) (0.998,0.658) (0.993,0.717) (0.985,0.779) (0.973,0.844) (0.957,0.911) (0.937,0.982) (0.912,1.054) (0.880,1.129) (0.842,1.206) (0.797,1.285) (0.743,1.364) (0.681,1.443) (0.609,1.521) (0.526,1.597) (0.432,1.669) (0.328,1.735) (0.212,1.794) (0.085,1.843) (-0.051,1.880) (-0.196,1.904) (-0.347,1.911) (-0.502,1.900) (-0.658,1.871) (-0.812,1.822) (-0.959,1.755) (-1.098,1.670) (-1.225,1.569) (-1.337,1.455) (-1.434,1.330) (-1.513,1.199) (-1.575,1.063) (-1.620,0.926) (-1.650,0.790) (-1.664,0.657) (-1.666,0.530) (-1.656,0.409) (-1.636,0.295) (-1.609,0.189) (-1.575,0.091) (-1.535,0.000) (-1.492,-0.082) (-1.446,-0.158) (-1.397,-0.226) (-1.348,-0.288) (-1.297,-0.344) (-1.246,-0.394) (-1.196,-0.440) (-1.146,-0.480) (-1.096,-0.516) (-1.048,-0.549) (-1.000,-0.577) (-0.954,-0.603) (-0.908,-0.626) (-0.864,-0.646) (-0.821,-0.663) (-0.780,-0.679) (-0.739,-0.693) (-0.700,-0.705) (-0.662,-0.715) (-0.625,-0.724) (-0.589,-0.731) (-0.555,-0.737) (-0.521,-0.743) (-0.488,-0.747) (-0.456,-0.750) (-0.426,-0.753) (-0.396,-0.755) (-0.367,-0.756) (-0.338,-0.756) (-0.311,-0.756) (-0.284,-0.755) (-0.258,-0.754) (-0.232,-0.752) (-0.207,-0.750) (-0.183,-0.747) (-0.159,-0.745) (-0.136,-0.741) (-0.114,-0.738) (-0.092,-0.734) (-0.070,-0.730) (-0.049,-0.725) (-0.028,-0.721) (-0.007,-0.716) (0.013,-0.710) (0.032,-0.705) (0.052,-0.699) (0.071,-0.693) (0.089,-0.687) (0.108,-0.681) (0.126,-0.674) (0.144,-0.668) (0.161,-0.661) (0.179,-0.654) (0.196,-0.646) (0.213,-0.639) (0.230,-0.631) (0.246,-0.623) (0.263,-0.615) (0.279,-0.607) (0.295,-0.599) (0.311,-0.590) (0.327,-0.581) (0.342,-0.572) (0.358,-0.563) (0.373,-0.553) (0.389,-0.544) (0.404,-0.534) (0.419,-0.523) (0.434,-0.513) (0.449,-0.502) (0.464,-0.491) (0.464,-0.491) (0.473,-0.504) (0.483,-0.517) (0.493,-0.531) (0.503,-0.545) (0.514,-0.561) (0.526,-0.577) (0.537,-0.594) (0.550,-0.612) (0.562,-0.632) (0.576,-0.652) (0.590,-0.674) (0.605,-0.697) (0.620,-0.722) (0.636,-0.749) (0.653,-0.778) (0.670,-0.809) (0.689,-0.842) (0.708,-0.878) (0.728,-0.917) (0.749,-0.960) (0.772,-1.006) (0.795,-1.057) (0.819,-1.113) (0.844,-1.175) (0.869,-1.243) (0.896,-1.320) (0.923,-1.405) (0.949,-1.501) (0.976,-1.609)} -- cycle;
            \draw[axis] (-1.9,0) -- (2.3,0) node[right, text=black] {$\Re$};
            \draw[axis] (0,-1.95) -- (0,2.15) node[above, text=black] {$\Im$};
            \draw[NavyBlue, very thick, dashed] plot coordinates {(1.000,-1.732) (1.123,-1.678) (1.240,-1.606) (1.346,-1.520) (1.440,-1.420) (1.518,-1.310) (1.580,-1.193) (1.624,-1.072) (1.653,-0.952) (1.667,-0.834) (1.667,-0.721) (1.655,-0.615) (1.634,-0.517) (1.605,-0.426) (1.571,-0.344) (1.532,-0.270) (1.491,-0.204) (1.447,-0.145) (1.403,-0.094) (1.359,-0.048) (1.315,-0.008) (1.271,0.027) (1.229,0.058) (1.188,0.085) (1.149,0.108) (1.111,0.128) (1.074,0.146) (1.040,0.161) (1.006,0.174) (0.974,0.185) (0.944,0.195) (0.954,0.232) (0.964,0.271) (0.973,0.312) (0.981,0.354) (0.988,0.399) (0.994,0.446) (0.997,0.495) (1.000,0.547) (1.000,0.601) (0.998,0.658) (0.993,0.717) (0.985,0.779) (0.973,0.844) (0.957,0.911) (0.937,0.982) (0.912,1.054) (0.880,1.129) (0.842,1.206) (0.797,1.285) (0.743,1.364) (0.681,1.443) (0.609,1.521) (0.526,1.597) (0.432,1.669) (0.328,1.735) (0.212,1.794) (0.085,1.843) (-0.051,1.880) (-0.196,1.904) (-0.347,1.911) (-0.502,1.900) (-0.658,1.871) (-0.812,1.822) (-0.959,1.755) (-1.098,1.670) (-1.225,1.569) (-1.337,1.455) (-1.434,1.330) (-1.513,1.199) (-1.575,1.063) (-1.620,0.926) (-1.650,0.790) (-1.664,0.657) (-1.666,0.530) (-1.656,0.409) (-1.636,0.295) (-1.609,0.189) (-1.575,0.091) (-1.535,0.000) (-1.492,-0.082) (-1.446,-0.158) (-1.397,-0.226) (-1.348,-0.288) (-1.297,-0.344) (-1.246,-0.394) (-1.196,-0.440) (-1.146,-0.480) (-1.096,-0.516) (-1.048,-0.549) (-1.000,-0.577) (-0.954,-0.603) (-0.908,-0.626) (-0.864,-0.646) (-0.821,-0.663) (-0.780,-0.679) (-0.739,-0.693) (-0.700,-0.705) (-0.662,-0.715) (-0.625,-0.724) (-0.589,-0.731) (-0.555,-0.737) (-0.521,-0.743) (-0.488,-0.747) (-0.456,-0.750) (-0.426,-0.753) (-0.396,-0.755) (-0.367,-0.756) (-0.338,-0.756) (-0.311,-0.756) (-0.284,-0.755) (-0.258,-0.754) (-0.232,-0.752) (-0.207,-0.750) (-0.183,-0.747) (-0.159,-0.745) (-0.136,-0.741) (-0.114,-0.738) (-0.092,-0.734) (-0.070,-0.730) (-0.049,-0.725) (-0.028,-0.721) (-0.007,-0.716) (0.013,-0.710) (0.032,-0.705) (0.052,-0.699) (0.071,-0.693) (0.089,-0.687) (0.108,-0.681) (0.126,-0.674) (0.144,-0.668) (0.161,-0.661) (0.179,-0.654) (0.196,-0.646) (0.213,-0.639) (0.230,-0.631) (0.246,-0.623) (0.263,-0.615) (0.279,-0.607) (0.295,-0.599) (0.311,-0.590) (0.327,-0.581) (0.342,-0.572) (0.358,-0.563) (0.373,-0.553) (0.389,-0.544) (0.404,-0.534) (0.419,-0.523) (0.434,-0.513) (0.449,-0.502) (0.464,-0.491) (0.464,-0.491) (0.473,-0.504) (0.483,-0.517) (0.493,-0.531) (0.503,-0.545) (0.514,-0.561) (0.526,-0.577) (0.537,-0.594) (0.550,-0.612) (0.562,-0.632) (0.576,-0.652) (0.590,-0.674) (0.605,-0.697) (0.620,-0.722) (0.636,-0.749) (0.653,-0.778) (0.670,-0.809) (0.689,-0.842) (0.708,-0.878) (0.728,-0.917) (0.749,-0.960) (0.772,-1.006) (0.795,-1.057) (0.819,-1.113) (0.844,-1.175) (0.869,-1.243) (0.896,-1.320) (0.923,-1.405) (0.949,-1.501) (0.976,-1.609)} -- cycle;

            \coordinate (cw) at (1.000,-0.577);
            \coordinate (cv) at (0.05,-1.001);
\def\rRed{0.42}
\coordinate (fzC) at (cw);
\coordinate (pz) at ($(cw)+(35:0.25)$);

            \filldraw[pt] (0,0) circle (1.54pt);
            \node[backing, below left=2pt] at (0,0) {$0$};
            \filldraw[pt] (cv) circle (1.54pt);
            \node[left=2pt, fill=white, inner sep=0.8pt, rounded corners=1.2pt] at (cv) {$(z-v)^{-1}$};
            \filldraw[zetacolor] (cw) circle (1.54pt);
            \draw[zetacolor] (cw) -- ++(0.8,-0.9)
                node[right, text=zetacolor, fill=white, inner sep=1pt] {$f_z(\zeta) = (z-w)^{-1}$};
            \filldraw[BrickRed] (pz) circle (1.54pt);
            \draw[BrickRed] (pz) -- ++(0.65,0.0)
                node[right, inner sep=1pt, fill=white, rounded corners=1.2pt] {$p(z)$};
            \node[backing, text=PineGreen!90!black] at (-0.5,0.85) {\large $f_z(\cD)$};
        \end{scope}

\draw[map] (1.0,2.6) to[bend left=20] node[midway, above] {\large $\psi$} (5.3,2.6);
        \draw[map] (7.4,2.6) to[bend left=20]
            node[midway, above] {\large $(z-\cdot)^{-1}$} (11.9,2.6);
        \draw[map] (0.6,-2.65) to[bend right=8]
            node[midway, below] {\large $f_z(u) = \frac{1}{z-\psi(u)}$} (13.0,-2.65);
    \end{tikzpicture}
    \caption{
        Visualization of the various elements at play in the proof of \cref{cor:approx-1/z-w}.
        We are given a point \(w\) in the complement domain \(\cQ = \widehat\bbC \setminus \cS\).
        \(\psi\) maps the inside of the unit disk \cD to \cQ, and \(\zeta\) is defined to be the preimage of \(w\) under \(\psi\).
For each fixed point \(z \in \cS\), we construct the univalent function \(1/(z-w)\) with domain \cQ.
        The function \(f_z(u) \defeq 1/(z-\psi(u))\) is therefore a univalent function with domain \cD.
        We then use \cref{impthm:univalent-approx} to construct a polynomial \(p\), with the guarantee that the error \(|p(z) - 1/(z-w)|\) is 9 times smaller than the distance \(|(z-w)^{-1} - (z-v)^{-1}|\) for all \(v\in\cS\).
    }
    \label{fig:univalent-proof}
\end{figure}
 
With this result in hand, we prove \cref{cor:approx-1/z-w}.

\begin{proof}[Proof of \cref{cor:approx-1/z-w}]
    Let \(\chi\) be the normalized conformal map from \(\cQ \defeq \widehat\bbC \setminus \cS\) to the outside of the unit disk \(\widehat\bbC\setminus \overline\cD\).
    Then \(\psi(u)\defeq \chi^{-1}(1/u)\) defines a conformal map from \cD to \cQ. Choose $\zeta \in \cD$ such that $\psi(\zeta) = w$.
    For each \(z\in\cS\), define the following univalent function on the unit disk:
    \[
        f_{z}(u) 
        \defeq \frac{1}{z-\psi(u)},
        \quad
        \text{so that }
        f_z(0)=0
        \text{ and }
        f_z(\zeta) = \frac{1}{z-w}.
    \]
See \cref{fig:univalent-proof} for an illustration of the setup.

    We shall construct the polynomial approximation $(z-w)^{-1}$ by using the power series coefficients of the functions in the univalent family $\{f_z\}_{z\in\cS}$.
    Namely, since each function \(f_z\) is analytic and satisfies \(f_z(0)=0\), it may be expanded as a power series \(f_z(u) = \sum_{j=1}^\infty a_j(z) u^j\).
    Note that the coefficients \(a_j(z)\) are themselves polynomials of degree at most \(j\).
    Indeed, since \(\chi\) is a normalized conformal map, it holds that \(\psi(0) = \chi^{-1}(\infty) =\infty\) and
\begin{equation*}
        \lim_{u\to0} u\psi(u) = \lim_{u\to\infty} u/\chi(u) = \alpha\quad \text{for some real } \alpha>0.
    \end{equation*}
In other words, \(\psi\) has a simple pole at 0, so we may decompose it as
    \[
        \psi(u) = \frac\alpha u + h(u)
        \quad
        \text{
            for some analytic function \(h\).
        }
    \]
    Therefore,  we can expand $h$ as a power series \(h(u) = \sum_{\ell=0}^\infty c_\ell u^\ell\) and write
    \[
        f_z(u)
        = \frac1{z-\frac \alpha u - h(u)}
        = -\frac u\alpha \frac1{1  - \frac u\alpha (z - h(u))}
        = -\frac u\alpha \sum_{k=0}^{\infty} \left(\frac u\alpha (z-h(u))\right)^k.
    \]
    Hence, if we substitute in the power series for \(h\) into this expression and collect terms, we see that the coefficient \(a_j(z)\) in front of \(u^j\) is a polynomial in \(z\) of degree at most \(j\).

    To create the polynomial approximation, we invoke \cref{impthm:univalent-approx}, giving us coefficients $\beta_1,\ldots,\beta_m$ for some \(m \lesssim (1-|\zeta|)^{-1}\).
We have that the polynomial
    \[
        p(z) \defeq \sum_{j=1}^{m} \beta_j a_j(z)
    \]
satisfies
\begin{equation} \label{eq:univalent-conclusion}
        \biggl| p(z) - \frac{1}{z-w} \biggr| \leq \frac19 \biggl| \frac{1}{z-w} - s\biggr|
        \quad \text{for all } z\in\cS \text{ and } s\notin f_z(\cD).
    \end{equation}
Since \(|\zeta| = \e^{-G_{\cS}(w)}\), the degree of this polynomial is \(m \lesssim (1-|\zeta|)^{-1} \lesssim 1 + 1/G_{\cS}(w)\).

    Now let $v \in \cS$.
    By construction, $s \coloneqq (z-v)^{-1} \notin f_z(\cD)$, so \cref{eq:univalent-conclusion} yields
    \[
        \biggl| p(z) - \frac{1}{z-w} \biggr| \leq \frac19 \biggl| \frac{1}{z-w} - \frac{1}{z-v}\biggr| = \frac19 \biggl|\frac{v-w}{z-v}\biggr| \cdot \biggl|\frac{1}{z-w}\biggr|.
    \]
    Conclude the proof by minimizing over $v$.
\end{proof}

With \cref{cor:approx-1/z-w} in tow, completing the proof of \cref{thm:backward-green} is now straightforward.

\begin{proof}[Proof of \cref{thm:backward-green}.]
    Fix any point \(|w| \leq \eps R/2\) with \(G_{\cS}(w) \geq \frac12 \max_{|z|\leq \eps R /2} G_{\cS}(z)\), and set $f(z) \coloneqq (z-w)^{-1}$.
    \Cref{cor:approx-1/z-w} furnishes a polynomial of degree \(t_{\rm b} \lesssim 1 + 1/G_{\cS}(w)\) for which
    \begin{equation}
        \label{eq:backward-poly-univalent}
        \left|p(z) - f(z) \right| \leq \frac{1}{9} \cdot \min_{v\in \cS}\biggl|\frac{w-v}{z-v}\biggr| \cdot \left|f(z)\right|
        \quad
        \text{for all } z\in\cS.
    \end{equation}
Taking \(v=0\) and rearranging gives
    \begin{equation*}
        |z||p(z) - f(z)| \leq \frac19 |w||f(z)|
        \quad \text{for all } z\in\cS.
    \end{equation*}
    Further, choosing $v = 0$ when $|z| \ge R/2$ and taking $|v| = R$ when $|z| \le R/2$ gives 
\begin{equation*}
        \left|p(z) - f(z) \right| \leq \frac{1}{3}\left|f(z)\right|
        \quad
        \text{for all } z\in\cS.
    \end{equation*}
Invoking the identity $(z-w)f(z) = 1$, we conclude
    \[
        \left|z - \frac1{p(z)}\right|
        = \frac{|zp(z)-1|}{|p(z)|}
= \frac{|z(p(z)-f(z))+w f(z)|}{|p(z)|}
        \leq \frac{\frac{10}9|w||f(z)|}{\frac23 |f(z)|}
        \leq \frac{5}{6} \eps R < \eps \norm{z}_{\cS}.\qedhere 
    \]
\end{proof}

\subsection{Bounding the new Green's function}
\label{sec:backward-green}

Having produced an upper bound on the degree of approximation \(t_{\rm b}\) in terms of the Green's function in \cref{sec:backward-poly-to-greens}, our goal is now to explicitly bound the Green's function.
Fortunately, the machinery we used to lower bound the Green's function in \cref{sec:harmonic} immediately yields a result for the \sector.

\begin{theorem}[Green's function lower bound for the circular sector]
    \label{thm:harmonic-backward}
    Let \(\theta\in(0,2\pi]\) and \(\eps\in(0,1)\).
    Then, the Green's function for the circular sector \(\cT_{\theta}\) has \(G_{\cT_{\theta}}(-\eps) \geq (\pi\e)^{-1}\theta\eps^{\pi/\theta}\).
\end{theorem}

\begin{proof}
    Evaluate \cref{thm:greens-to-wedge} at \(z=-\eps\), yielding
    \(
        G_{\cT_\theta}(-\eps) \geq \frac{1}{\pi \e} \theta \eps^{\pi/\theta}.
    \)
\end{proof}

Combining \cref{thm:backward-green,thm:harmonic-backward} proves the backward error upper bound in \cref{thm:poly-backward-upper}.

 \section*{AI methodology}

Independent of any AI tool use, the authors became interested in this research problem and developed a proof showing that the asymptotic rate of convergence for GMRES was $\exp(-\Theta(\kappa)^{\pi/\theta})$.
Afterwards, the authors used Gemini 3 Pro and 4 Argon as well as ChatGPT 5.6, 6.0, and 6.1 models to sharpen this proof outline into non-asymptotic results with explicit constants.
Essentially all final results in this paper were improved by the use of AI, and the tools from geometric function theory were new to the authors and surfaced to us by working with AI systems.
The paper is written by the authors.
AI tools were further used for identifying and correcting mistakes, line editing, and generating diagrams.
The authors have verified the correctness and originality of all content, including references, and take full responsibility for the paper's contents.

  \section*{Acknowledgments}

We thank Chris Cama\~no, Micha\l\  Derezi\'nski, and Anne Greenbaum for valuable conversations.
We especially thank Nick Trefethen for providing us with a very nice note explaining the $\Theta(\theta^{-2})$ complexity of GMRES for eigenvalues in a circular arc.
CC is supported by the National Science Foundation Graduate Research Fellowship under Grant No. 2146752.
ENWE is supported by the Miller Institute for Basic Research in Science at the University of California Berkeley. 
MWM and RAM would like to acknowledge the DARPA DIAL and DARPA AIQ programs for providing partial support of this work.
RAM and ENWE's research was supported in part by a grant of access to OpenAI models through the ChatGPT for Academic Researchers program. JU thanks Alex Cohen for many conversations regarding logarithmic potential theory, harmonic functions, and a much cruder version of \cref{thm:greens-function-bound-residual}. This material is based upon work supported by the National Science Foundation under grant no. DMS-2513687.

\bibliographystyle{halpha}
{\footnotesize\newcommand{\etalchar}[1]{$^{#1}$}

 }

\end{document}